\documentclass[11pt]{article}
\usepackage{amsthm}
\usepackage{enumerate}
\usepackage{color}
\usepackage{mathtools}
\usepackage{amsmath}
\usepackage{amssymb}
\usepackage{amsfonts}
\usepackage{siunitx}
\usepackage[utf8]{inputenc} 
\usepackage[T1]{fontenc}    
\usepackage{hyperref}       
\usepackage{url}            
\usepackage{booktabs}       
\usepackage{amsfonts}       
\usepackage{nicefrac}       
\usepackage{microtype}      
\usepackage{lipsum}
\usepackage{fancyhdr}       
\usepackage{graphicx,graphbox}       
\usepackage{algorithm}
\usepackage{algpseudocode}
\usepackage{caption}
\usepackage{subcaption}
\usepackage{cite}
\usepackage{lineno}
\usepackage{soul}
\usepackage{multirow}
\usepackage[normalem]{ulem}
\usepackage{CJKutf8}

\UseRawInputEncoding
\let\Item\item

\begin{document}

\centerline{{\LARGE Recurrent neural chemical reaction networks trained to}}
\centerline{{\LARGE switch dynamical behaviours through learned bifurcations}}

\medskip
\bigskip

\centerline{
{\renewcommand{\thefootnote}{\arabic{footnote}}
\large Alexander Dack
\renewcommand{\thefootnote}{\arabic{footnote}}\footnote[1]{
  Department of Bioengineering and Imperial College Centre for Synthetic Biology, Imperial College London, Exhibition Road, London SW7 2AZ, UK. Email of co-corresponding authors: alex.dack14@imperial.ac.uk or  t.ouldridge@imperial.ac.uk. }
\qquad
Tomislav Plesa\footnote[2]{
Department of Applied Mathematics and Theoretical Physics, University of Cambridge, Centre for Mathematical Sciences, Wilberforce Road, Cambridge, CB3 0WA, UK.
}
\qquad 
Thomas E. Ouldridge\renewcommand{\thefootnote}{\arabic{footnote}}\footnotemark[1]
}
}
\medskip
\bigskip

\noindent
{\bf Abstract}: Both natural and synthetic chemical systems not only exhibit a range of non-trivial dynamics, but also transition between qualitatively different dynamical behaviours as environmental parameters change. Such transitions are called bifurcations. Here, we show that recurrent neural chemical reaction networks (RNCRNs), a class of chemical reaction networks based on recurrent artificial neural networks that can be trained to reproduce a given dynamical behaviour, can also be trained to exhibit bifurcations. First, we show that RNCRNs can inherit some bifurcations defined by smooth ordinary differential equations (ODEs). Second, we demonstrate that the RNCRN can be trained to infer bifurcations that allow it to approximate different target behaviours within different regions of parameter space, without explicitly providing the bifurcation itself in the training. These behaviours can be specified using target ODEs that are discontinuous with respect to the parameters, or even simply by specifying certain desired dynamical features in certain regions of the parameter space. To achieve the latter, we introduce an ODE-free algorithm for training the RNCRN to display designer oscillations, such as a heart-shaped limit cycle or two coexisting limit cycles.

\section{Introduction}

It is common to observe biological entities transition between dynamical behaviours as they process information from their environment~\cite{lev_bar-or_generation_2000} or as they proceed through their developmental cycle~\cite{manser_mathematical_2024}. For example, it is well-known in cellular biology that cells differentiate, i.e. change type, to specialise into particular functions~\cite{cerneckis_induced_2024}. Expressed in the nomenclature of dynamical systems, cells are moving between different dynamical regimes capable of different behaviours that govern their function. In some dynamical systems, it is  external factors, called \textit{parameters}, that determine the dynamical regime. In biochemistry, such parameters may be interpreted as the reaction rate coefficients or the concentrations set by exogenous chemostat.

Switching of dynamical behaviours as parameters are changed is studied in a branch of mathematics called 
\textit{bifurcation theory}~\cite{strogatz_nonlinear_2018}.
This general theory has categorized the types of transitions into canonical forms~\cite{strogatz_nonlinear_2018} and been applied broadly to understand a range of biological phenomena including snapping mechanism in the Venus fly trap~\cite{forterre_how_2005}, predator-prey relationships in ecosystems~\cite{fussmann_crossing_2000}, and to model neural networks in the brain~\cite{borisyuk_bifurcation_1992, erdi_dynamics_1993, izhikevich_multiple_1998,  ma_multistability_2009, eisenmann_bifurcations_2023}. Of particular relevance to this work is the investigation of cellular regulation at the biochemical level through the lens of bifurcation theory~\cite{palsson_mathematical_1988, borisuk_bifurcation_1998, Kar_2009_explore, marco_bifurcation_2014,ferrell_bistability_2012}.

Aside from being used to model natural biological phenomena,
dynamical systems are also used in the 
fields of synthetic biology and nucleic-acid nanotechnology, 
which have been successful at engineering biological systems with desired dynamical behaviours. Examples include 
neural network classification algorithms~\cite{qian_neural_2011, kieffer_molecular_2023, samaniego_neural_2024, van_der_linden_dna_2022, cherry_scaling_2018, xiong_molecular_2022,
lopez_molecular_2018, baltussen_chemical_2024} and attractor orbits~\cite{stricker_fast_2008, fujii_predatorprey_2013, srinivas_enzyme-free_2017}, which have been built 
in a range of molecular media. Given that individual dynamical behaviours have been realised successfully, a natural question arises:  can one combine these components to engineer biological systems with the complicated switching behaviour, or bifurcations, seen in the native systems?

In this context, on the one hand, 
some research 
has focused on eliminating undesirable bifurcations 
in engineered systems~\cite{bury_deep_2021, deb_machine_2022},
and studying how sudden and unexpected changes in dynamical behaviours  can lead to catastrophic system failures~\cite{demazure_catastrophe_2000}. For example, such bifurcations are known to lead to hazardous events in synthetic biology control systems~\cite{plesa_integral_2023}.
On the other hand, there has also been a growing interest in bifurcation-inspired design paradigms~\cite{yang_bifurcation_2018, yang_bifurcation_2023}, given that bifurcations are potentially a highly efficient solution to achieving large changes in output for only a small change in input. Indeed, there has been both experimental~\cite{genot_high-resolution_2016, lobato-dauzier_neural_2024} and theoretical work~\cite{otero-muras_optimization-based_2018, otero-muras_automated_2023, szep_parameter_2021, fucho-rius_local_2024, melot_multi-parametric_2024, otero-muras_method_2014, banaji_bifurcations_2024} investigating the construction of bifurcations in biochemical systems. 
In particular, mathematical methods for designing 
\textit{chemical reaction networks} (CRNs), mathematical models of chemistry, that implement a given bifurcation 
are developed in~\cite{plesa_chemical_2016, plesa_mapping_2024}. All of these methods take as 
input a sufficiently smooth dynamical system with desired
bifurcation structure, which is then converted into a CRN~\cite{kerner_universal_1981, samardzija_nonlinear_1989, kowalski_universal_1993, poland_cooperative_1993,  hangos_mass_2011, plesa_test_2017, plesa_integral_2023}.

In this paper, we present methods to construct CRNs that use bifurcations to interpolate between regions of desired dynamics, for which only the desired dynamics need to be specified ahead of training. In particular, we base our methods on the \textit{recurrent neural chemical reaction network} (RNCRN) - a class of CRNs that operates as a recurrent neural network, whose fast-acting chemical perceptrons can be trained to approximate arbitrary dynamics~\cite{dack_recurrent_2025}.  

We first demonstrate that the RNCRN with sufficiently many species can be adapted for training to reproduce classical bifurcations from a given smooth parameter-dependent dynamical system, in line with the capabilities of more minimal methods~\cite{plesa_chemical_2016, plesa_mapping_2024}. We then demonstrate that the RNCRN can be further adapted and trained to approximate and merge different behaviours from a given dynamical system which depends on its parameters discontinuously. Finally, we introduce an algorithm for training the RNCRN to approximate limit cycles specified only by the shape of the limit cycle itself, rather than via ODEs that generate a limit cycle, which enables the construction of a variety of behaviours from sparse data. Motivated by the success of chemical classifiers~\cite{hjelmfelt_chemical_1991, kim_neural_2004, chiang_reconfigurable_2015, poole_chemical_2017, anderson_reaction_2021, linder_robust_2021, nagipogu_neuralcrns_2024, vasic_programming_2022, moorman_dynamical_2019}, we use this algorithm to develop an extended RNCRN that can sense its local environment, perform a non-linear classification, and then toggle between dynamical regimes accordingly. We note that these discontinuous and sparsely defined target systems showcase the RNCRN's ability to extrapolate outside, and interpolate between, training data to create a self-consistent vector field that matches the intended dynamical behaviour.

This paper is structured in five parts. In Section~\ref{sec:background}, some background theory is presented. In Section~\ref{sec:inherit_bifs_main} and Appendix~\ref{sec:app_bif}, we demonstrate that RNCRNs can approximate bifurcations of a given smooth dynamical system. In Section~\ref{sec:dynamical_piecewises}, we design RNCRNs that approximate piecewise systems, i.e. non-smooth 
systems that have different dynamical behaviours 
according to some externally imposed condition.
Finally, in Section~\ref{sec:data_defined_dynamics} and Appendix~\ref{app:data_defined_attractors}, we demonstrate that RNCRNs can be used to construct dynamical behaviours entirely from relatively sparse data and demonstrate switching between dynamical regimes in this context.

\section{Background}
\label{sec:background}
{\bf Chemical reaction networks (CRNs).} 
A \emph{chemical reaction network} (CRN) under mass-action kinetics
is a set of chemical species which interact via a set of reactions whose speed is parametrized by rate coefficients~\cite{feinberg_lecture_1979}.
This mathematical abstraction of chemistry
is commonly used to model (bio)chemical systems in synthetic biology and DNA nanotechnology, including toggle switches in gene regulatory networks~\cite{gardner_construction_2000}, chemotaxis in E. coli~\cite{yi_robust_2000} and biological pattern formation~\cite{turing_chemical_1952}. 

For example, CRN 
\begin{align}
    X_1 \xrightarrow[]{\alpha_1} \varnothing, \;\;\; X_1 \xrightarrow[]{\alpha_2} X_1 + X_1, \;\;\; X_1 + X_1 \xrightarrow[]{\alpha_3} \varnothing, 
    \label{eq:an_exampe_reaction}
\end{align}
describes a single chemical species $X_1$ that is involved in degradation, auto-catalytic production and self-annihilation
with the (dimensionless) rate coefficients respectively given by 
$\alpha_1, \alpha_2, \alpha_3 > 0$, and where 
$\varnothing$ represents some chemical species 
that are of no interest. Under the assumption that chemical reactions occur in a well-mixed container, and that the chemical species 
are in high abundance, CRNs can be modelled deterministically with mass-action \emph{reaction-rate equations} (RREs) - a system of ordinary-differential equations (ODEs) with polynomials on the right-hand side~\cite{feinberg_lecture_1979}. For example, CRN~(\ref{eq:an_exampe_reaction}) has the corresponding RRE
\begin{align}
    \frac{\mathrm{d}x_1}{\mathrm{d}t} &= (\alpha_2-\alpha_1) x_1 - \alpha_3x_1^2, 
    \label{eq:example_RRE}
\end{align}
where $x_1=x_1(t) \geq 0$ is the time-varying molecular concentration of the chemical species $X_1$.

{\bf Recurrent neural chemical reaction networks (RNCRNs).}  In prior work, the authors have introduced a CRN based on artificial neural networks (ANNs), called the \emph{recurrent neural chemical reaction network} (RNCRN)~\cite{dack_recurrent_2025}, 
and used it to approximate the dynamics of well-behaved ODE systems. In particular, given a target system of ODEs with suitably smooth right-hand side (vector field), given by
\begin{align}
    \frac{\mathrm{d} \bar{x}_i}{\mathrm{d}t} &= f_i(\bar{x}_1, \dots, \bar{x}_n), \;\; \bar{x}_i(0)  \geq 0, 
    \; \; \;  i = 1, 2, \dots, N,
    \label{eq:target_ODE}
\end{align}
there exists an RNCRN with the RREs
\begin{align}
\frac{\mathrm{d} x_i}{\mathrm{d} t} & = \beta_{i} +  x_i \sum_{j=1}^M \alpha_{i,j} y_j,  
&&x_i(0)\geq 0, 
\; \; \; i = 1, 2, \ldots, N, \nonumber \\
\frac{\mathrm{d} y_j}{\mathrm{d} t} & =  \frac{\gamma}{\mu }
+ y_j \left(\sum_{i=1}^N \frac{\omega_{j,i}}{\mu } x_i + \frac{\theta_{j}}{\mu } \right) -  \frac{1}{\mu}y_j^2, 
&&y_j(0) \geq 0, 
\; \; \; j = 1, 2, \ldots, M,
\label{eq:single_layer_RRE}
\end{align}
such that the concentration $x_i(t)$ 
approximates the target variable $\bar{x}_i(t)$ from~(\ref{eq:target_ODE}) arbitrarily well. 
In particular, parameters $\beta_{i}, \gamma > 0$ 
from~(\ref{eq:single_layer_RRE}) are positive, 
while $\alpha_{i,j}, \omega_{j,i}, \theta_{j} \in \mathbb{R}$ 
can take any real value; furthermore, parameter $\mu > 0$ 
controls how fast the variables $y_j(t)$ 
are with respect to $x_i(t)$. 
For suitable values of these parameters, with $\mu > 0$
sufficiently small, i.e. with $y_j(t)$ sufficiently fast
compared to $x_i(t)$, (\ref{eq:single_layer_RRE}) 
approximates a given target system~(\ref{eq:target_ODE})
arbitrarily well. In order to attempt to find such parameter values,
Algorithm 1 is presented in~\cite{dack_recurrent_2025}.

\begin{figure}[hbt!]
    \centering
    \includegraphics[trim={8.0cm 6.5cm 8.0cm 6.5cm}, clip,width=\linewidth]{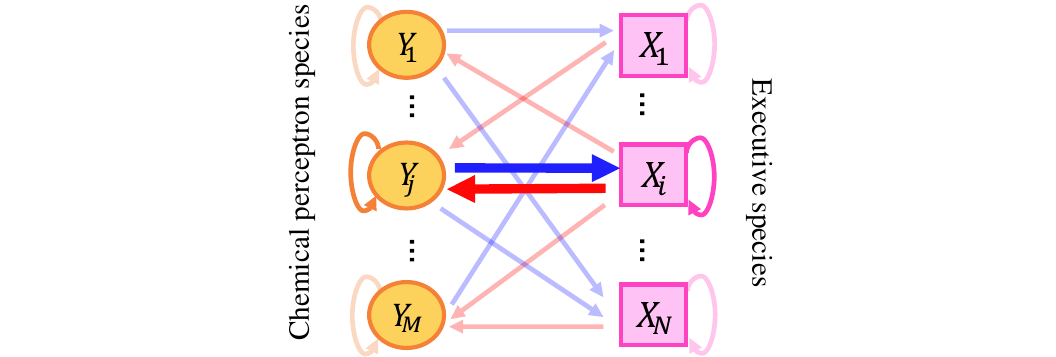}
    \caption{A graphical visualisation of the recurrent neural chemical reaction network (RNCRN) underlying the reaction-rate equations (RREs)~(\ref{eq:single_layer_RRE}), as introduced in~\cite{dack_recurrent_2025}. Orange circles represent chemical perceptron species $Y_1, \dots, Y_M$, while magenta squares represent the executive species $X_1, \dots, X_N$. Coloured arrows represent chemical reactions consistent with the RREs: blue arrows represent $X_i + Y_j \xrightarrow{|\alpha_{i,j}|} 2X_i + Y_j$ or $X_i + Y_j \xrightarrow{|\alpha_{i,j}| }Y_j$ (according to the sign of $\alpha_{i,j}$), red arrows represent $X_i + Y_j \xrightarrow{|\omega_{j,i}|/\mu} X_i + 2Y_j$ or $X_i + Y_j \xrightarrow{|\omega_{j,i}|/ \mu} X_i$ (according to the sign of $\omega_{j,i}$), magenta arrows represent $\varnothing \xrightarrow{\beta_i} X_i$, and orange arrows represent three reactions $\varnothing \xrightarrow{\gamma/ \mu} Y_j$, $2Y_j \xrightarrow{1/ \mu} Y_j$, and one of $Y_j \xrightarrow{|\theta_j|/ \mu} 2Y_j$ or $Y_j \xrightarrow{|\theta_j|/ \mu} \varnothing$ (according to the sign of $\theta_j$). 
  }
    \label{fig:intro_rncrn}
\end{figure}

The concentrations $x_i(t)$ and $y_j(t)$
respectively correspond to two 
sets of chemical species: the \emph{executive} species $X_i$ which manifest the target dynamical behaviour, and the \emph{chemical perceptron} species $Y_j$ which control the dynamics of the executive species through auxiliary reactions, which we illustate in Figure~\ref{fig:intro_rncrn}. Chemical perceptrons $Y_j$
are initially introduced in~\cite{anderson_reaction_2021}
and proposed as a chemical realisation of a feed-forward chemical neural network used for classification tasks, processing static input concentrations $x_i$ to reach an equilibrium concentration $y_j$ that corresponds to a classification decision. In the RNCRN, however, we introduce feedback reactions to the executive species $X_i$ (see caption of Figure \ref{fig:intro_rncrn}) that allow the output of a single layer of chemical perceptrons to dictate the dynamics of  the executive species as they all evolve together. 

The parameter $\mu  > 0$ in~(\ref{eq:single_layer_RRE}) controls the time-scale separation between the fast perceptrons and the slow executive species. Due to the stability of the underlying chemical perceptrons~\cite{anderson_reaction_2021}, in the formal
limit $\mu \to 0$, the dynamics are 
governed by the \textit{reduced} RNCRN with RREs
\begin{align}
\frac{\mathrm{d} \tilde{x}_i}{\mathrm{d} t} & = 
g_i(\tilde{x}_1,\ldots,\tilde{x}_N) = \beta_i + \tilde{x}_i \sum_{j=1}^M \alpha_{i,j} 
\sigma_{\gamma} \left(\sum_{k=1}^N \omega_{j,k} \tilde{x}_k + \theta_{j} \right), \;\; \tilde{x}(0) \geq 0,
\; \; \; i = 1, 2, \ldots, N,
\label{eq:single_layer_RRE_reduced}
\end{align}
where $\sigma_\gamma$ is a chemical activation function defined as $\sigma_\gamma(x) = (x+\sqrt{x^2+4\gamma})/2$. In this reduced limit, the chemical perceptrons operate as a conventional ANN with an activation function for each perceptron that approximates the commonly used ReLU activation function, allowing training  of the RNCRN's parameters by standard methods. 
Loosely, Algorithm 1 in~\cite{dack_recurrent_2025} uses standard machine learning methods to find an ANN such that $g_i(\tilde{x}_1,\ldots,\tilde{x}_N) \approx f_i(\tilde{x}_1, \dots, \tilde{x}_n)$, and then relaxes the assumption that $\mu=0$ to find a non-zero $\mu > 0$ such that the full RNCRN~(\ref{eq:single_layer_RRE}) approximates the target system~(\ref{eq:target_ODE}) 
to a sufficient accuracy.

{\bf Bifurcations.} In~(\ref{eq:target_ODE}), 
we have assumed that the target vector field 
$\mathbf{f} = (f_1,f_2,\ldots,f_N)$ does not depend
on any parameters. More broadly, we can 
relax this assumption by generalizing~(\ref{eq:target_ODE}) 
to include a real parameter $\bar{\lambda }$,
and consider the parameter-dependent target system
\begin{align}
    \frac{\mathrm{d} \bar{x}_i}{\mathrm{d}t} &= f_i(\bar{x}_1, \dots, \bar{x}_n; \bar{\lambda}), \; \; \; i = 1, 2, \dots, N,
    \label{eq:target_ODE_w_bif}
\end{align}
where $f_i$ depends on $\bar{\lambda }$ sufficiently smoothly.
The value of $\bar{\lambda}$ can have a profound impact on the dynamical behaviour of~(\ref{eq:target_ODE_w_bif}).  
In particular, a \emph{bifurcation}~\cite{strogatz_nonlinear_2018} is said to occur in~(\ref{eq:target_ODE_w_bif})  if, as $\bar{\lambda }$ is changed smoothly through a critical value $\bar{\lambda }=\bar{\lambda }^*$, the resulting long-time behaviour is qualitatively different below and above that critical value.

The example CRN~(\ref{eq:an_exampe_reaction}) demonstrates that bifurcations are present even in relatively simple molecular systems. In particular, by writing the difference in rate coefficient as a single parameter  $\bar{\lambda } = (\alpha_2 - \alpha_1)$, and setting $\alpha_3 = 1$ such that the other coefficients are defined relative to $\alpha_3$, the associated RRE~(\ref{eq:example_RRE}) is of the form~(\ref{eq:target_ODE_w_bif}). We observe that if $\bar{\lambda } > 0$ (i.e. $\alpha_2 > \alpha_1$), then molecular concentrations starting near zero, i.e. with $0 < x_1(0) \ll 1$, will equilibrate to $\bar{\lambda }$. However, if $\bar{\lambda } < 0$ (i.e. $\alpha_2 < \alpha_1$), then molecular concentrations starting near zero will equilibrate to zero. In the terminology of dynamical systems, the equilibrium $x_1^* = 0$ goes from stable to unstable 
as the parameter $\bar{\lambda }$ changes from negative to positive,
while the other equilibrium $x_1^* = \lambda$
then goes from unstable to stable. This change 
of stability is an example of a bifurcation; in fact, 
ODE~(\ref{eq:example_RRE}) is the normal form for \textit{transcritical bifurcation}~\cite{strogatz_nonlinear_2018}.

\section{Bifurcations with parametrized RNCRNs}
\label{sec:inherit_bifs_main}
The RNCRN is introduced in~\cite{dack_recurrent_2025}
to approximate target systems of the form~(\ref{eq:target_ODE}),
with vector fields independent of parameters. 
In this section, we generalize the RNCRN
to approximate parameter-dependent target systems. 
In particular, let us now consider as input data
the target system
\begin{align}
    \frac{\mathrm{d} \bar{x}_i}{\mathrm{d}t} &= f_i(\bar{x}_1, \dots, \bar{x}_n, \bar{\lambda}_1, \dots, \bar{\lambda}_L), \;\; &&\text{ for } i = 1, 2, \dots, N, \nonumber \\
    \frac{\mathrm{d} \bar{\lambda}_l}{\mathrm{d}t} &= 0, \;\; &&\text{ for } l = 1, 2, \dots, L,
    \label{eq:target_ODE_w_general_static_species}
\end{align}
whose vector field $\mathbf{f} = (f_1,f_2,\ldots,f_N)$
depends sufficiently smoothly on $L$ positive parameters
$\bar{\lambda}_1, \bar{\lambda}_2, \dots, \bar{\lambda}_L > 0$, which 
can be seen as a generalization of~(\ref{eq:target_ODE_w_bif});
note that, if the parameters are not positive, then one
can define new positive parameters by suitably shifting the old ones.
In~(\ref{eq:target_ODE_w_general_static_species}), we append the trivial ODEs for the parameters, allowing one to interpret them as auxiliary dependent variables. Provided that the initial
conditions for these auxiliary variables match the target parameter values, the ODEs with parametrized vector fields, and 
those with vector fields depending on the auxiliary 
variables subject to trivial ODEs, are equivalent. 

{\bf Parametrized RNCRNs.} Let us now present 
a generalized RNCRN which can approximate the dynamics of 
the parameter-dependent 
target systems, represented in the form~(\ref{eq:target_ODE_w_general_static_species}), 
over a range of parameter values. 
In particular, consider the \emph{parametrized} RNCRN
defined via the RREs
\begin{align}
    \frac{\mathrm{d}x_i}{\mathrm{d}t} &= \beta_i + x_i\sum_{j=1}^M\alpha_{i,j} y_j,   &&\text{ for } i = 1, \dots, N, \nonumber\\
    \frac{\mathrm{d}\lambda_k}{\mathrm{d}t} &= 0,  && \text{ for } l= 1, \dots, L, \nonumber \\
     \frac{\mathrm{d}y_j}{\mathrm{d}t} &= \frac{\gamma_j}{\mu} + \left(\sum_{i=1}^N\frac{\omega_{j,i}}{\mu}x_i + \sum_{l=1}^L\frac{\psi_{j,l}}{\mu}\lambda_l+ \frac{\theta_{j}}{\mu} \right)y_j - \frac{1}{\mu}y_j^2, && \text{ for } j = 1, \dots, M,
    \label{eq:single_layer_RNCRN_static_param}
\end{align}
where $x_i = x_i(t) \geq 0$, $y_j = y_j(t) \geq 0$,  $\lambda_l = \lambda_l(t) = \lambda_l(0) \geq 0$, $\beta_i, \gamma_j \geq0$ 
and $\alpha_{i,j}, \omega_{ji}, \theta_{j}, \psi_{j,l} \in \mathbb{R}$. 
In particular, compared to the original RNCRN~(\ref{eq:single_layer_RRE}), 
the parametrized RNCRN~(\ref{eq:single_layer_RNCRN_static_param}) contains additional executive species
$\Lambda_l$, called \textit{parameter-species}, 
which are static in the sense that they are
governed by trivial ODEs $\mathrm{d}\lambda_l/\mathrm{d}t = 0$.
Let us stress that, by representing parameters 
as species $\Lambda_l$ in~(\ref{eq:single_layer_RNCRN_static_param}), 
their values are controlled via initial concentrations,
as opposed to rate coefficients. 
Species $\Lambda_l$ influence chemical perceptrons
 $Y_j$, but not vice-versa, and do so via the
 following catalytic reactions:
\begin{align}
    &\Lambda_l + Y_j \xrightarrow[]{|\psi_{j,l}|/\mu} \Lambda_l + Y_j + \text{sign}(\psi_{j,l})Y_j, \; \; \; j = 1, 2, \dots, M, \;\; l= 1, 2, \dots, L,  
    \label{eq:crn_static_exec}
\end{align}
where, for brevity, $\text{sign}(\psi_{j,l})$ is used to indicate that the products contain two $Y_j$ molecules when $\psi_{j,l}$ is positive, and none when it is negative. The parametrized RNCRN can be trained with a slight modification to Algorithm~1 from~\cite{dack_recurrent_2025}; for completeness, 
we present this algorithm for~(\ref{eq:single_layer_RNCRN_static_param}) in Appendix~\ref{app:training_with_static_species}. 

Since parameters can be represented as dependent 
variables with trivial dynamics, as in~(\ref{eq:single_layer_RNCRN_static_param}), 
one may expect that the RNCRN, 
which can approximate dependent variables 
undergoing dynamics resulting from arbitrary smooth vector fields~\cite{dack_recurrent_2025}, 
can also reproduce some bifurcations. In what follows, 
we numerically demonstrate via examples that the RNCRN can approximate two classical bifurcations, namely a Hopf bifurcation~\cite{strogatz_nonlinear_2018} and a homoclinic bifurcation~\cite{Glendinning1987, plesa_mapping_2024}. 
We note that both of these types of bifurcations have been observed in biochemical contexts~\cite{borisuk_bifurcation_1998, Kar_2009_explore}.
\subsection{Hopf bifurcation}
\label{sec:hopf_bif}
Let us apply RNCRNs to approximate the target system
\begin{align}
    \frac{\mathrm{d} \bar{x}_1}{\mathrm{d} t} &= \left(\bar{\lambda}_1 - 2 - (\bar{x}_1-5)^2 - (\bar{x}_2-5)^2\right)(\bar{x}_1-5) -  (\bar{x}_2-5),\nonumber \\
     \frac{\mathrm{d} \bar{x}_2}{\mathrm{d} t} &= \left(\bar{\lambda}_1- 2 - (\bar{x}_1-5)^2 - (\bar{x}_2-5)^2\right)(\bar{x}_2-5) +  (\bar{x}_1-5), \nonumber \\
     \frac{\mathrm{d} \bar{\lambda}_1}{\mathrm{d} t} &= 0,
     \label{eq:target_hopf}
\end{align}
which undergoes a Hopf bifurcation~\cite{strogatz_nonlinear_2018}.
In particular, (\ref{eq:target_hopf}) is obtained from 
the normal form of a Hopf bifurcation by shifting the target equilibrium to $(\bar{x}_1, \bar{x}_2) = (5,5)$, and the critical parameter value, at which the Hopf bifurcation occurs, to $\bar{\lambda}_1^*=2$, 
so that the dynamics of interest are within the positive orthant.
The $(\bar{x}_1,\bar{x}_2)$ state-space of~(\ref{eq:target_hopf}) is displayed in Figure~\ref{fig:rncrn_inhert_bifs}(a) below, at and above the critical parameter value. In particular, when 
$\bar{\lambda}_1 = 1 < 2$, one can notice that the trajectories are attracted to the stable equilibrium $(\bar{x}_1, \bar{x}_2) = (5,5)$, 
while for $\bar{\lambda}_1 = 3 > 2$ they are attracted 
to an isolated periodic orbit, called a limit cycle, enclosing the 
now unstable equilibrium $(\bar{x}_1, \bar{x}_2) = (5,5)$.

One way to replicate this behaviour locally
in the parameter space is to train 
the original RNCRN~(\ref{eq:single_layer_RRE})
to approximate~(\ref{eq:target_hopf}) 
specifically at the bifurcation point $\bar{\lambda}=1$. 
The results of such a training are presented in Appendix~\ref{sec:app_hopf_bif_critical}, showing that 
the RNCRN numerically replicates the dynamics 
of~(\ref{eq:target_hopf}) for parameter values
in a small neighbourhood of $\bar{\lambda}_1^*=2$.
In particular, even though it is trained only at a single value 
$\bar{\lambda}_1^*=2$, the RNCRN automatically 
reproduces the transition from stable equilibrium 
to stable limit cycle locally in the parameter space. 

To replicate the dynamics of~(\ref{eq:target_hopf})
over a larger set in the parameter space, 
we train the parametrized RNCRN of the form~(\ref{eq:single_layer_RNCRN_static_param}) to approximate~(\ref{eq:target_hopf}) for all $1 \le \bar{\lambda}_1 \le 3$ 
with $M=5$ chemical perceptrons; 
see Appendix~\ref{sec:app_hopf_bif} for details.
We display the state-space of the resulting RNCRN in Figure~\ref{fig:rncrn_inhert_bifs}(b), demonstrating that 
qualitatively similar dynamics and bifurcation 
are inherited. Let us note that the RNCRN undergoes the 
Hopf bifurcation, not necessarily exactly at $\lambda_1 = 2$, 
but at a slightly different critical value, numerically found to be $\lambda_1 \approx 2.14$. This difference in the critical parameter values is a consequence of the approximating nature of the RNCRN, 
and can be reduced if the number and speed of chemical perceptrons is  increased. However, in this paper, we focus on the RNCRNs with a smaller number of chemical perceptrons that still produce adequate behaviours, as CRNs with more species are, in general, more challenging to build experimentally.

\subsection{Homoclinic bifurcation}
\begin{figure}[hbt!]
    \includegraphics[width=\linewidth]{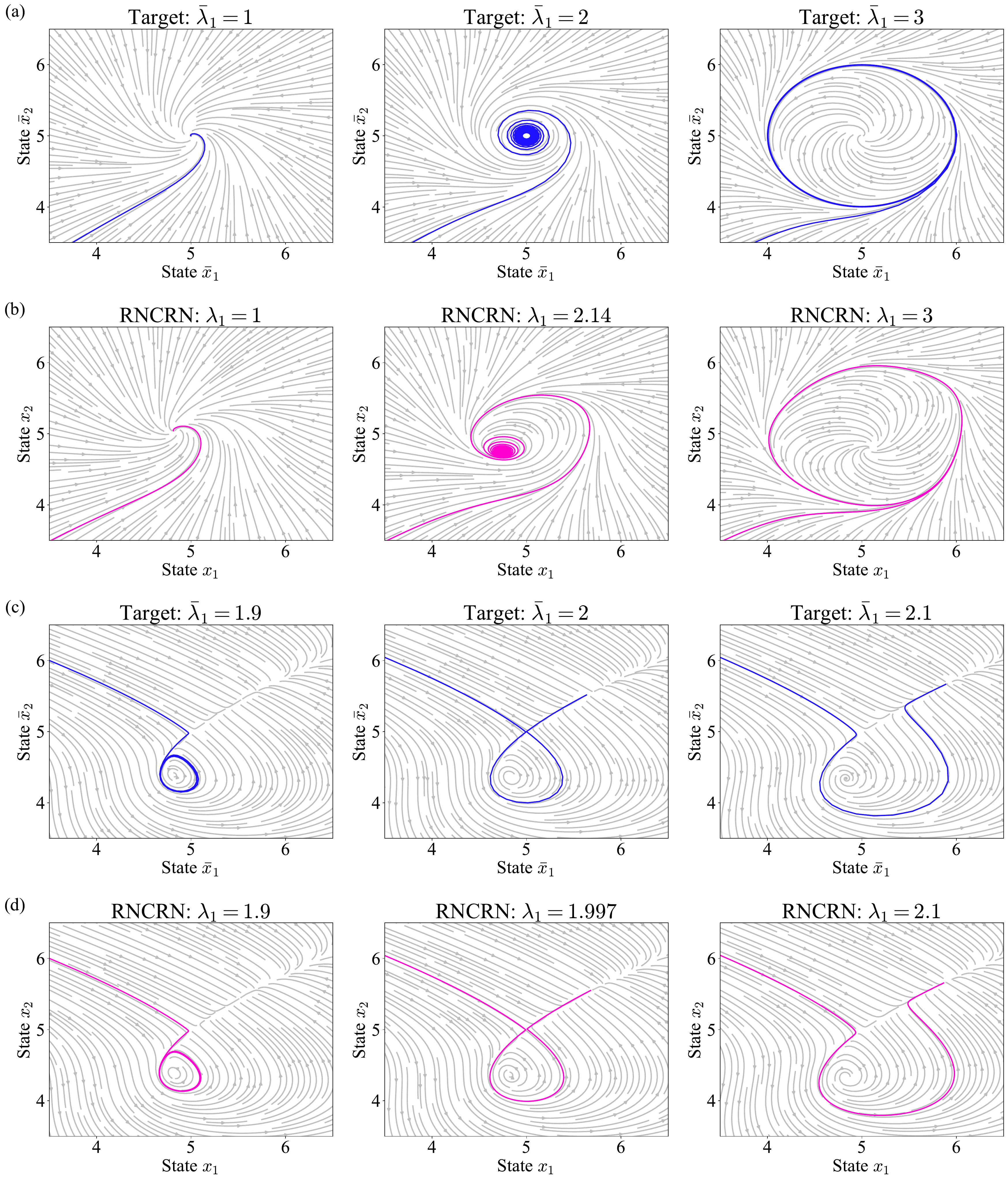}
    \caption{Approximation of two bifurcation-exhibiting ODEs using RNCRNs. 
    Row (a) shows the target ODE~(\ref{eq:target_hopf}) undergoing a Hopf bifurcation as the parameter $\bar{r}_1$ is varied through its critical value. Row (b) shows the $M=5$ chemical perceptron RNCRN approximation from Appendix~\ref{sec:app_hopf_bif} with $\mu=0.01$ undergoing the Hopf bifurcation as its equivalent control species $\lambda_1$ is varied. Row (c) shows the target ODE~(\ref{eq:homoclinic_orbit_RRE}) undergoing a homoclinic bifurcation as the parameter $\bar{r}_1$ is varied. Row (d) shows the $M=10$ chemical perceptron RNCRN approximation from Appendix~\ref{sec:app_homoclinic_bif} with $\mu=0.0075$ undergoing a homoclinic bifurcation. The gray arrows show the vector field of the target ODEs for rows (a) and (b) while for rows (b) and (d) the arrows show the vector field of the reduced RNCRN (i.e. $\mu=0)$. }
    \label{fig:rncrn_inhert_bifs}
\end{figure}

Let us now consider the target system
\begin{align}
    \frac{\mathrm{d}\bar{x}_1}{\mathrm{d}t} &= \left( \bar{\lambda}_1 - 2 -\frac{4}{5} \right)(\bar{x}_1-5) + (\bar{x}_2-5) - \frac{6}{5}(\bar{x}_1-5)(\bar{x}_2-5) + \frac{3}{2}(\bar{x}_2-5)^2 \nonumber \\
    \frac{\mathrm{d}\bar{x}_2}{\mathrm{d}t} &= (\bar{x}_1-5) - \frac{4}{5}(\bar{x}_2-5) -\frac{4}{5}(\bar{x}_2-5)^2,   \nonumber \\
     \frac{\mathrm{d} \bar{\lambda}_1}{\mathrm{d} t} &= 0,\label{eq:homoclinic_orbit_RRE}
\end{align}
which undergoes a homoclinic bifurcation~\cite{Glendinning1987, plesa_mapping_2024}
in the vicinity of the equilibrium $(\bar{x}_1, \bar{x}_2) = (5,5)$
at the critical parameter value $\bar{\lambda}_1^*=2$.
The state-space of~(\ref{eq:homoclinic_orbit_RRE}) is displayed in Figure~\ref{fig:rncrn_inhert_bifs}(c) below, at and above the critical parameter value. In particular, when $\bar{\lambda}_1 = 1.9 < 2$, 
there is a stable limit cycles in the state-space, 
which is absent when $\bar{\lambda}_1 = 2.1 > 2$. 
This transition happens at $\bar{\lambda}_1 = 2$
via a homoclinic orbit - a solution in the state-space 
that connects the equilibrium $(\bar{x}_1, \bar{x}_2) = (5,5)$ to itself.
To replicate these dynamical behaviours, 
we train the parametrized RNCRN~(\ref{eq:single_layer_RNCRN_static_param}) with 
$M=10$ chemical perceptron to approximate~(\ref{eq:target_hopf}) for all $1.9 \le \bar{\lambda}_1 \le 2.1$; 
see Appendix~\ref{sec:app_homoclinic_bif} for details.
We show in Figure~\ref{fig:rncrn_inhert_bifs}(d)
that the RNCRN displays the desired dynamics. 

Before closing this section, let us note that 
there are other systematic methods
for designing CRNs with bifurcations~\cite{plesa_chemical_2016,plesa_mapping_2024}, 
such as Hopf or homoclinic bifurcations, 
which in general contain a smaller number of species and reactions 
compared to the RNCRNs. In particular, being based on ANNs, 
the strength of the RNCRNs is not necessarily in producing 
relatively small systems, but instead in being 
more versatile with the types of input data
needed to achieve desired objectives. In the next
two sections, we further explore this ability.

\clearpage
\section{Piecewise systems with classification-controlled RNCRNs}
\label{sec:dynamical_piecewises}
In Section~\ref{sec:inherit_bifs_main}, 
we have trained RNCRNs to exhibit predefined bifurcations
by using as input data the parametrized target ODEs of the form~(\ref{eq:target_ODE_w_general_static_species}) which already display 
the desired bifurcations.
Importantly, for classical bifurcations to exist in such ODEs, 
the vector field from~(\ref{eq:target_ODE_w_general_static_species})
is required to be sufficiently smooth with respect to 
the parameters~\cite{perko_differential_2001}.
In this section, we relax this smoothness assumption by 
allowing the vector field from~(\ref{eq:target_ODE_w_general_static_species}) 
to be discontinuous with respect to the parameters.
The resulting ODEs are then known 
as \emph{piecewise systems}~\cite{filippov_differential_1988}, 
and provide more flexible training data for the RNCRN.
In particular, on the one hand, this approach may allow one
to simply specify suitable dynamics piecewise 
in the parameter space, thereby implicitly 
enforcing desired types of bifurcations at the associated boundaries,
which the RNCRN is allowed to automatically interpolate.
On the other hand, in some applications, more important
may be that a system displays certain dynamical features 
for certain parameters values, rather than 
how it transitions between 
the different regimes, and the piecewise 
approach allows for this without imposing unnecessary
regularity conditions on the dynamics.
In what follows, we adapt the parametrized RNCRN
from Section~\ref{sec:inherit_bifs_main} to 
be trainable on such piecewise systems, 
replicating the desired dynamics for desired 
parameter values, while automatically connecting
the different dynamical behaviours 
via suitable bifurcations.

\subsection{Simple unistable-bistable piecewise system}
\label{sec:point_controlled_piecewises}
As a motivation, let us consider the target system 
\begin{align}
\frac{\mathrm{d} \bar{x}_1}{\mathrm{d} t} = 
f_1(\bar{x}_1, \bar{\lambda}_1) = 
\begin{cases}
30 - 6\bar{x}_1, &\text{ if }  \bar{\lambda}_1 = 1,\\ 
-\left(\bar{x}_1 - 2\right)(\bar{x}_1 - 5)(\bar{x}_1 - 8), &\text{ if }\bar{\lambda}_1 = 0.
\end{cases}
\label{eq:point_con_piecewise}
\end{align}
In particular, the vector field has a unique stable equilibrium (unistability) when $\bar{\lambda}_1 = 1$,
while it has two stable equilibria (bistability) when 
$\bar{\lambda}_1 = 0$. Being defined at only two isolated
points in the parameter space, (\ref{eq:point_con_piecewise})
is a piecewise system.

\begin{figure}[hbt!]
    \centering
    \includegraphics[width=\linewidth]{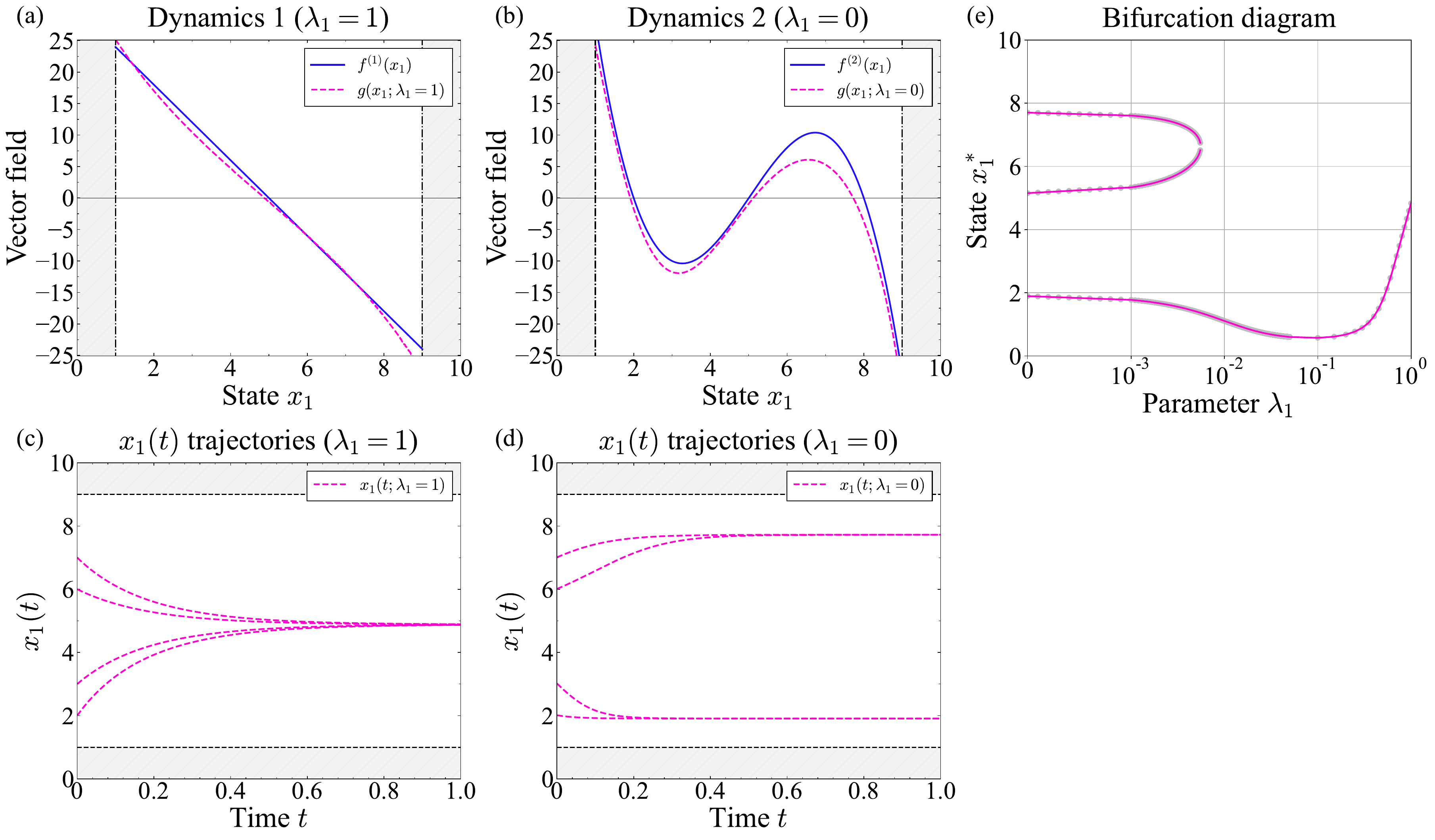}
\caption{Emergence of a bifurcation via interpolation when an RNCRN is trained at discrete concentrations of a parameter species. Approximation of~(\ref{eq:point_con_piecewise}) by the parametrized RNCRN~(\ref{eq:toogle_piecewise_ex_rncrn}) specified in Appendix~\ref{sec:simple_uni_bi_piecewise}. Panel (a) shows the dynamics in the first dynamical regime ($\lambda_1 = 1$). The target dynamics $d \bar{x}_1 / d t = 30 - 6 \bar{x}_1$ are shown in blue while the reduced vector field of the RNCRN approximation, i.e. when $\mu=0$, are shown in magenta. Panel (b) shows the dynamics in the second dynamical regime ($\lambda_1 = 0$). In blue the target dynamics $d \bar{x}_1 / d t =  -\left(\bar{x}_1 - 2\right)(\bar{x}_1 - 5)(\bar{x}_1 - 8)$ while in magenta the reduced vector field of the RNCRN approximation. Panel (c)--(d) show $x_1(t)$ time-trajectories over a variety of executive species initial conditions, $x_1(0) = 2, 3, 6, 7$, for the full RNCRN in each dynamical regime with $\mu=0.1$ and all chemical perceptron initial conditions at zero $y_1(0)=y_2(0)=y_3(0)=0$. Panel (e) shows the bifurcation diagram for the RNCRN approximation as the parameter $\lambda_1$ is varied through intermediate values in magenta. A bifurcation is observed at around $\lambda_1 =10^{-2}$.   } 
    \label{fig:toggle_example}
\end{figure}

We train the parametrized RNCRN~(\ref{eq:single_layer_RNCRN_static_param}) with $3$ chemical perceptrons to replicate the dynamics of~(\ref{eq:point_con_piecewise}),
with details shown in Appendix~\ref{sec:simple_uni_bi_piecewise} and the training algorithm in Appendix~\ref{app:alg_para_piece},
and with the results presented in Figure~\ref{fig:toggle_example}.
In particular, in Figure~\ref{fig:toggle_example}(a)--(d), 
we demonstrate that the RNCRN is unistable at 
$\lambda_1 = 1$ and bistable at $\lambda_1 = 0$, respectively, 
with the equilibria relatively close to those of~(\ref{eq:point_con_piecewise}).

As opposed to the target system~(\ref{eq:point_con_piecewise}),
which is defined only at $\bar{\lambda}_1 = 1$ and $\bar{\lambda}_1 = 0$, 
the RNCRN is defined for all $\lambda_1 \ge 0$, 
and depends smoothly on this parameter. 
To analyse this dependence, we display 
in Figure~\ref{fig:toggle_example}(e) a bifurcation diagram
- the equilibria of the RNCRN as a function of the parameter $\lambda_1$.
One can notice that the RNCRN automatically
interpolates unistability at $\lambda_1 = 1$ and bistability at 
$\lambda_1 = 0$ via a classical bifurcation.
In particular, one of the equilibria exists for all
the parameter values shown in Figure~\ref{fig:toggle_example}(e), 
while the other two exist for all $\lambda_1 \in [0,\lambda_1^*)$
for some $\lambda_1^* < 10^{-2}$, after which they 
disappear via saddle-node bifurcation~\cite{strogatz_nonlinear_2018}. 

\subsection{Classification-controlled RNCRNs}
\label{sec:piecewise_defn}
Let us now consider a general target piecewise system
taking the form
\begin{align}
    \frac{\mathrm{d} \bar{x}_i}{\mathrm{d} t} =
        f^{(p)}_i(\bar{x}_1, \dots, \bar{x}_N)  \text{ for all }  \bar{\boldsymbol{\lambda}} \in \mathbb{L}_p,
        \; i = 1, \dots, N, \; p = 1, \dots, P,
    \label{eq:target_general_piecewise_dynamics}
\end{align}
where $\bar{\boldsymbol{\lambda}} = 
(\bar{\lambda}_1, \bar{\lambda}_2, \dots, 
\bar{\lambda}_L) \in \mathbb{R}^L$ 
is the vector of parameters, $\mathbb{L}_1,
\mathbb{L}_2, \ldots,\mathbb{L}_P \subset \mathbb{R}^L$
are suitable disjoint sets in the parameter space, 
and $f^{(p)}_i$ is a suitably smooth function.
In particular, when the parameters $\bar{\boldsymbol{\lambda}}$
are chosen to lie in the set $\mathbb{L}_p$,
the ODE is governed by the vector field $\mathbf{f}^{(p)}$.
Let us note that set $\mathbb{L}_p$ can be a single point.
For example, system~(\ref{eq:point_con_piecewise}) 
is a special case of~(\ref{eq:target_general_piecewise_dynamics}) 
with a single parameter $\bar{\lambda}_1$, 
and two single-point parameter regions 
$\mathbb{L}_1 = \{1\}$ and $\mathbb{L}_2 = \{0\}$.

\begin{figure}[hbt!]
    \centering
    \includegraphics[trim={4.5cm 5cm 4.5cm 3.5cm},clip, width=\linewidth]{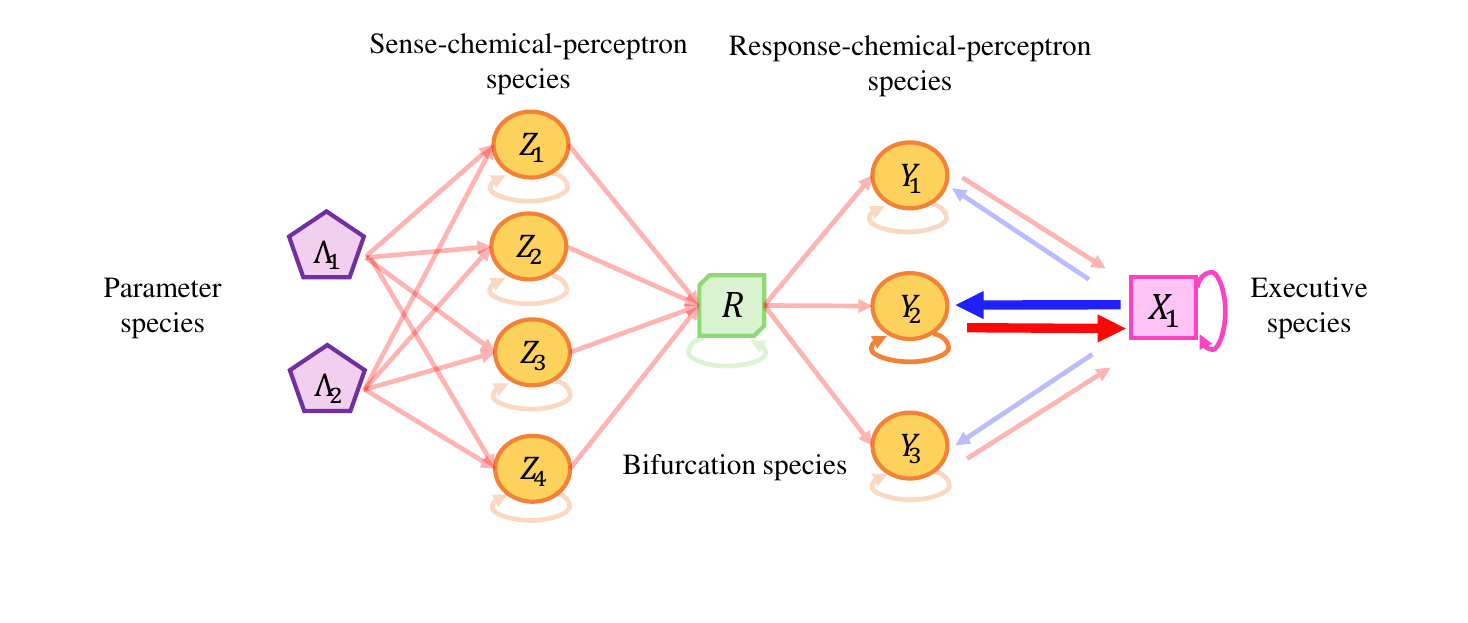}
    \caption{ A schematic of the classification-controlled RNCRN used to approximate equation~(\ref{eq:class_con_piecewise}). In purple the parameter-species $\Lambda_1$ and $\Lambda_2$ are shown to interact with the sense-chemical-perceptron species that perform the non-linear classification presenting the output as the magnitude of the bifurcation species $R$. The bifurcation species $R$ then acts as a parameter in the response-chemical-perceptron species that are trained to induce the intended dynamics in the executive species. } 
    \label{fig:sar_rncrn}
\end{figure}

The parametrized RNCRN~(\ref{eq:single_layer_RNCRN_static_param})
was sufficient to approximate piecewise system~(\ref{eq:point_con_piecewise}), which 
is defined at isolated parameter values.
In order to approximate more general piecewise systems~(\ref{eq:target_general_piecewise_dynamics}), 
which have more complicated parameter-space structure,
we now modify the RNCRN. In particular, 
consider the parametrized RNCRN with RREs
\begin{align}
\frac{\mathrm{d} x_i}{\mathrm{d} t} & = \beta_{i} +  x_i \sum_{j=1}^M \alpha_{i,j} y_j,  
&& i = 1, 2, \ldots, N, \nonumber \\
 \frac{\mathrm{d} y_j}{\mathrm{d} t} & =  \frac{\gamma_j}{\mu} 
+ y_j \left(\sum_{i=1}^N \frac{\omega^{(y)}_{j,i} }{\mu} x_i + \frac{\theta^{(y)}_{j}}{\mu}  + \frac{\psi_j}{\mu} r \right) - \frac{\tau_j}{\mu}  y_j^2, 
&&  j = 1, 2, \ldots, M,
\label{eq:modular_sac_RRE}
\end{align}
which are of the form~(\ref{eq:single_layer_RNCRN_static_param})
with a single parameter $r$, which itself is 
governed by the RREs
\begin{align}
\frac{\mathrm{d} \lambda_l}{\mathrm{d} t} & =  0, 
&& l = 1, 2, \ldots, L, \nonumber \\
\frac{\mathrm{d} z_k}{\mathrm{d} t} & =  \frac{\gamma}{\mu}  
+ z_k \left(\sum_{l=1}^L \frac{\omega^{(z)}_{k,l} }{\mu} \lambda_l + \frac{\theta_{l}^{(z)} }{\mu} \right) - \frac{1}{\mu}  z_k^2, 
&& k = 1, 2, \ldots, K, \nonumber \\
 \frac{\mathrm{d} r}{\mathrm{d} t} & =  \frac{\gamma}{\mu} 
+ r \left(\sum_{k=1}^K \frac{\omega_{k}^{(r)}}{\mu} z_k + \frac{\theta_{0}^{(r)}}{\mu} \right) - \frac{1}{\mu} r^2,
\label{eq:modular_sac_RRE_0}
\end{align}
which involve auxiliary chemical perceptrons 
with concentrations $z_1,z_2,\ldots,z_K$.
We call the RNCRN corresponding to the  
RREs~(\ref{eq:modular_sac_RRE})--(\ref{eq:modular_sac_RRE_0})
the \emph{classification-controlled} RNCRN, 
and illustrate it in Figure~\ref{fig:sar_rncrn}.
Let us note that this RNCRN, like the original~(\ref{eq:single_layer_RRE}), 
contains only two time-scales: 
the slow species $X_i$ and $\Lambda_l$, 
and the fast species $Y_j$, $Z_k$ and $R$.

Loosely, (\ref{eq:modular_sac_RRE})
is designed to implement the dynamics of~(\ref{eq:target_general_piecewise_dynamics}) point-wise, 
while~(\ref{eq:modular_sac_RRE_0}) then assembles/classifies
this dynamics to the appropriate parts of the parameter space.
In particular, in the parametrized RNCRN~(\ref{eq:single_layer_RNCRN_static_param}), 
the ODEs for the chemical perceptron
depend linearly on the parameters $\lambda_l$,
while the perceptrons in~(\ref{eq:modular_sac_RRE})
depend on $\lambda_l$ only implicitly via the
auxiliary variable $r$. Importantly, it follows 
from~(\ref{eq:modular_sac_RRE_0}) that, 
for every sufficiently small $\mu > 0$,
\begin{align}
z_k & \approx  \sigma_\gamma  \left(\sum_{l=1}^L \omega^{(z)}_{k,l} \lambda_l(0) + \theta_{k}^{(z)} \right) 
\; \; \textrm{and} \; \;  
r \approx \sigma_\gamma \left(\sum_{k=1}^K \omega_{k}^{(r)} z_k + \theta_{0}^{(r)} \right), 
\label{eq:r_ANN}
\end{align}
where $\sigma_\gamma(x)$ is the activation function
defined in~(\ref{eq:single_layer_RRE_reduced}). (\ref{eq:r_ANN}) therefore operates as a two-layer ANN, 
with a single perceptron in the second layer, 
that maps between parameters $\boldsymbol{\lambda}$ 
and the point-wise control parameter $r$. 

Based on this modular design, we train 
the classification-controlled RNCRN in two steps. 
In the first step, we disregard the precise 
structure of the parameter space,
and simply train the parametrized RNCRN~(\ref{eq:modular_sac_RRE})
point-wise to display at some $r = r_p $ 
the dynamical behaviour induced by the vector field
$\mathbf{f}^{(p)}$ for every $p = 1,2,\ldots, P$.
Then, in the second step, we factor in 
the precise structure of the parameter space
by training~(\ref{eq:modular_sac_RRE_0})
to map from a given $\mathbb{L}_p$ set
in $(\lambda_1,\lambda_2,\ldots,\lambda_L)$-space to the associated $r_p$ value 
by using standard feed-forward ANN training methods
via~(\ref{eq:r_ANN}).
We present an algorithm for this training
in Appendix~\ref{app:class_controlled_algo}.

\subsection{XOR-gated unistable-bistable piecewise system}
\label{fig:classification_piecewises}
Let us now consider a target piecewise system 
with more complicated parameter-space structure
than~(\ref{eq:point_con_piecewise}), 
depending on two parameters $\bar{\lambda}_1, \bar{\lambda}_2$
and given by
\begin{align}
\frac{\mathrm{d} \bar{x}_1}{\mathrm{d} t} = \begin{cases}
    30 - 6\bar{x}_1, & \text{ for all } 
    (\bar{\lambda}_1, \bar{\lambda}_2) \in \mathbb{L}_1 = \left(\mathbb{K}_1 \times \mathbb{K}_2 \right) \cup \left(\mathbb{K}_2 \times \mathbb{K}_1 \right), \\ 
    -\left(\bar{x}_1 - 2\right)(\bar{x}_1 - 5)(\bar{x}_1 - 8), & \text{ for all } 
    (\bar{\lambda}_1, \bar{\lambda}_2) \in \mathbb{L}_2 = \left(\mathbb{K}_1 \times \mathbb{K}_1 \right) \cup \left(\mathbb{K}_2 \times \mathbb{K}_2 \right),
    \end{cases}
    \label{eq:class_con_piecewise}
\end{align}
where $\mathbb{K}_1 = [0, 1)$ and $\mathbb{K}_2 = [1, 2]$. 
Loosely, if on the domain $[0,2] \times [0,2]$ and
one of the parameters is in the larger region,
while the other is in the smaller region,
then the piecewise system~(\ref{eq:class_con_piecewise}) is unistable;
otherwise, it is bistable.
This dependence can be interpreted as 
a XOR operator, and creates in the parameter space 
a non-linear boundary between different dynamical regions.

To approximate the dynamics of~(\ref{eq:class_con_piecewise}), 
we use Algorithm~\ref{algh:class_controlled_RNCRN} to train the 
classification-controlled RNCRN~(\ref{eq:modular_sac_RRE})--(\ref{eq:modular_sac_RRE_0}), with details presented
in Appendix~\ref{app:sar_piecewise_struct}. In particular, 
in the first step of the algorithm, we simply 
re-use the parametrized RNCRN from Appendix~\ref{sec:simple_uni_bi_piecewise}, 
which was previously constructed in Section~\ref{sec:point_controlled_piecewises} to 
approximate system~(\ref{eq:point_con_piecewise}). 
In the second step, we then train the classification
module~(\ref{eq:modular_sac_RRE_0}) to implement
the parameter-space structure from~(\ref{eq:class_con_piecewise}).

\begin{figure}[hbt!]
    \centering
    \includegraphics[width=\linewidth]{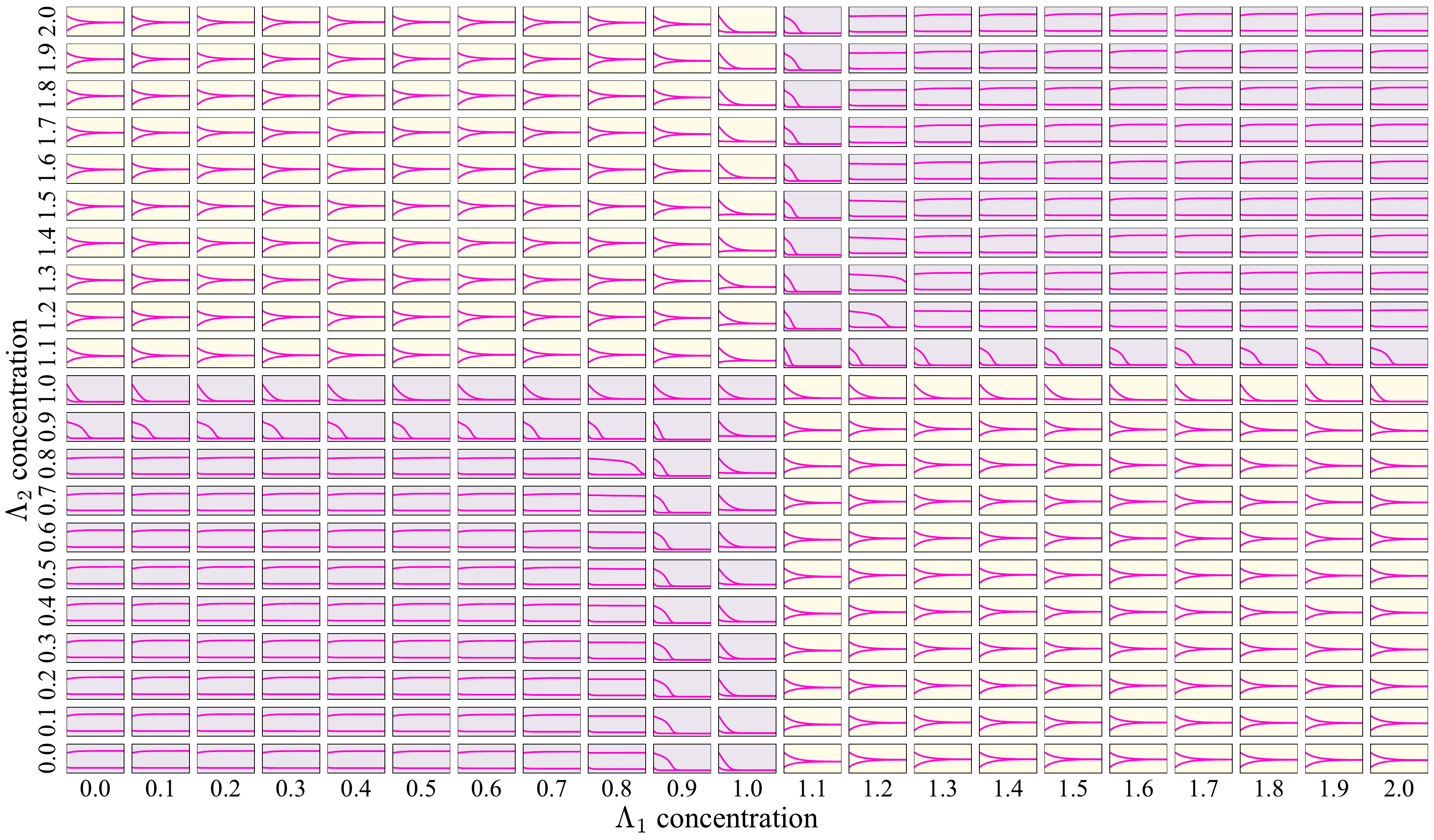}
    \caption{ A panel of $x_1(t)$ trajectories for different concentrations of $\Lambda_1$ and $\Lambda_2$ showing the non-linear XOR classification boundary separating bistability and unistability for an RNCRN approximating the target piecewise system in equation~(\ref{eq:class_con_piecewise}). The associated classification-controlled RNCRN, i.e. equation~(\ref{eq:modular_sac_RRE}),  uses four sense chemical-perceptrons, one bifurcation chemical-perceptron, and three response chemical-perceptrons to induce the intended dynamics in one executive species according to the states of two parameter species $\Lambda_1$ and $\Lambda_2$. Each panel has two time-trajectories simulated for a total time of $\Delta t=1$ with $\mu=0.001$, all chemical perceptron values starting at zero and the only executive species taking alternative values of $x_1(0) = 2$ and $x_1(0)=7$. The hue of each panel shows the indented dynamical behaviour in the piecewise system: yellow should be unistable while pink should be bistable.  Full details of the reactions in this example are given in Appendix~\ref{app:sar_piecewise_struct}. } 
    \label{fig:sac_results}
\end{figure}

In Figure~\ref{fig:sac_results}, we plot a two-parameter
bifurcation diagram for the resulting RNCRN, 
displaying representative trajectories
$x_1(t)$ at different points in the 
$(\lambda_1,\lambda_2)$ parameter-space.
One can notice a relatively good match 
with the target piecewise system~(\ref{eq:class_con_piecewise}): 
the RNCRN is unistable approximately when $(\lambda_1,\lambda_2) 
\in ([0, 1.0]\times[1.1, 2.0]) \cup ([1.1, 2.0]\times[0, 1.0])$, 
and bistable when $(\lambda_1,\lambda_2) 
\in ([0, 1.0]\times[0, 1.0]) \cup ([1.1, 2.0]\times[1.1, 2.0])$.
Furthermore, the RNCRN depends smoothly on the parameters 
$(\lambda_1,\lambda_2)$, and interpolates 
the dynamics across the boundaries of
the parameter space via saddle-node bifurcation (as shown in  Figure~\ref{fig:toggle_example} (e)).

\clearpage
\section{Data-defined limit cycles and switching}
\label{sec:data_defined_dynamics}
In both Sections~\ref{sec:inherit_bifs_main}
and~\ref{sec:dynamical_piecewises},
the RNCRNs are trained on the data taking the form 
of ODEs with suitably regular vector fields. 
In this section, we exploit the interpolation and extrapolation abilities of the underlying neural network-based methodology to show that the RNCRN 
can also be trained in an ODE-free manner
from relatively sparse data: a set of points in the state-space. 
More precisely, we focus on 
generating the data that may allow the RNCRN
to incorporate into its state-space two-dimensional 
limit cycles defined in an ODE-free manner. 

To this end, we sample a finite number of points on,
and in a neighbourhood of, a given closed curve
in the state-space. We then assign a single vector
to each of these points with appropriate direction:
the vectors on the closed curve point along it 
(corresponding to the limit cycle), 
while the remaining vectors all point either
towards the closed curve or away from it
(corresponding to respectively stability or instability).
One thus obtains a local vector field, defined only on a finite
number of states, which can be used to train the RNCRN, 
allowing it to extrapolate the behaviour beyond the grid. 
The algorithm to generate this data is given 
as Algorithm~\ref{algh:define_attractor}. 
Let us note that the vector field 
near the limit cycle varies exponentially with states
in the algorithm; in Appendix~\ref{sec:app_single_cycle_linear},
we show that a linear variation can also be used.
Let us also note that this algorithm can be generalized
to higher dimensions; see Appendix~\ref{app:torus_orbit}
for a three-dimensional version and an example.

\begin{table}[ht]
\renewcommand\tablename{Algorithm}
    \centering
    \begin{minipage}{\textwidth}
    \caption{{\it Algorithm for generating data to train \emph{RNCRN}s to display two-dimensional stable limit cycles defined via the set of points $\{ P_1, P_2, \dots, P_D, P_{D+1}\}$.}}
    \label{algh:define_attractor}
    \end{minipage}
        \begin{minipage}{\textwidth}
        \hrule
        \vskip 2.5 mm
        Fix an ordered set of points in the executive-species space 
        $\{ P_1, P_2, \dots, P_D, P_{D+1}\}$, where $P_d \in \mathbb{R}^2_{>}$ for $d= 1, 2, \dots, D, D+1$ and such that $P_{D+1} = P_1$, where $\mathbb{R}^2_{>}$ is the positive quadrant. Fix also the padding number $K$, padding distance $\delta > 0$, decay scale $\kappa > 0$, and vector magnitude $\eta > 0$.
        \end{minipage}
    \begin{minipage}{0.5\textwidth}
        \centering
        \vskip 2.5 mm
\begin{enumerate}
\item[\textbf{(a)}] \textbf{Padding points}.
For each point $P_d$ compute the unit normal vector $\hat{n}_{d, d+1}$ which is perpendicular to $\hat{v}_{d, d+1}$, the unit vector pointing from $P_d$ to $P_{d+1}$. Create $2K$ symmetric padding points $Q_{d,k}^{\pm} = P_d {\pm} k\delta \hat{n}_{d, d+1}$ 
for $k=1,2, \dots, K$.
\item[\textbf{(b)}] \textbf{Local vector field}. 
Assign vector $\eta\hat{v}_{d, d+1}$ to $P_d$, 
and vector $\left(\eta\hat{v}_{d, d+1} \mp \eta \hat{n}_{d, d+1}\exp{(\kappa \delta k)} \right)$ to $Q_{d,k}^{\pm}$.
\end{enumerate}
\end{minipage}
\hfill
\begin{minipage}{0.45\textwidth}
    \centering
    \includegraphics[trim={10.0cm 4.0cm 12.0cm 3.75cm}, clip, width=\linewidth]{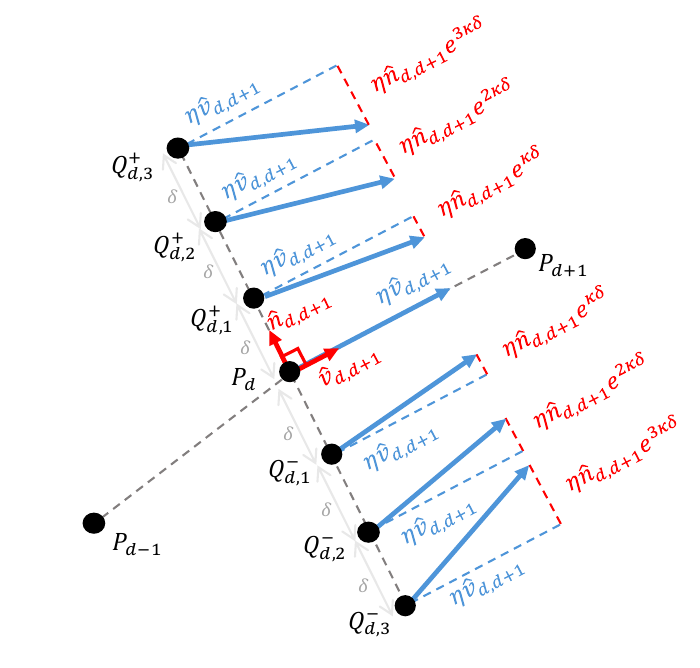}
\end{minipage}
    
\begin{minipage}{\textwidth}
\hfill
\vskip 2.5 mm
\hrule
\end{minipage}
\end{table}
 
\subsection{Circular limit cycle}
\label{sec:single_cycle}
Consider a circle of radius $a > 0$ center at 
$(b_1,b_2) \in \mathbb{R}_{>}^2$, 
with parametric equation given by
\begin{align}
    x_1(s) &= a\cos(s) + b_1, \nonumber \\
    x_2(s) &= a\sin(s) + b_2, 
    \label{eq:parametric_circle}
\end{align}
where $s \in [0, 2\pi)$. 
In what follows, we choose
$a = 1$ and $(b_1,b_2) = (2,2)$. 
We sample an ordered set of
points $P_d = (x_1(s_d), x_2(s_d))$ by taking 
$s = s_d = 0.01(d-1)$ in~(\ref{eq:parametric_circle}) 
for $d=1,\dots, 629, 630$. Using Algorithm~\ref{algh:define_attractor} with $K=20$, $\delta=0.01$,  $\kappa=1$, and $\eta=1$, we generate the input data to train the RNCRN~(\ref{eq:single_layer_RRE}) 
with $M=5$ chemical perceptrons to display in the state-space 
a stable two-dimensional limit cycle which 
is close to the circle. This training is performed by 
suitably adapting Algorithm 1 from~\cite{dack_recurrent_2025}, 
as presented in Algorithm~\ref{algh:data_def_RNCRN_with_AD} and Appendix~\ref{sec:app_single_cycle}; 
see also Appendix~\ref{sec:app_single_cycle_repeller} for an example RNCRN with approximately circular limit cycle which is unstable.

We show the training data, together with the state-space of the trained RNCRN, in Figure~\ref{fig:attract_defined_dynamics}(a).
One can notice that the RNCRN with $\mu=0.1$ numerically displays a 
stable limit cycle whose shape is relatively close to the desired circle. Furthermore, the RNCRN extends the vector consistently beyond this limit cycle, producing inside it an unstable equilibrium.

\subsection{Heart-shaped limit cycle}
\label{sec:heart_shaped_cycle}
Let us now consider the parametric equation  
\begin{align}
    x_1(s) &= 2+ 1.12\sin(s)^3, \nonumber \\
    x_2(s) &= 2+0.91\cos(s)- 0.35\cos(2s)-0.14\cos(3s)-0.07\cos(4s), 
    \label{eq:parametric_heart}
\end{align}
which for $s \in [0, 2\pi)$ produces a heart shape 
in the $(x_1,x_2)$-space. In this example, 
we generate the training data for each lobe of the heart separately,
$s_d = 0.01(d-1)$ for $d=1, 2, \dots, 261, 262, $ and $d= 314, 315 \dots, 575, 576$, and then use Algorithm~\ref{algh:define_attractor}
with $K=20$, $\delta=0.01$, $\kappa=1$, and $\eta=1$; 
see Appendix~\ref{sec:app_heart_cycle} for more details.
We display the results in Figure~\ref{fig:attract_defined_dynamics}(b),
numerically demonstrating that the RNCRN displays a stable limit cycle
that is approximately heart-shaped and, consistently with the given local vector field, creates an unstable equilibrium inside the cycle. 

\subsection{Multiple limit cycles}
\label{sec:multiple_cycle}
\begin{figure}[hbt!]
    \centering
    \includegraphics[width=\linewidth]{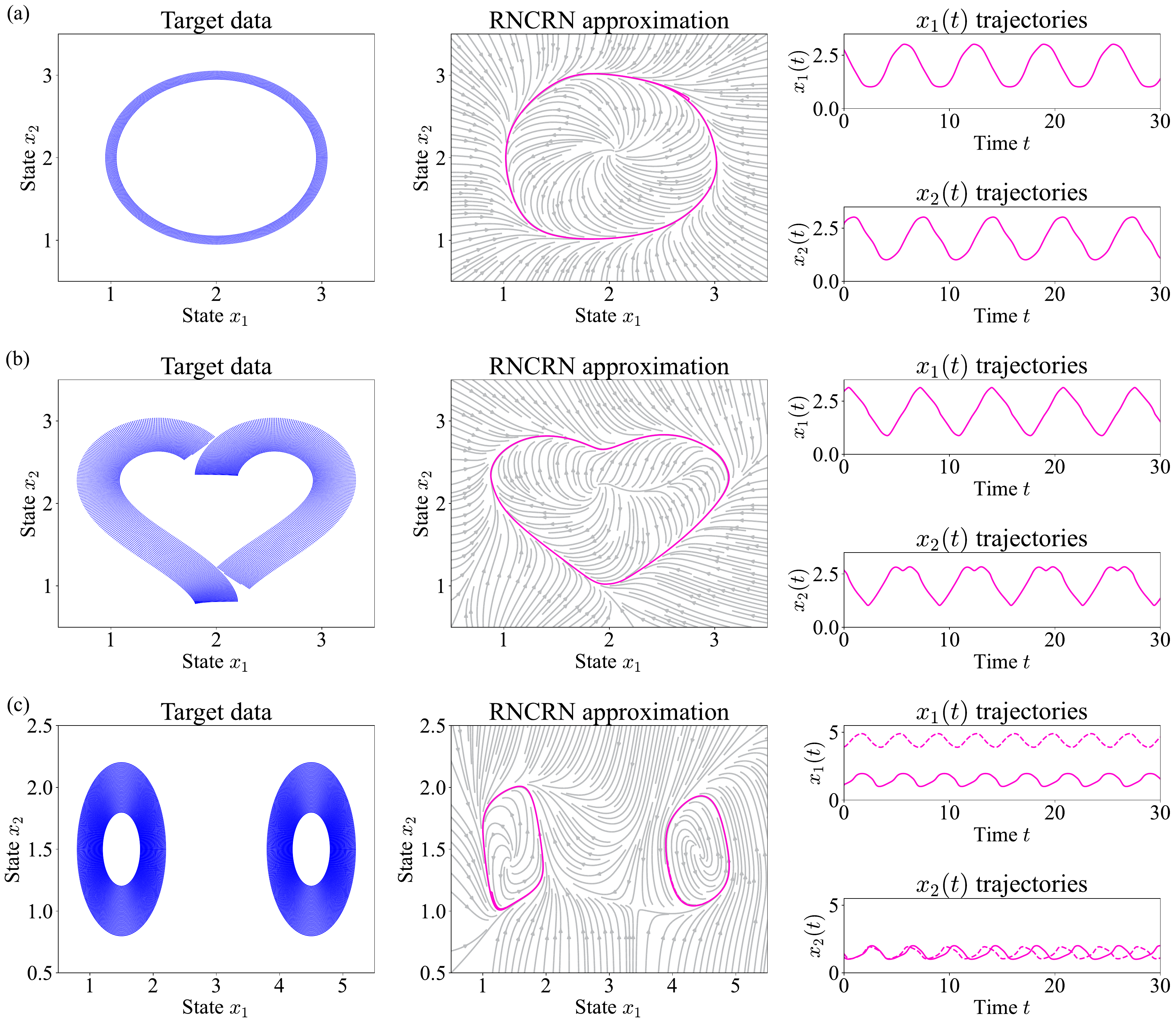}
    \caption{Example RNCRNs trained on data-defined limit cycles. Column 1 shows the points generated by Algorithm~\ref{algh:define_attractor} at which the training data is given for different target attractor shapes. A close-up of the training data is shown in Appendix~\ref{algh:define_attractor}. Column 2 shows the RNCRN approximation trained on the data in column 1. The phase plane of the reduced RNCRN approximations, i.e. the RNCRN with $\mu =0$, are shown as grey arrows, while example ($x_1,x_2$)-trajectories of the full RNCRNs with $\mu > 0$ are shown in magenta. Column 3 shows time-trajectories of the full RNCRN approximation. Row (a) shows a $5$-chemical-perceptron RNCRN (full parameters are given in  Appendix~\ref{sec:app_single_cycle}) used to approximate a data-defined circular limit cycle. The time-trajectory was simulated with $x_1(0)=2.744, x_2(0)=2.693$, $\mu=0.1$, and $y_1(0)=\dots, y_5(0)=0$.  Row (b) shows a $10$-chemical-perceptron RNCRN (as specified in Appendix~\ref{sec:app_heart_cycle}) approximating the data-defined heart-shaped limit cycle. The time-trajectory was simulated with $x_1(0)=2.957, x_2(0)=2.665$, $\mu=0.1$, and $y_1(0)=\dots, y_{10}(0)=0$. Row (c) shows a $15$-chemical-perceptron RNCRN (specified in Appendix~\ref{sec:app_multiple_cycle}) approximating a system with two limit cycles. The two time-trajectories were both simulated with $\mu=0.1$ and $y_1(0)=\dots, y_{15}(0)=0$, the executive species' initial concentrations differed with $x_1(0)= 1.118, x_2(0)=1.154$ or $x_1(0)= 3.916, x_2(0)=1.392$.  }
    \label{fig:attract_defined_dynamics}
\end{figure}

Algorithm~\ref{algh:define_attractor} can also be used
to train RNCRNs to display multiple coexisting limit cycles. 
For example, let us apply Algorithm~\ref{algh:define_attractor} 
to generate data for two coexisting stable limit cycle of 
the form~(\ref{eq:parametric_circle}), one 
with $a=0.5$ and $(b_1,b_2) = (1.5,1.5)$, 
and the other with $a=0.5$ and $(b_1,b_2) = (4.5,1.5)$,
as described in detail in Appendix~\ref{sec:app_multiple_cycle}.
The result of training an RNCRN on the concatenated training data is shown in Figure~\ref{fig:attract_defined_dynamics}(c), 
numerically demonstrating two stable limit cycles. 
Let us note that, in this example, the RNCRN not only 
generates an unstable equilibrium inside each of the limit cycles, 
but also creates an equilibrium of saddle type 
as a way to connect the two regions 
specified by the training data.

\subsection{Equilibrium-oscillation piecewise system with data-defined oscillations}
\label{sec:targ_pulse_drug}
Let us now combine the results from this section
and Section~\ref{sec:dynamical_piecewises},
by considering the target piecewise system
\begin{align}
\frac{\mathrm{d} \mathbf{x}}{\mathrm{d} t} = 
\begin{cases}
\mathbf{f}^1(\mathbf{x}), &\text{ for all } \bar{\lambda}_1 \in [0, 0.25],\\ 
\mathbf{f}^2(\mathbf{x}), &\text{ for all }\bar{\lambda}_1 > 0.25,
\label{eq:equ_osc}
\end{cases}
\end{align}
with $\mathbf{x} \in \mathbb{R}^2$, 
where $\mathbf{f}^1(\mathbf{x}) \in \mathbb{R}^2$ 
is a smooth vector field with a unique stable equilibrium at $(0.5, 0.5)$, while 
$\mathbf{f}^2(\mathbf{x})$ is a vector field 
defined only on a finite number of states in $\mathbb{R}_{>}^2$
via Algorithm~\ref{algh:define_attractor} 
to display a circular limit cycle. We might imagine chemical systems that implement dynamics such as~(\ref{eq:equ_osc}) are implementing environment-specific oscillations. That is, when an environmental signal,  $\Lambda_1$, is abundant the system will produce oscillations of an executive species, $X_1$, but when the environmental signal is weak, trace amounts of the executive species will be produced.

We now consider using a classification-controlled RNCRN,~(\ref{eq:modular_sac_RRE})-(\ref{eq:modular_sac_RRE_0}), to implement~(\ref{eq:equ_osc}). Following the steps in Algorithm~\ref{algh:class_controlled_RNCRN}, modified as in Algorithm~\ref{algh:data_def_RNCRN_with_AD} to use the data-defined vector field,  we train a classification-controlled RNCRN to implement~(\ref{eq:equ_osc}); see Appendix~\ref{app:pulse_drug} for details ($5$ $Y_j$-chemical perceptrons, $3$ $Z_k$-chemical perceptrons, and a single bifurcation species was used). 
 
\begin{figure}[hbt!]
    \centering
    \includegraphics[width=\linewidth]{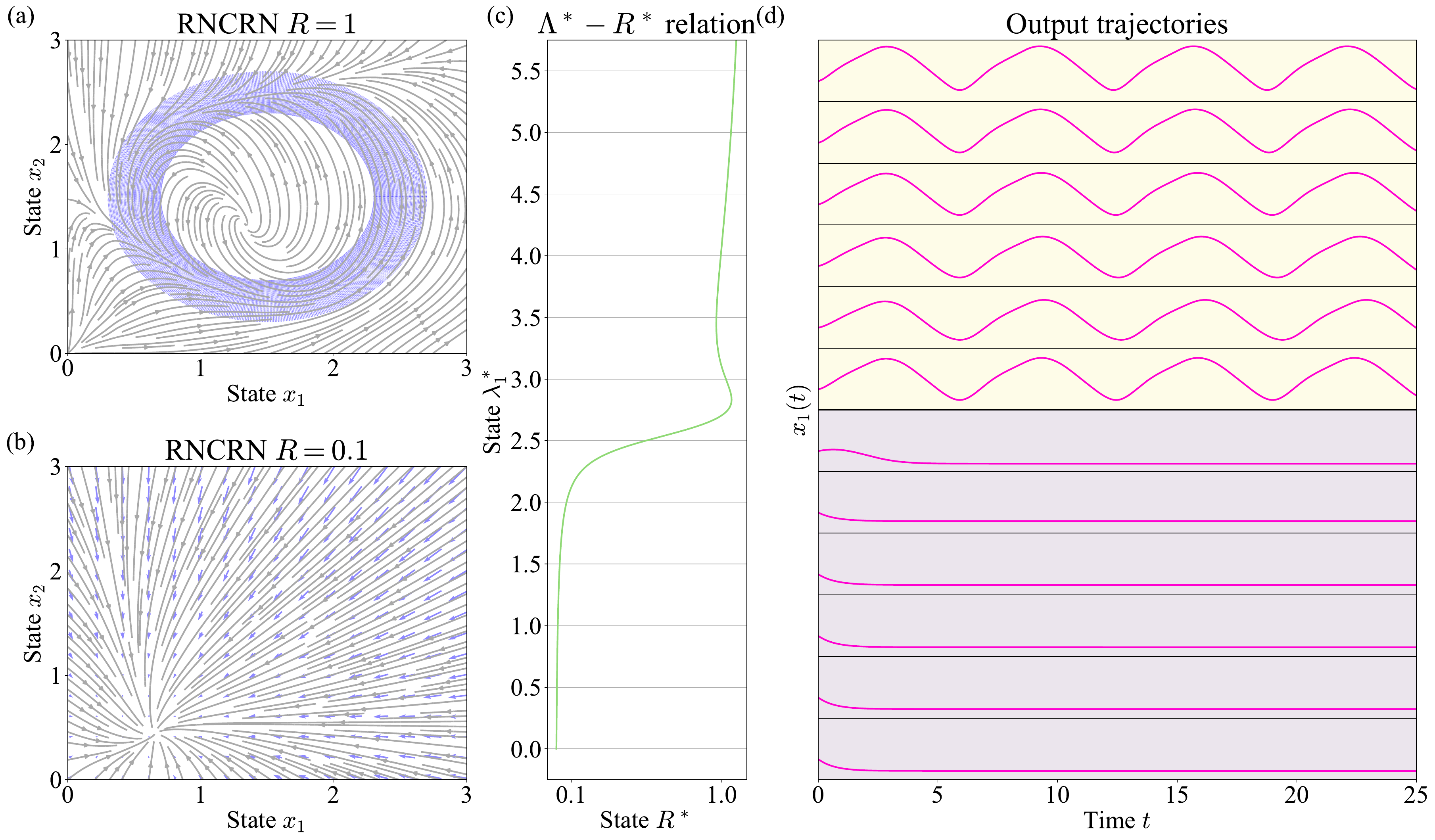}
    \caption{A classification-controlled RNCRN designed to approximate a piecewise system that either produces trace amounts of executive species $X_1$ or oscillations in $X_1$ according to the concentration of environmental species $\Lambda_1$. Panel (a)--(b) shows the data-defined target dynamics (in blue) as well as the reduced vector field of the point-defined RNCRN approximation as grey arrows. Panel (c) shows the input-output relationship between $\Lambda_1$ and $R$ (an intermediary species). Panel (d) shows numerically integrated time-trajectories of output species, $x_1(t)$, for the full RRE of the classification-controlled RNCRN with different $\lambda_1(0)$ values (according to the alignment with panel (c) $\lambda_1^*$-axis) and $\mu=0.1$.  Full details of the reactions in this system are given in Appendix~\ref{app:pulse_drug}.} 
    \label{fig:sac_pulse_target}
\end{figure}

 Figure~\ref{fig:sac_pulse_target}(a)--(b) shows the data-defined oscillations and the stable equilibrium as blue arrows. The gray streamlines in Figure~\ref{fig:sac_pulse_target}(a)--(b) shows the vector field of the reduced parametrized RNCRN for two particular values of the bifurcation species, $R=0.1$ and $R=1$, which were used in training. The input-output relationship between the parameter species $\Lambda_1$ and the bifurcation species $R$ is shown in Figure~\ref{fig:sac_pulse_target}(c).

To show the dependence of the $X_1$ behaviour on the parameter species $\Lambda_1$, sub-panels in Figure~\ref{fig:sac_pulse_target}(d) have been aligned with the $\lambda_1^*$-axis in Figure~\ref{fig:sac_pulse_target}(c). Each sub-panel in Figure~\ref{fig:sac_pulse_target}(d) corresponds to a different $\lambda_1(0)$ concentration and plots the resulting $x_1(t)$ time-trajectories. The full classification-controlled RNCRN is defined in Appendix~\ref{app:pulse_drug}, and we simulate it using $\mu = 0.01$ and all chemical-perceptron species concentrations starting from zero. 
We observe that the classification-controlled RNCRN produces the desired behaviour. Below a critical value of $\lambda_1\approx 2.5$ we observe that the $x_1(t)$ concentrations tends to a low equilibrium (the magenta panels in Figure~\ref{fig:sac_pulse_target}(d)), but above that point we observe oscillations of a similar frequency and amplitude (the yellow panels in Figure~\ref{fig:sac_pulse_target}(d)). This demonstrates an effective approximation of a piecewise systems as different regions of $\lambda_1$-space are being mapping to different functions in $(x_1, x_2)$-space (and these functions are largely independent of $\lambda_1$ in these regions).

\section{Discussion}
In this work we use a recurrent neural chemical reaction network (RNCRN) architecture to approximate atypical dynamical behaviours and demonstrate that different dynamical regimes can be instantiated in a single system, along with bifurcations as the system moves between those regimes. Mapping target dynamical behaviours, in the form of a non-chemical target ODE, into a CRN has an extensive literature~\cite{kerner_universal_1981, samardzija_nonlinear_1989, kowalski_universal_1993, poland_cooperative_1993,  hangos_mass_2011, plesa_chemical_2016, plesa_test_2017, plesa_integral_2023, plesa_mapping_2024}. These predominately analytical techniques can produce small CRNs in terms of reaction and species number. For example, the target ODE with a homoclinic bifurcation considered here~(\ref{eq:homoclinic_orbit_RRE}) was realised with only 2 chemical species and 9 chemical reactions~\cite{plesa_mapping_2024} compared to the 12 chemical species and 82 chemical reactions required by the RNCRN approximation identified here. Nonetheless, showing that RNCRNs can accurately replicate these bifurcations served as a motivation for the later sections where such analytical mapping techniques do not apply.

Sections~\ref{sec:dynamical_piecewises} and~\ref{sec:data_defined_dynamics} demonstrate the major advantage of the RNCRN over the aforementioned analytical techniques: RNCRNs are mathematically derived from artificial neural networks (ANNs) which are well-known for their ability to learn functions via data-driven training techniques. Exploiting this fact, we introduced Algorithm~\ref{algh:define_attractor} for generating artificial data to train an RNCRN to reproduce isolated dynamical features defined on small regions of phase space. We were able to use Algorithm~\ref{algh:define_attractor} to define exotic attractor shapes including a heart-shaped limit cycle, multiple limit cycles, and a three executive species toroidal limit cycle (see Appendix~\ref{app:torus_orbit}). 

A key feature of RNCRN-based methods was the ability of the RNCRN to extrapolate and interpolate dynamical behaviour to achieve dynamics in the whole space consistent with the features of interest.  For example, we observe the presence of multiple equilibria that were not provided in the training data in the reduced vector fields of Figure~\ref{fig:attract_defined_dynamics}. Similarly, bifurcations naturally emerge -- without their location or type being specified explicitly -- when RNCRNs are made to interpolate between qualitatively distinct behaviours at different values of the input parameters species.
This ability for the RNCRN to fill in the gaps of the vector field makes it much easier to design complicated dynamics as there is no requirement to find an ODE with such dynamics.

In Section~\ref{sec:targ_pulse_drug} we presented a classification-controlled RNCRN inspired by applications where it is necessary to sense the chemical environmental signals and act accordingly - in this case, turn on an oscillating output. Several major challenges with modern therapeutics could be addressed by systems with similar behaviour. Most straightforwardly, environmental-specific activation would allow for the localisation of therapeutic activity within pathologically relevant environments thus reducing off-target effects. It is well-known, for example, that off-target killing of benign cells limits the success of chemotherapies\cite{anand_cancer_2023} and antibiotics therapies\cite{patangia_impact_2022}. More interestingly for our purposes, switching to a new dynamical regime would allow fine-tuned control over the temporal concentration of a therapeutic species and would allow for the design of non-trivial dosing profiles. For example, it has been suggested that pulsatile dosing schedules are beneficial for preventing the development of resistance in a wide range of contexts including antibiotic resistance \cite{baker_beyond_2018
}, cancer treatments \cite{wang_drug_2019
}, and preventing insulin resistance in diabetes treatments\cite{satin_pulsatile_2015}. It should be noted that experimental chemical neural networks have already been built that are able to classify RNAs from cancer\cite{zhang_cancer_2020} and bacterial infections \cite{lopez_molecular_2018}, highlighting that classification-controlled RNCRNs might one day be developed in such contexts.

There are several outstanding challenges that remain to be explored in the context of RNCRNs. While the underlying reactions can, in principle, be encoded using DNA strand displacement~\cite{dack_recurrent_2025}, it is possible that other contexts such enzyme-aided DNA nanotechnology~\cite{thachuk_implementing_2019} or transcription factor networks~\cite{zhu_synthetic_2022} may be more convenient in practice. In that vein, the principles behind the construction of the RNCRN are more general than the particular CRNs used to instantiate the chemical perceptron. We anticipate that many alternative CRNs that play the equivalent role, \textit{i.e. } weighted summation and non-linear activation,  might be engineered to instantiate RNCRN-like systems in a range of biochemical media. We would like to explore systematic compilation methods that take the RNCRN's chemical reaction network and converts them into a dynamically equivalent reaction network with alternative kinetic laws (such as enzyme-based synthetic biology). 

An open question is whether interesting RNCRNs are feasible in practice. The number of reacting components in the RNCRNs presented in this work is lower than some other complex engineered molecular networks~\cite{cherry_scaling_2018, xiong_molecular_2022}, but a feature of RNCRNs that is more challenging to engineer is the necessary degree of connectivity between components and control over kinetics. We would therefore like to investigate compression or pruning algorithms to systematically reduce the number of auxiliary chemical species and chemical reactions within the resulting CRNs.

\section{Declarations}
\noindent
{\small {\bf Author Contributions}:
AD, TEO, and TP conceptualized the study; AD performed the simulations and wrote the original draft; TEO and TP reviewed and edited the final submission.}
\\
\noindent
{\small
{\bf Funding}:
Alexander Dack acknowledges funding from the Department of Bioengineering at Imperial College London. Thomas E. Ouldridge would like to thank the Royal Society for a University Research Fellowship. Tomislav Plesa would like to thank Peterhouse, 
University of Cambridge, for a Fellowship.}
\\
\noindent
{\small
{\bf Conflict of interest}:
The authors declare that they have no competing interests.
}

\bibliographystyle{ieeetr}  
\bibliography{references.bib}

\appendix
\newpage
\section{Appendix: Example bifurcations}
\label{sec:app_bif}
\subsection{Hopf bifurcation}
\label{sec:app_hopf_bif}
We use Algorithm~\ref{algh:para_RNCRN}, a modified version of Algorithm 1 from~\cite{dack_recurrent_2025} with static executive species,  to approximate the target system~(\ref{eq:target_hopf}) on $\mathbb{K}_1= [3.5,6.5],  \mathbb{K}_2=  [3.5,6.5]$ and $\mathbb{L} = [1,3]$ with a parametrized RNCRN. The absolute tolerance $\varepsilon \approx 10^{-1}$ is met with an RNCRN with $M=5$ chemical perceptrons and coefficients $\beta_1=\beta_2 = 260$, $\gamma = 1$, and  
\begin{align}
    \boldsymbol{\alpha}_1 &= \begin{pmatrix}
    -4.315 \\
-5.861 \\
-9.788 \\
3.796 \\
-1.209 \\
    \end{pmatrix}, \; \; 
    \boldsymbol{\alpha}_2 = \begin{pmatrix} 
    -1.505 \\
-4.968 \\
0.149 \\
-25.501 \\
0.824 \\
    \end{pmatrix}, \;\;
    \boldsymbol{\theta} = \begin{pmatrix}
    10.656 \\
8.520 \\
5.385 \\
3.369 \\
4.653 \\
    \end{pmatrix}, \;\; \nonumber \\
    \boldsymbol{\omega_1} &= \begin{pmatrix}
    -1.298 \\
-0.228 \\
-1.545 \\
0.103 \\
-2.239 \\
    \end{pmatrix}, \;\;
    \boldsymbol{\omega_2} = \begin{pmatrix}
    -0.599 \\
-0.131 \\
0.061 \\
-0.940 \\
1.676 \\
    \end{pmatrix},
    \boldsymbol{\psi} = \begin{pmatrix}
0.105 \\
-0.080 \\
0.070 \\
0.039 \\
0.069 \\
    \end{pmatrix},
    \label{eq:coefficents_hopf}
\end{align}
where $\boldsymbol{\alpha}_i = (\alpha_{i,1},\alpha_{i,2},\ldots, \alpha_{i,5})^{\top}$ for $i = 1,2$, $\boldsymbol{\omega_i} = 
(\omega_{1,i},\omega_{2,i},\ldots,\omega_{5,i})^{\top}$
for $i = 1, 2$, $\boldsymbol{\theta} = 
(\theta_{1},\theta_{2},\ldots,\theta_{5})^{\top}$, and $\boldsymbol{\psi} = 
(\psi_{1},\psi_{2},\ldots,\psi_{5})^{\top}$.
The reduced ODEs are given by 
\begin{align}
\frac{\mathrm{d} \tilde{x}_1}{\mathrm{d} t} & = 
g_1(\tilde{x}_1, \tilde{x}_2; \lambda_1) = 260 + \tilde{x}_1 \sum_{j=1}^{5}\alpha_{1,j} 
\sigma_{1} \left(\omega_{j,1} \tilde{x}_1 + \omega_{j,2} \tilde{x}_2 + \psi_j \lambda_1 + \theta_{j} \right), \nonumber \\
\frac{\mathrm{d} \tilde{x}_2}{\mathrm{d} t} & = 
g_2(\tilde{x}_1, \tilde{x}_2; \lambda_1) = 260 + \tilde{x}_2 \sum_{j=1}^{5}\alpha_{2,j} 
\sigma_{1} \left(\omega_{j,1} \tilde{x}_1 + \omega_{j,2} \tilde{x}_2 + \psi_j \lambda_1+ \theta_{j} \right), 
\label{eq:single_layer_RRE_reduced_hopf}
\end{align}
 while the full ODEs read
 \small
\begin{align}
    \frac{\mathrm{d} x_1}{\mathrm{d}t} &= 260+ x_1\left(\sum_{j=1}^{5} \alpha_{1,j}y_j\right), \;\;
    &&\frac{\mathrm{d} x_2}{\mathrm{d}t} = 260+ x_2\left(\sum_{j=1}^{5} \alpha_{2,j}y_j\right), \nonumber \\
    \mu\frac{\mathrm{d} y_1}{\mathrm{d}t} &= 1 + \left( \sum_{i=1}^2 \omega_{1,i}x_i + \psi_1 \lambda_1 +\theta_{1} \right)y_1 -y_1^2, 
    \;\;&&\mu\frac{\mathrm{d} y_2}{\mathrm{d}t} = 1 + \left( \sum_{i=1}^2\omega_{2,i}x_i + \psi_2 \lambda_1 +  \theta_{2}\right)y_2 -y_2^2, \nonumber \\
    \mu\frac{\mathrm{d} y_3}{\mathrm{d}t} &= 1 + \left( \sum_{i=1}^2\omega_{3,i}x_i +  \psi_3 \lambda_1 + \theta_{3}\right)y_3 -y_3^2,
    \;\;&&\mu\frac{\mathrm{d} y_4}{\mathrm{d}t} = 1 + \left( \sum_{i=1}^2\omega_{4,i}x_i + \psi_4\lambda_1 + \theta_{4}\right)y_4 -y_4^2, \nonumber \\
    \mu\frac{\mathrm{d} y_5}{\mathrm{d}t} &= 1 + \left( \sum_{i=1}^2\omega_{5,i}x_i +\psi_5 \lambda_1 + \theta_{5}\right)y_5 -y_5^2, 
    \;\;&& \frac{\mathrm{d} \lambda_1}{\mathrm{d}t} = 0 ,
    \label{eq:full_rncrn_hopf}
\end{align}
\normalsize
Here, we assume general initial concentrations: $x_1(0)= a_1 \in [3.5, 6.5]$, $x_2(0)= a_2 \in [3.5, 6.5]$, and $y_1(0)= b_1 \geq 0$, $y_2(0)= b_2 \geq 0$, $y_3(0)= b_3 \geq 0$, $y_4(0)= b_4 \geq 0$, and $y_5(0)= b_5 \geq 0$. The coefficients were rounded to 3 decimal places to quote in the text, the full precision coefficients are available in the code repository. 

In Figure~\ref{fig:rncrn_inhert_bifs}, we used a range of  initial conditions for the executive species, but the chemical perceptron species always started at zero. For Figure~\ref{fig:rncrn_inhert_bifs} (a)  we used $\bar{x}_1(0) = 2, \bar{x}_2(0)=2$, and for the RNCRN approximation in  Figure~\ref{fig:rncrn_inhert_bifs} (b) we used $x_1(0) = 2, x_2(0)=2$. 

\subsection{Homoclinic bifurcation}
\label{sec:app_homoclinic_bif}
We use Algorithm~\ref{algh:para_RNCRN}, a modified version of Algorithm 1 from~\cite{dack_recurrent_2025} with static executive species, to approximate the target system~(\ref{eq:homoclinic_orbit_RRE}) on $\mathbb{K}_1 = [3.5,6.5], \mathbb{K}_2 =[3.5,6.5]$ and $\mathbb{L}= [1.9, 2.1]$ with a parametrized RNCRN. Tolerance $\varepsilon \approx 10^{-3}$ is met with an RNCRN with $M=10$ chemical perceptrons and coefficients $\beta_1=\beta_2 = 1$, $\gamma = 1$, and  
\begin{align}
    \boldsymbol{\alpha}_1 &= \begin{pmatrix}
    0.307 \\
    -0.533 \\
    0.324 \\
    0.999 \\
    4.348 \\
    1.820 \\
    -2.823 \\
    -0.744 \\
    -1.270 \\
    0.837 \\
    \end{pmatrix}, \; \; 
    \boldsymbol{\alpha}_2 = \begin{pmatrix} 
    0.812 \\
    0.923 \\
    -0.425 \\
    -0.260 \\
    -0.191 \\
    -2.821 \\
    -0.174 \\
    0.228 \\
    -0.349 \\
    -0.636 \\
    \end{pmatrix}, \;\;
    \boldsymbol{\theta} = \begin{pmatrix}
    0.790 \\
    1.455 \\
    -1.391 \\
    3.440 \\
    -1.901 \\
    3.461 \\
    0.657 \\
    0.634 \\
    1.932 \\
    -3.335 \\
    \end{pmatrix}, \;\; \nonumber \\
    \boldsymbol{\omega_1} &= \begin{pmatrix}
    0.372 \\
    -0.183 \\
    -0.205 \\
    0.448 \\
    -0.986 \\
    -0.056 \\
    -0.763 \\
    1.043 \\
    -0.253 \\
    0.271 \\
    \end{pmatrix}, \;\;
    \boldsymbol{\omega_2} = \begin{pmatrix}
    -0.697 \\
    0.084 \\
    0.835 \\
    -1.034 \\
    0.797 \\
    -1.153 \\
    0.025 \\
    -0.515 \\
    -0.296 \\
    0.335 \\
    \end{pmatrix},
    \boldsymbol{\psi} = \begin{pmatrix}
0.364 \\
-0.022 \\
-0.781 \\
-0.085 \\
0.062 \\
-0.023 \\
0.586 \\
-0.556 \\
0.323 \\
0.385 \\
    \end{pmatrix},
    \label{eq:coefficents_homoclinic}
\end{align}
where $\boldsymbol{\alpha}_i = (\alpha_{i,1},\alpha_{i,2},\ldots, \alpha_{i,10})^{\top}$ for $i = 1,2$, $\boldsymbol{\omega_i} = 
(\omega_{1,i},\omega_{2,i},\ldots,\omega_{10,i})^{\top}$
for $i = 1, 2$, $\boldsymbol{\theta} = 
(\theta_{1},\theta_{2},\ldots,\theta_{10})^{\top}$, and $\boldsymbol{\psi} = 
(\psi_{1},\psi_{2},\ldots,\psi_{10})^{\top}$. The reduced ODEs are given by 
\begin{align}
\frac{\mathrm{d} \tilde{x}_1}{\mathrm{d} t} & = 
g_1(\tilde{x}_1, \tilde{x}_2; \lambda_1) = 1 + \tilde{x}_1 \sum_{j=1}^{10}\alpha_{1,j} 
\sigma_{1} \left(\omega_{j,1} \tilde{x}_1 + \omega_{j,2} \tilde{x}_2 + \psi_j \lambda_1 + \theta_{j} \right), \nonumber \\
\frac{\mathrm{d} \tilde{x}_2}{\mathrm{d} t} & = 
g_2(\tilde{x}_1, \tilde{x}_2; \lambda_1) = 1 + \tilde{x}_2 \sum_{j=1}^{10}\alpha_{2,j} 
\sigma_{1} \left(\omega_{j,1} \tilde{x}_1 + \omega_{j,2} \tilde{x}_2 + \psi_j \lambda_1+ \theta_{j} \right), 
\label{eq:single_layer_RRE_reduced_homoclinic}
\end{align}
 while the full ODEs read
 \small
\begin{align}
    \frac{\mathrm{d} x_1}{\mathrm{d}t} &= 1+ x_1\left(\sum_{j=1}^{10} \alpha_{1,j}y_j\right), \;\;
    &&\frac{\mathrm{d} x_2}{\mathrm{d}t} = 1+ x_2\left(\sum_{j=1}^{10} \alpha_{2,j}y_j\right), \nonumber \\
    \mu\frac{\mathrm{d} y_1}{\mathrm{d}t} &= 1 + \left( \sum_{i=1}^2 \omega_{1,i}x_i + \psi_1 \lambda_1 +\theta_{1} \right)y_1 -y_1^2, 
    \;\;&&\mu\frac{\mathrm{d} y_2}{\mathrm{d}t} = 1 + \left( \sum_{i=1}^2\omega_{2,i}x_i + \psi_2 \lambda_1 +  \theta_{2}\right)y_2 -y_2^2, \nonumber \\
    \mu\frac{\mathrm{d} y_3}{\mathrm{d}t} &= 1 + \left( \sum_{i=1}^2\omega_{3,i}x_i +  \psi_3 \lambda_1 + \theta_{3}\right)y_3 -y_3^2,
    \;\;&&\mu\frac{\mathrm{d} y_4}{\mathrm{d}t} = 1 + \left( \sum_{i=1}^2\omega_{4,i}x_i + \psi_4\lambda_1 + \theta_{4}\right)y_4 -y_4^2, \nonumber \\
    \mu\frac{\mathrm{d} y_5}{\mathrm{d}t} &= 1 + \left( \sum_{i=1}^2\omega_{5,i}x_i +\psi_5 \lambda_1 + \theta_{5}\right)y_5 -y_5^2, 
    \;\;&&\mu\frac{\mathrm{d} y_6}{\mathrm{d}t} = 1 + \left( \sum_{i=1}^2\omega_{6,i}x_i + \psi_6 \lambda_1  + \theta_{6}\right)y_6 -y_6^2,  \nonumber \\
     \mu\frac{\mathrm{d} y_7}{\mathrm{d}t} &= 1 + \left( \sum_{i=1}^2\omega_{7,i}x_i +\psi_7 \lambda_1 + \theta_{7}\right)y_7 -y_7^2, 
    \;\;&&\mu\frac{\mathrm{d} y_8}{\mathrm{d}t} = 1 + \left( \sum_{i=1}^2\omega_{8,i}x_i + \psi_8 \lambda_1  + \theta_{8}\right)y_8 -y_8^2,  \nonumber \\
    \mu\frac{\mathrm{d} y_9}{\mathrm{d}t} &= 1 + \left( \sum_{i=1}^2\omega_{9,i}x_i +\psi_9 \lambda_1 + \theta_{9}\right)y_9 -y_9^2, 
    \;\;&&\mu\frac{\mathrm{d} y_{10}}{\mathrm{d}t} = 1 + \left( \sum_{i=1}^2\omega_{10,i}x_i + \psi_{10} \lambda_1  + \theta_{10}\right)y_{10} -y_{10}^2, 
    \nonumber \\
    \frac{\mathrm{d} \lambda_1}{\mathrm{d}t} &= 0
    \label{eq:full_rncrn_homoclinic}
\end{align}
\normalsize
Here, we assume general initial concentrations: $x_1(0)= a_1 \in [3, 7]$, $x_2(0)= a_2 \in [3, 7]$, and $y_1(0)= b_1 \geq 0$, $y_2(0)= b_2 \geq 0$, $y_3(0)= b_3 \geq 0$, $y_4(0)= b_4 \geq 0$, $y_5(0)= b_5 \geq 0$, $y_6(0)= b_6 \geq 0$, $y_7(0)= b_7 \geq 0$, $y_8(0)= b_8 \geq 0$, $y_9(0)= b_9 \geq 0$, and $y_{10}(0)= b_{10} \geq 0$ . The coefficients were rounded to 3 decimal places to quote in the text, the full precision coefficients are available in the code repository. A lower absolute tolerance was required in this example as the homoclinic orbit is a delicate dynamical behaviour to approximate effectively. 

In Figure~\ref{fig:rncrn_inhert_bifs}, we used a range of  initial conditions for the executive species, but the chemical perceptron species always started at zero. For the homoclinic bifurcation we used slightly different executive species concentrations due to the sensitivity of the homoclinic orbit. In Figure~\ref{fig:rncrn_inhert_bifs} (c) we used: $\bar{x}_1(0) = 3.5$, $\bar{x}_2(0)=6.0$ for the sub-panel with $\bar{\lambda}_1=1.9$; and  $\bar{x}_1(0) = 3.5$, $\bar{x}_2(0)=6.0435 $ for the sub-panel with $\bar{\lambda}_1=2$ and  $\bar{\lambda}_1=2.1$. In Figure~\ref{fig:rncrn_inhert_bifs} (d) we used:  $x_1(0) = 3.5$, $x_2(0)=6.0$ for the sub-panel with $\lambda_1=1.9$; $x_1(0) = 3.5$, $x_2(0)=6.04$ for the sub-panel with $\lambda_1=1.997$; and  $x_1(0) = 3.5$, $x_2(0)=6.0435$ for the sub-panel with $\lambda_1=2.1$.

\subsection{Hopf bifurcation by training at criticality alone}
\label{sec:app_hopf_bif_critical}
We now demonstrate that by training a RNCRN at the critical point of a Hopf bifurcation it is possible to find random perturbation of the RNCRN's parameters that recreate the qualitative behaviour of a Hopf bifurcation without any data about either side of the bifurcation being used during training. 

\begin{figure}[hbt!]
    \centering
    \includegraphics[width=\linewidth]{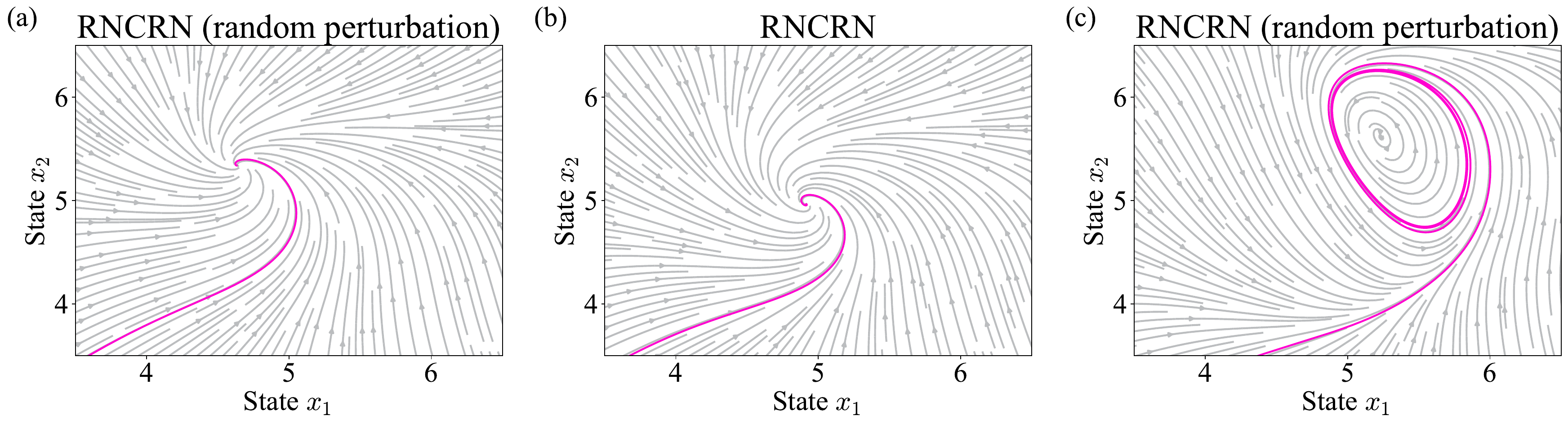}
    \caption{ An RNCRN trained at the critical point of a Hopf bifurcation, i.e. ~(\ref{eq:target_hopf_critical}), demonstrating extrapolation of dynamics to behaviours outside of the training data. Panel (b) shows the $M=6$ chemical perceptron RNCRN trained to approximate~(\ref{eq:target_hopf_critical}) while panels (a) and (c) show the same RNCRN approximation with the rates randomly perturbed (see Appendix~\ref{sec:app_hopf_bif_critical} for the rates of the RNCRNs). All panels show their respective reduced RNCRN vector field as grey arrows and a $(x_1(t), x_2(t))$-time-trajectories with  $\mu=0.01$, and initial concentrations $x_1(0)=x_2(0) = 2$ and $y_1(0)=\dots = y_6(0) = 0$ in magenta.  } 
    \label{fig:inherit_bif_from_random_bif}
\end{figure}

Consider the target ODE for a Hopf bifurcation at its critical point with an equilibrium at $(x_1, x_2) = (5,5)$ 
\begin{align}
    \frac{\mathrm{d} \bar{x}_1}{\mathrm{d} t} &= ( - (\bar{x}_1-5)^2 - (\bar{x}_2-5)^2)(\bar{x}_1-5) -  (\bar{x}_2-5),\nonumber \\
     \frac{\mathrm{d} \bar{x}_2}{\mathrm{d} t} &= (- (\bar{x}_1-5)^2 - (\bar{x}_2-5)^2)(\bar{x}_2-5) +  (\bar{x}_1-5).
     \label{eq:target_hopf_critical}
\end{align}
In Figure~\ref{fig:rncrn_inhert_bifs}(a) the $\bar{\lambda}=2$ plot visualises the dynamics of~(\ref{eq:target_hopf_critical}). We train an RNCRN to approximate~(\ref{eq:target_hopf_critical}) and show the associated vector field of the reduced RNCRN and an example $(x_1(t), x_2(t))$ trajectory in Figure~\ref{fig:inherit_bif_from_random_bif}(b). 
In Figures~\ref{fig:inherit_bif_from_random_bif}(a) and (c) we show that by inducing a particular but randomly sampled perturbation of the rate constants we can induce unistable or oscillatory behaviour. Unlike the parametrized RNCRN the training of an RNCRN at criticality alone does not use data about bifurcation parameter, instead this behaviour emerges, showing that the RNCRN is reproducing key features of the mathematical structure of the underlying ODE (\ref{eq:target_hopf_critical}). 

We use Algorithm~1 from~\cite{dack_recurrent_2025} to approximate the target system~(\ref{eq:target_hopf_critical}) on $\mathbb{K}_1 = [3.5,6.5], \mathbb{K}_2 = [3.5,6.5]$ with a single-layer RNCRN (not parameterized). Tolerance $\varepsilon \approx 10^{0}$ is met with an RNCRN with $M=6$ chemical perceptrons and coefficients $\beta_1=\beta_2 = 1$, $\gamma = 1$, and 
\begin{align}
    \boldsymbol{\alpha}_1 &= \begin{pmatrix}
    0.492 \\
-0.117 \\
2.229 \\
-0.708 \\
-1.146 \\
1.363 \\
    \end{pmatrix}, \; \; 
    \boldsymbol{\alpha}_2 = \begin{pmatrix} 
    -0.283 \\
0.657 \\
2.199 \\
-1.294 \\
-2.632 \\
3.118 \\
    \end{pmatrix}, \;\;
    \boldsymbol{\theta} = \begin{pmatrix}
    2.243 \\
10.838 \\
10.718 \\
-3.164 \\
7.959 \\
3.002 \\
    \end{pmatrix}, \;\; \nonumber \\
    \boldsymbol{\omega_1} &= \begin{pmatrix}
    -3.672 \\
0.911 \\
-1.391 \\
0.241 \\
-0.454 \\
-0.077 \\
    \end{pmatrix}, \;\;
    \boldsymbol{\omega_2} = \begin{pmatrix}
    2.091 \\
-3.981 \\
-1.433 \\
1.058 \\
-0.922 \\
-0.073 \\
    \end{pmatrix},
    \label{eq:coefficents_hopf_critical}
\end{align}
where $\boldsymbol{\alpha}_i = (\alpha_{i,1},\alpha_{i,2},\ldots, \alpha_{i,6})^{\top}$ for $i = 1,2$, $\boldsymbol{\omega_i} = 
(\omega_{1,i},\omega_{2,i},\ldots,\omega_{6,i})^{\top}$
for $i = 1, 2$, and $\boldsymbol{\theta} = 
(\theta_{1},\theta_{2},\ldots,\theta_{6})^{\top}$. The reduced ODEs are given by 
\begin{align}
\frac{\mathrm{d} \tilde{x}_1}{\mathrm{d} t} & = 
g_1(\tilde{x}_1, \tilde{x}_2) = 1 + \tilde{x}_1 \sum_{j=1}^{6}\alpha_{1,j} 
\sigma_{1} \left(\omega_{j,1} \tilde{x}_1 + \omega_{j,2} \tilde{x}_2 + \theta_{j} \right), \nonumber \\
\frac{\mathrm{d} \tilde{x}_2}{\mathrm{d} t} & = 
g_2(\tilde{x}_1, \tilde{x}_2) = 1 + \tilde{x}_2 \sum_{j=1}^{6}\alpha_{2,j} 
\sigma_{1} \left(\omega_{j,1} \tilde{x}_1 + \omega_{j,2} \tilde{x}_2 + \theta_{j} \right), 
\label{eq:single_layer_RRE_reduced_hopf_critical}
\end{align}
 while the full ODEs read
 \small
\begin{align}
    \frac{\mathrm{d} x_1}{\mathrm{d}t} &= 1+ x_1\left(\sum_{j=1}^{6} \alpha_{1,j}y_j\right), \;\;
    &&\frac{\mathrm{d} x_2}{\mathrm{d}t} = 1+ x_2\left(\sum_{j=1}^{6} \alpha_{2,j}y_j\right), \nonumber \\
    \mu\frac{\mathrm{d} y_1}{\mathrm{d}t} &= 1 + \left( \sum_{i=1}^2 \omega_{1,i}x_i +\theta_{1} \right)y_1 -y_1^2, 
    \;\;&&\mu\frac{\mathrm{d} y_2}{\mathrm{d}t} = 1 + \left( \sum_{i=1}^2\omega_{2,i}x_i +  \theta_{2}\right)y_2 -y_2^2, \nonumber \\
    \mu\frac{\mathrm{d} y_3}{\mathrm{d}t} &= 1 + \left( \sum_{i=1}^2\omega_{3,i}x_i + \theta_{3}\right)y_3 -y_3^2,
    \;\;&&\mu\frac{\mathrm{d} y_4}{\mathrm{d}t} = 1 + \left( \sum_{i=1}^2\omega_{4,i}x_i + \theta_{4}\right)y_4 -y_4^2, \nonumber \\
    \mu\frac{\mathrm{d} y_5}{\mathrm{d}t} &= 1 + \left( \sum_{i=1}^2\omega_{5,i}x_i + \theta_{5}\right)y_5 -y_5^2, 
    \;\;&&\mu\frac{\mathrm{d} y_6}{\mathrm{d}t} = 1 + \left( \sum_{i=1}^2\omega_{6,i}x_i + \theta_{6}\right)y_6 -y_6^2, 
    \label{eq:full_rncrn_hopf_critial}
\end{align}
\normalsize
Here, we assume general initial concentrations: $x_1(0)= a_1 \in [3.5,6.5]$, $x_2(0)= a_2 \in [3.5,6.5]$, and $y_1(0)= b_1 \geq 0$, $y_2(0)= b_2 \geq 0$, $y_3(0)= b_3 \geq 0$, $y_4(0)= b_4 \geq 0$, $y_5(0)= b_5 \geq 0$, and $y_6(0)= b_6 \geq 0$. The coefficients were rounded to 3 decimal places to quote in text, the full precision coefficients are available in the code repository.

{\textbf{Randomly perturbed rates.}} By randomly perturbing the rate constants in equation~(\ref{eq:coefficents_hopf_critical}) with $10\%$ uniform noise we found sets of rate constants that lead to the oscillator dynamical behaviour in a Hopf bifurcation e.g., 
\begin{align}
    \boldsymbol{\alpha}_1 &= \begin{pmatrix}
    0.532 \\
-0.106 \\
2.315 \\
-0.693 \\
-1.116 \\
1.467 \\
    \end{pmatrix}, \; \; 
    \boldsymbol{\alpha}_2 = \begin{pmatrix} 
    -0.286 \\
0.645 \\
1.989 \\
-1.231 \\
-2.614 \\
3.228 \\
    \end{pmatrix}, \;\;
    \boldsymbol{\theta} = \begin{pmatrix}
    2.025 \\
10.419 \\
11.140 \\
-3.029 \\
8.754 \\
3.069 \\
    \end{pmatrix}, \;\; \nonumber \\
    \boldsymbol{\omega_1} &= \begin{pmatrix}
    -3.912 \\
0.843 \\
-1.345 \\
0.222 \\
-0.481 \\
-0.083 \\
    \end{pmatrix}, \;\;
    \boldsymbol{\omega_2} = \begin{pmatrix}
    2.059 \\
-3.679 \\
-1.552 \\
1.070 \\
-0.929 \\
-0.075 \\
    \end{pmatrix},
    \label{eq:coefficents_hopf_critical_rand_osc}
\end{align}
and, for completeness, we found examples of unistable behaviour including
\begin{align}
    \boldsymbol{\alpha}_1 &= \begin{pmatrix}
0.493 \\
-0.117 \\
2.223 \\
-0.711 \\
-1.147 \\
1.355 \\
    \end{pmatrix}, \; \; 
    \boldsymbol{\alpha}_2 = \begin{pmatrix} 
-0.284 \\
0.657 \\
2.192 \\
-1.299 \\
-2.609 \\
3.110 \\
    \end{pmatrix}, \;\;
    \boldsymbol{\theta} = \begin{pmatrix}
2.220 \\
10.930 \\
10.815 \\
-3.142 \\
7.890 \\
2.991 \\
    \end{pmatrix}, \;\; \nonumber \\
    \boldsymbol{\omega_1} &= \begin{pmatrix}
-3.667 \\
0.920 \\
-1.396 \\
0.241 \\
-0.451 \\
-0.077 \\
    \end{pmatrix}, \;\;
    \boldsymbol{\omega_2} = \begin{pmatrix}
2.085 \\
-4.016 \\
-1.432 \\
1.065 \\
-0.924 \\
-0.073 \\
    \end{pmatrix}.
\label{eq:coefficents_hopf_critical_rand_stable}
\end{align}

\newpage
\section{Appendix: Piecewise systems}
\label{app:dynamical_piecewises}
We now present the RNCRNs associated with Section~\ref{app:dynamical_piecewises}.

\subsection{Simple unistable-bistable piecewise system}
\label{sec:simple_uni_bi_piecewise}

We use Algorithm~\ref{algh:para_RNCRN_with_AD}, a modified version of Algorithm 1 from~\cite{dack_recurrent_2025} with static executive species augmented with automatic differentiation techniques, to approximate the target system~(\ref{eq:point_con_piecewise}) on $\mathbb{K}_{1} = [1,9]$ and $\mathbb{L}_{1} = \{1\}, \mathbb{L}_{2} = \{0\}$ with a parametrized RNCRN. Tolerance $\varepsilon \approx 10^{1}$ is met with an RNCRN with $M=3$ chemical perceptrons and coefficients
\begin{align}
    \frac{\mathrm{d} x_1}{\mathrm{d}t} &= 35.700 -54.403x_1y_1  +188.696x_1y_2  -112.261x_1y_3, &&x_1(0) = a_1 \in \mathbb{K}_1,\nonumber\\
    \mu\frac{\mathrm{d} y_1}{\mathrm{d}t} &= 52.618 +137.501x_1y_1 -477.661\lambda_1y_1 -1513.810y_1 -90.515y_1^2, &&y_1(0) = b_1 \geq 0, \nonumber \\
    \mu\frac{\mathrm{d} y_2}{\mathrm{d}t} &= 179.420 -249.394x_1y_2 -21494.744\lambda_1y_2 -220.319y_2 -12.931y_2^2, &&y_2(0) = b_2  \geq 0, \nonumber \\
    \mu\frac{\mathrm{d} y_3}{\mathrm{d}t} &= 130.184 -225.249x_1y_3 -1876.261\lambda_1y_3 + 854.494 y_3 -1122.973y_3^2, &&y_3(0) = b_3  \geq 0,  \nonumber \\
    \frac{\mathrm{d} \lambda_1}{\mathrm{d}t} &= 0,  && \lambda_1(0) =   c_1 \in \mathbb{K}_{\boldsymbol{r}}.
    \label{eq:toogle_piecewise_ex_rncrn}
\end{align}
The reduced vector field of equation~(\ref{eq:toogle_piecewise_ex_rncrn}) is
\begin{align}
    \frac{\mathrm{d} \tilde{x}_1}{\mathrm{d}t} &= 35.700 -54.403\tilde{x}_1\tilde{y}_1  +188.696\tilde{x}_1\tilde{y}_2  -112.261\tilde{x}_1\tilde{y}_3, &&x_1(0) = a_1 \in \mathbb{K}_1,\nonumber\\
    \tilde{y}_1 &=  \sigma_{52.618, 90.515}(137.501\tilde{x}_1 -477.661\tilde{\lambda}_1 -1513.810), \nonumber \\
    \tilde{y}_2 &=  \sigma_{179.420, 12.931}(-249.394\tilde{x}_1 -21494.744\tilde{\lambda}_1 -220.319), \nonumber \\
    \tilde{y}_3 &=  \sigma_{130.184, 1122.973}(-225.249\tilde{x}_1 -1876.261\tilde{\lambda}_1 + 854.494 ), \nonumber \\
    \tilde{\lambda}_1 &= \lambda(0) = c_1
    \label{eq:reduced_toogle_piecewise_ex_rncrn}
\end{align}
where the chemical activation function is defined as in equation~(\ref{eq:full_chemical_activation_fn}).

\subsection{XOR-gated unistable-bistable piecewise system}
\label{app:sar_piecewise_struct}
We used Algorithm~\ref{algh:class_controlled_RNCRN} to train a classification-controlled RNCRN, i.e. equation~(\ref{eq:modular_sac_RRE})-(\ref{eq:modular_sac_RRE_0}), to approximate the more complicated unistable-bistable piecewise systems~(\ref{eq:class_con_piecewise}). Intuitively, Algorithm ~\ref{algh:class_controlled_RNCRN} first trains a parametrized RNCRN to approximate a bifurcations between the desired dynamical dynamical regimes. Then, a feed-forward network of chemical perceptrons is trained to control the chemical concentration of the bifurcation parameter according to the desired classification boundaries of the environmental species. \\

In this appendix we walk through the steps of Algorithm~\ref{algh:class_controlled_RNCRN} in the context of the example target system~(\ref{eq:class_con_piecewise}). 

{\textbf{\textit{Step 1. }Point-wise parametrized RNCRN.}} For ease of demonstration, we use the RNCRN found in Section~\ref{sec:point_controlled_piecewises}. We note from Figure~\ref{fig:toggle_example}(e) that the bifurcation point is approximately at $10^{-2}$. Otherwise, apply the step \textbf{(a)} from Algorithm~\ref{algh:class_controlled_RNCRN}.
\\

{\textbf{\textit{Step 2.} Quasi-static classification CRN.}} We want to create a feed-forward classification that embeds the XOR logic from the target equation~(\ref{eq:class_con_piecewise}) into a chemical reaction network and triggers the appropriate dynamical regime. That is, we train a neural CRN to approximate the function
\begin{align}
    \bar{r}(\bar{\lambda_1}, \bar{\lambda_2}) = \begin{cases}
        1 \text{ for all } (\bar{\lambda_1}, \bar{\lambda_2}) \in \mathbb{L}_1 = \left(\mathbb{K}_1 \times \mathbb{K}_2 \right) \cup \left(\mathbb{K}_2 \times \mathbb{K}_1 \right) \\
        0  \text{ for all } (\bar{\lambda_1}, \bar{\lambda_2}) \in \mathbb{L}_2 = \left(\mathbb{K}_1 \times \mathbb{K}_1 \right) \cup \left(\mathbb{K}_2 \times \mathbb{K}_2 \right) ,
    \end{cases} 
\end{align}
where $\mathbb{K}_1 = [0, 1)$ and $\mathbb{K}_2 = [1, 2]$. We define a structure of $K=4$  chemical perceptrons~\cite{anderson_reaction_2021} that catalytically senses the environmental species $\Lambda_1$ and $\Lambda_2$. These sense-chemical-perceptrons then feed into a bifurcation species $R$, which also has a chemical perceptron structure, that will later be connected to the parametrized RNCRN in place of the bifurcation parameter. The corresponding RRE of the aforementioned feed-forward chemical perceptron structure is defined as 
\begin{align}
    \mu\frac{\mathrm{d} z_1}{\mathrm{d}t} &= 0.01 +2.726\lambda_1z_1 -3.025\lambda_2 z_1 +0.296z_1 -z_1^2, &&z_1(0) = c_1 \geq 0, \nonumber \\
    \mu\frac{\mathrm{d} z_2}{\mathrm{d}t} &= 0.01 +2.259\lambda_1z_2 +0.057\lambda_2 z_2 -2.328z_2 -z_2^2, &&z_2(0) = c_2 \geq 0, \nonumber \\
    \mu\frac{\mathrm{d} z_3}{\mathrm{d}t} &= 0.01 -0.051\lambda_1z_3 -2.511\lambda_2 z_3 +2.582z_3-z_3^2, &&z_3(0) = c_3 \geq 0, \nonumber \\
    \mu\frac{\mathrm{d} z_4}{\mathrm{d}t} &= 0.01 +0.158\lambda_1z_4 -0.156\lambda_2 z_4 +0.666z_4 -z_4^2, &&z_4(0) = c_4 \geq 0, \nonumber \\
    \mu\frac{\mathrm{d} r}{\mathrm{d}t} &= 0.01 + 4.528 z_1r -5.719 z_2r -5.648z_3r + 0.777z_4r +0.682r -r^2, &&r(0) = d \geq 0,\nonumber \\
    \frac{\mathrm{d} \lambda_1}{\mathrm{d}t} &= 0 , &&\lambda_1(0) = e_1 \geq 0,\nonumber \\
    \frac{\mathrm{d} \lambda_2}{\mathrm{d}t} &= 0 , &&\lambda_2(0) = e_2 \geq 0,
    \label{eq:modular_sac}
\end{align}
where $z_1, \dots, z_4$ are the time-dependent molecular concentrations of the sense-chemical-perceptron species, $\lambda_1$ and $\lambda_2$ are the molecular concentrations of the environmental species (which are assumed to be static), and r is the molecular concentration of the bifurcation species. We found these rate constants by training an ANN with the chemical activation function, $\sigma_\gamma$, using standard machine learning methods then mapping the parameters back to the full chemical perceptron RRE with an additional timescale parameter $\mu$. 
\\

{\textbf{\textit{Step 3.} Dynamical approximation.}} The complete classification-controlled RNCRN is constructed from the combination of the chemical reaction networks that produce reaction-rate equation~(\ref{eq:toogle_piecewise_ex_rncrn}) and reaction-rate equation~(\ref{eq:modular_sac}) as shown in Figure~\ref{fig:sac_results} the $\mu = 0.001$ parameter was found to be sufficient. 

\begin{align}
    \frac{\mathrm{d} x_1}{\mathrm{d}t} &= 35.700 -54.403x_1y_1  +188.696x_1y_2  -112.261x_1y_3, &&x_1(0) = a_1 \in \mathbb{K}_1,\nonumber\\
    \mu\frac{\mathrm{d} y_1}{\mathrm{d}t} &= 52.618 +137.501x_1y_1 -477.661ry_1 -1513.810y_1 -90.515y_1^2, &&y_1(0) = b_1 \geq 0, \nonumber \\
    \mu\frac{\mathrm{d} y_2}{\mathrm{d}t} &= 179.420 -249.394x_1y_2 -21494.744ry_2 -220.319y_2 -12.931y_2^2, &&y_2(0) = b_2  \geq 0, \nonumber \\
    \mu\frac{\mathrm{d} y_3}{\mathrm{d}t} &= 130.184 -225.249x_1y_3 -1876.261ry_3 + 854.494 y_3 -1122.973y_3^2, &&y_3(0) = b_3  \geq 0,  \nonumber \\
    \mu\frac{\mathrm{d} z_1}{\mathrm{d}t} &= 0.01 +2.726\lambda_1z_1 -3.025\lambda_2 z_1 +0.296z_1 -z_1^2, &&z_1(0) = c_1 \geq 0, \nonumber \\
    \mu\frac{\mathrm{d} z_2}{\mathrm{d}t} &= 0.01 +2.259\lambda_1z_2 +0.057\lambda_2 z_2 -2.328z_2 -z_2^2, &&z_2(0) = c_2 \geq 0, \nonumber \\
    \mu\frac{\mathrm{d} z_3}{\mathrm{d}t} &= 0.01 -0.051\lambda_1z_3 -2.511\lambda_2 z_3 +2.582z_3-z_3^2, &&z_3(0) = c_3 \geq 0, \nonumber \\
    \mu\frac{\mathrm{d} z_4}{\mathrm{d}t} &= 0.01 +0.158\lambda_1z_4 -0.156\lambda_2 z_4 +0.666z_4 -z_4^2, &&z_4(0) = c_4 \geq 0, \nonumber \\
    \mu\frac{\mathrm{d} r}{\mathrm{d}t} &= 0.01 + 4.528 z_1r -5.719 z_2r -5.648z_3r + 0.777z_4r +0.682r -r^2, &&r(0) = d \geq 0,\nonumber \\
    \frac{\mathrm{d} \lambda_1}{\mathrm{d}t} &= 0 , &&\lambda_1(0) = e_1 \geq 0,\nonumber \\
    \frac{\mathrm{d} \lambda_2}{\mathrm{d}t} &= 0 , &&\lambda_2(0) = e_2 \geq 0,
    \label{eq:full_class_piecewise}
\end{align}

\section{Appendix: Example data-defined limit cycles and switching}
\label{app:data_defined_attractors}
\begin{figure}[hbt!]
    \centering
    \includegraphics[width=\linewidth]{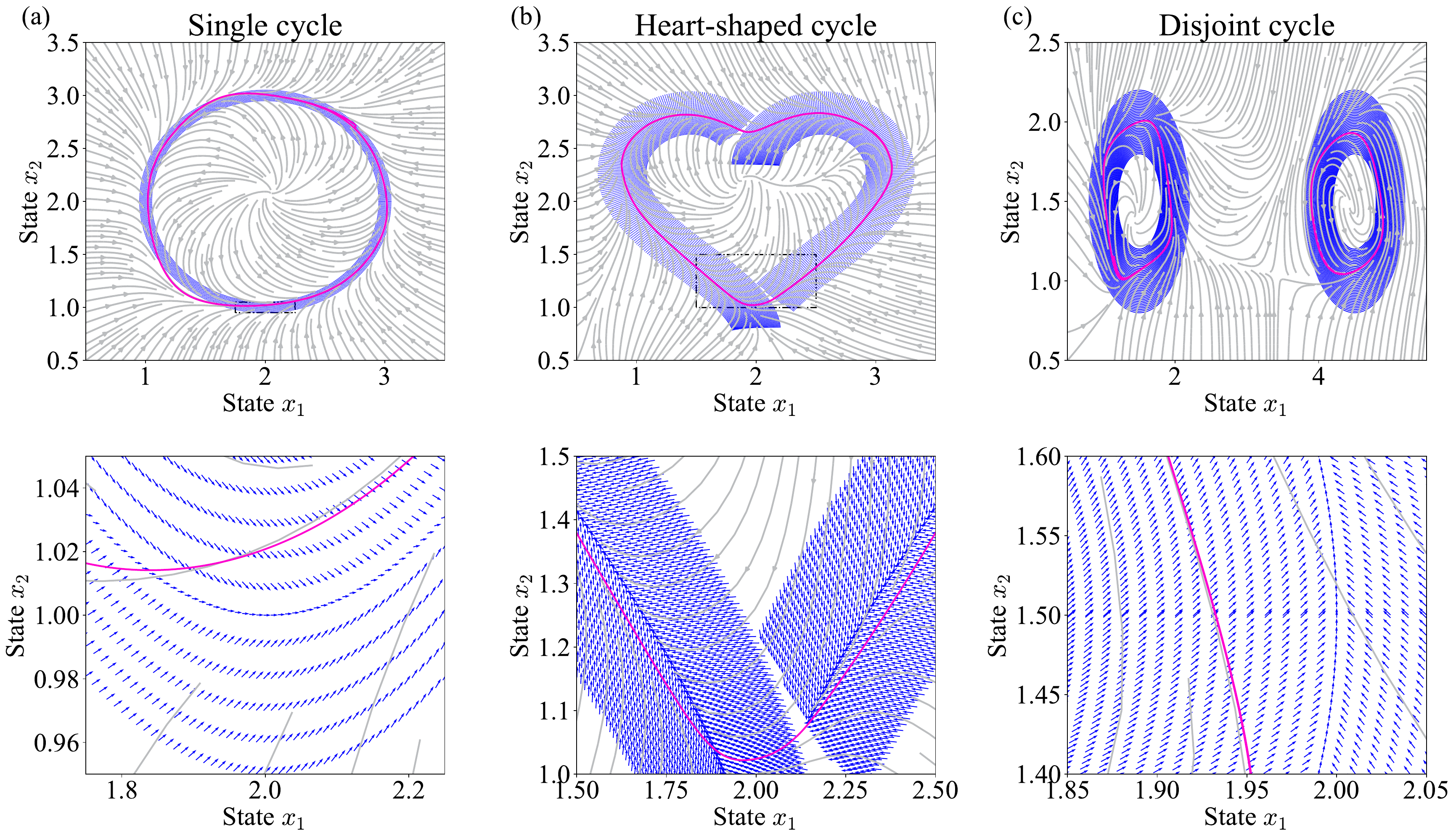}
    \caption{Artificial data used to train the data-defined attractors plotted alongside the reduced vector field (grey) and time-trajectories for the trained RNCRN shown in Figure~\ref{fig:attract_defined_dynamics}. }
    \label{fig:attract_defined_dynamics_close_up}
\end{figure}

We now train RNCRNs on training data generated by Algorithm~\ref{algh:define_attractor} to learn data-defined dynamics. In Figure~\ref{fig:attract_defined_dynamics_close_up} we emphasize the artificial data generated by Algorithm~\ref{algh:define_attractor} by showing close-ups of the data used in the examples in Section~\ref{sec:data_defined_dynamics}.

\label{sec:app_data_def_attractors}
\subsection{Single circular limit cycle (attractor)}
\label{sec:app_single_cycle}
Using the parametric equation defined in Section~\ref{sec:single_cycle}, Algorithm~\ref{algh:define_attractor}, and Algorithm~\ref{algh:data_def_RNCRN_with_AD} we found an RNCRN with $M=5$ chemical perceptrons and coefficients $\beta_1= 16.861$, $\beta_2 = 13.955$, and  
\begin{align}
    \boldsymbol{\alpha}_1 &= \begin{pmatrix}
    -22.642 \\
-3.214 \\
-13.344 \\
-5.983 \\
11.413 \\
    \end{pmatrix}, \; \; 
    \boldsymbol{\alpha}_2 = \begin{pmatrix} 
    20.600 \\
-31.984 \\
4.736 \\
-9.337 \\
-11.829 \\
    \end{pmatrix}, \;\;
    \boldsymbol{\theta} = \begin{pmatrix}
    70.437 \\
-86.509 \\
-62.502 \\
139.089 \\
77.599 \\
    \end{pmatrix}, \;\; \nonumber \\
    \boldsymbol{\omega_1} &= \begin{pmatrix}
    -46.744 \\
-21.405 \\
63.399 \\
-31.199 \\
20.670 \\
    \end{pmatrix}, \;\;
    \boldsymbol{\omega_2} = \begin{pmatrix}
    -1.875 \\
30.682 \\
-47.177 \\
-27.589 \\
-48.644 \\
    \end{pmatrix}, \nonumber \\
    \boldsymbol{\gamma} &= \begin{pmatrix}
    22.316 \\
23.786 \\
4.064 \\
0.484 \\
2.106 \\
    \end{pmatrix}, \;\;
    \boldsymbol{\tau} = \begin{pmatrix}
    111.885 \\
106.208 \\
113.115 \\
68.879 \\
94.832 \\
    \end{pmatrix},
    \label{eq:coefficents_data_cycle_att}
\end{align}
where $\boldsymbol{\alpha}_i = (\alpha_{i,1},\alpha_{i,2},\ldots, \alpha_{i,5})^{\top}$ for $i = 1,2$, $\boldsymbol{\omega_i} = 
(\omega_{1,i},\omega_{2,i},\ldots,\omega_{5,i})^{\top}$
for $i = 1, 2$, $\boldsymbol{\theta} = 
(\theta_{1},\theta_{2},\ldots,\theta_{5})^{\top}$,  $\boldsymbol{\gamma} = 
(\gamma_{1},\gamma_{2},\ldots,\gamma_{5})^{\top}$, and  $\boldsymbol{\tau} = 
(\tau_{1},\tau_{2},\ldots,\tau_{5})^{\top}$.
The reduced ODEs are given by 
\begin{align}
\frac{\mathrm{d} \tilde{x}_1}{\mathrm{d} t} & = 
g_1(\tilde{x}_1, \tilde{x}_2) = 16.861 + \tilde{x}_1 \sum_{j=1}^{5}\alpha_{1,j} 
\sigma_{\gamma_j,\tau_j} \left(\omega_{j,1} \tilde{x}_1 + \omega_{j,2} \tilde{x}_2 + \theta_{j} \right), \nonumber \\
\frac{\mathrm{d} \tilde{x}_2}{\mathrm{d} t} & = 
g_2(\tilde{x}_1, \tilde{x}_2) = 13.955 + \tilde{x}_2 \sum_{j=1}^{5}\alpha_{2,j} 
\sigma_{\gamma_j,\tau_j} \left(\omega_{j,1} \tilde{x}_1 + \omega_{j,2} \tilde{x}_2 + \theta_{j} \right), 
\label{eq:single_layer_RRE_reduced_cycle}
\end{align}
 while the full ODEs read
 \small
\begin{align}
    \frac{\mathrm{d} x_1}{\mathrm{d}t} &= 16.861 + x_1\left(\sum_{j=1}^{5} \alpha_{1,j}y_j\right), \;\;
    &&\frac{\mathrm{d} x_2}{\mathrm{d}t} = 13.955 + x_2\left(\sum_{j=1}^{5} \alpha_{2,j}y_j\right), \nonumber \\
    \mu\frac{\mathrm{d} y_1}{\mathrm{d}t} &= 22.316 + \left( \sum_{i=1}^2 \omega_{1,i}x_i + \theta_{1} \right)y_1 - 111.885y_1^2, 
    \;\;&&\mu\frac{\mathrm{d} y_2}{\mathrm{d}t} = 23.786 + \left( \sum_{i=1}^2\omega_{2,i}x_i +  \theta_{2}\right)y_2 -106.208y_2^2, \nonumber \\
    \mu\frac{\mathrm{d} y_3}{\mathrm{d}t} &= 4.064 + \left( \sum_{i=1}^2\omega_{3,i}x_i + \theta_{3}\right)y_3 -113.115y_3^2,
    \;\;&&\mu\frac{\mathrm{d} y_4}{\mathrm{d}t} = 0.484 + \left( \sum_{i=1}^2\omega_{4,i}x_i + \theta_{4}\right)y_4 -68.879y_4^2, \nonumber \\
    \mu\frac{\mathrm{d} y_5}{\mathrm{d}t} &= 2.106 + \left( \sum_{i=1}^2\omega_{5,i}x_i + \theta_{5}\right)y_5 -94.832y_5^2, 
\end{align}
\normalsize
Here, we assume general initial concentrations: $x_1(0)= a_1 > 0$, $x_2(0)= a_2 > 0$, and $y_1(0)= b_1 \geq 0$, $y_2(0)= b_2 \geq 0$, $y_3(0)= b_3 \geq 0$, $y_4(0)= b_4 \geq 0$, and $y_5(0)= b_5 \geq 0$. The coefficients were rounded to 3 decimal places to quote in text, the full precision coefficients are available in the code repository.

\subsection{Single circular limit cycle (repeller) }
\label{sec:app_single_cycle_repeller}
\begin{figure}[hbt!]
    \centering
    \includegraphics[width=0.67\linewidth]{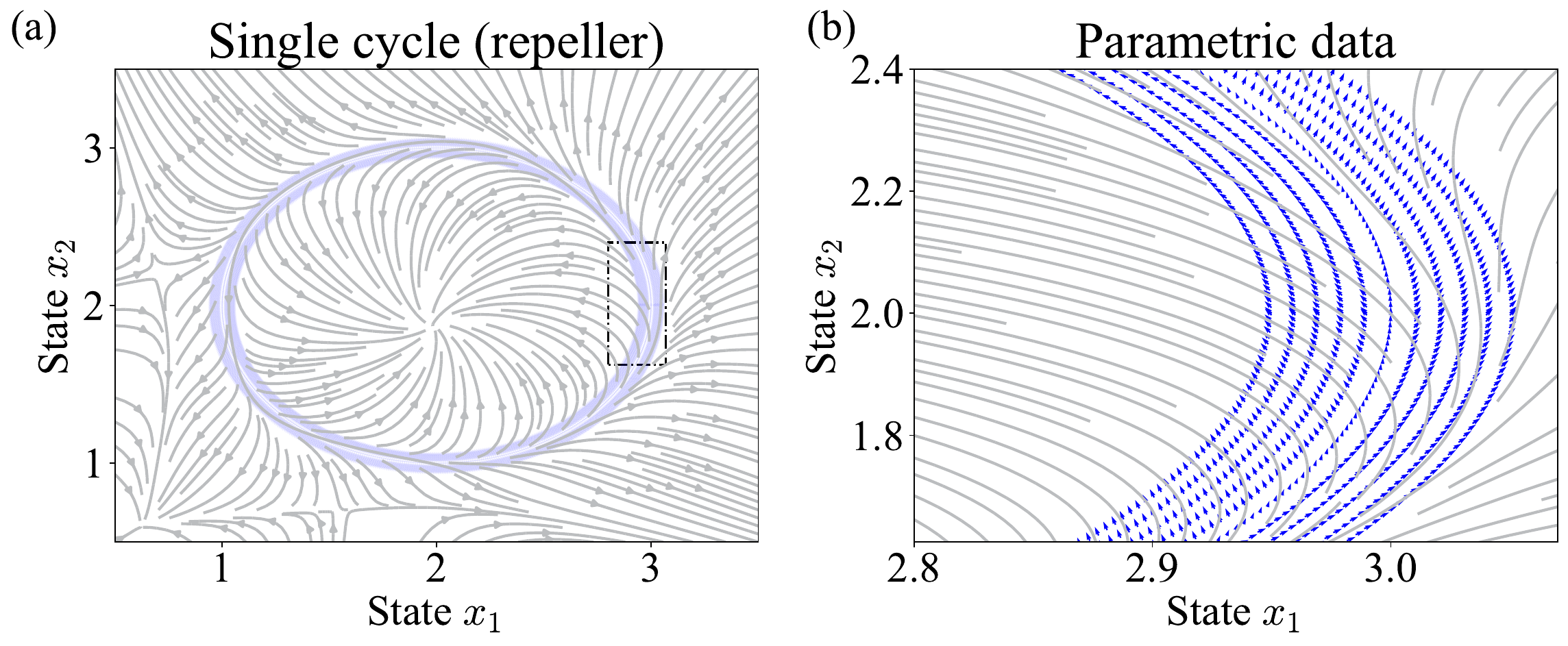}
    \caption{Example RNCRN trained on data to realise a circular repeller. In blue the target repeller data that was created by sampling points from parametric equation~(\ref{eq:parametric_circle}) then applying a vector field interpolation algorithm, i.e. Algorithm~\ref{algh:define_attractor}. In grey the reduced RNCRN vector field for a $5$ chemical perceptron RNCRN (see Appendix~\ref{sec:app_single_cycle_repeller} for parameters). Panel (a) shows the entire state-space of interest while panel (b) shows a close-up of the state-space to emphasise the training data.   }
    \label{fig:repeller_circle}
\end{figure}

As described in Section~\ref{sec:single_cycle} training data generated by Algorithm~\ref{algh:define_attractor} (with an inverted normal direction) and equation~(\ref{eq:parametric_circle}) was used to train an RNCRN using the approach laid out in and Algorithm~\ref{algh:data_def_RNCRN_with_AD}. Figure~\ref{fig:repeller_circle} shows the vector field of the reduced RNCRN approximation of the circular repeller orbit. 

This training produced an RNCRN with $M=5$ chemical perceptrons and coefficients $\beta_1= 81.948$, $\beta_2 = 77.952$, and  
\begin{align}
    \boldsymbol{\alpha}_1 &= \begin{pmatrix}
43.091 \\
-81.455 \\
-178.720 \\
-6.523 \\
-41.211 \\
    \end{pmatrix}, \; \; 
    \boldsymbol{\alpha}_2 = \begin{pmatrix} 
-174.745 \\
-5.587 \\
51.884 \\
-69.925 \\
-45.358 \\
    \end{pmatrix}, \;\;
    \boldsymbol{\theta} = \begin{pmatrix}
-306.069 \\
422.215 \\
-377.697 \\
445.634 \\
411.775 \\
    \end{pmatrix}, \;\; \nonumber \\
    \boldsymbol{\omega_1} &= \begin{pmatrix}
125.349 \\
-355.600 \\
-354.684 \\
-48.422 \\
-48.630 \\
    \end{pmatrix}, \;\;
    \boldsymbol{\omega_2} = \begin{pmatrix}
 -404.272 \\
-34.736 \\
130.538 \\
-344.728 \\
-43.372 \\
    \end{pmatrix}, \nonumber \\
    \boldsymbol{\gamma} &= \begin{pmatrix}
 87.125 \\
31.103 \\
86.664 \\
26.322 \\
1.338 \\
    \end{pmatrix}, \;\;
    \boldsymbol{\tau} = \begin{pmatrix}
1.719 \\
236.683 \\
2.733 \\
239.852 \\
488.075 \\
    \end{pmatrix},
    \label{eq:coefficents_data_cycle_repel}
\end{align}
where $\boldsymbol{\alpha}_i = (\alpha_{i,1},\alpha_{i,2},\ldots, \alpha_{i,5})^{\top}$ for $i = 1,2$, $\boldsymbol{\omega_i} = 
(\omega_{1,i},\omega_{2,i},\ldots,\omega_{5,i})^{\top}$
for $i = 1, 2$, $\boldsymbol{\theta} = 
(\theta_{1},\theta_{2},\ldots,\theta_{5})^{\top}$,  $\boldsymbol{\gamma} = 
(\gamma_{1},\gamma_{2},\ldots,\gamma_{5})^{\top}$, and  $\boldsymbol{\tau} = 
(\tau_{1},\tau_{2},\ldots,\tau_{5})^{\top}$.
The reduced ODEs are given by 
\begin{align}
\frac{\mathrm{d} \tilde{x}_1}{\mathrm{d} t} & = 
g_1(\tilde{x}_1, \tilde{x}_2) = 81.948  + \tilde{x}_1 \sum_{j=1}^{5}\alpha_{1,j} 
\sigma_{\gamma_j,\tau_j} \left(\omega_{j,1} \tilde{x}_1 + \omega_{j,2} \tilde{x}_2 + \theta_{j} \right), \nonumber \\
\frac{\mathrm{d} \tilde{x}_2}{\mathrm{d} t} & = 
g_2(\tilde{x}_1, \tilde{x}_2) = 77.952  + \tilde{x}_2 \sum_{j=1}^{5}\alpha_{2,j} 
\sigma_{\gamma_j,\tau_j} \left(\omega_{j,1} \tilde{x}_1 + \omega_{j,2} \tilde{x}_2 + \theta_{j} \right), 
\label{eq:single_layer_RRE_reduced_cycle_repel}
\end{align}
 while the full ODEs read
 \small
\begin{align}
    \frac{\mathrm{d} x_1}{\mathrm{d}t} &= 81.948 + x_1\left(\sum_{j=1}^{5} \alpha_{1,j}y_j\right), \;\;
    &&\frac{\mathrm{d} x_2}{\mathrm{d}t} = 77.952  + x_2\left(\sum_{j=1}^{5} \alpha_{2,j}y_j\right), \nonumber \\
    \mu\frac{\mathrm{d} y_1}{\mathrm{d}t} &=  87.125 + \left( \sum_{i=1}^2 \omega_{1,i}x_i + \theta_{1} \right)y_1 - 1.719 y_1^2, 
    \;\;&&\mu\frac{\mathrm{d} y_2}{\mathrm{d}t} = 31.103  + \left( \sum_{i=1}^2\omega_{2,i}x_i +  \theta_{2}\right)y_2 -236.683 y_2^2, \nonumber \\
    \mu\frac{\mathrm{d} y_3}{\mathrm{d}t} &= 86.664  + \left( \sum_{i=1}^2\omega_{3,i}x_i + \theta_{3}\right)y_3 -2.733 y_3^2,
    \;\;&&\mu\frac{\mathrm{d} y_4}{\mathrm{d}t} = 26.322  + \left( \sum_{i=1}^2\omega_{4,i}x_i + \theta_{4}\right)y_4 -239.852 y_4^2, \nonumber \\
    \mu\frac{\mathrm{d} y_5}{\mathrm{d}t} &= 1.338  + \left( \sum_{i=1}^2\omega_{5,i}x_i + \theta_{5}\right)y_5 -488.075 y_5^2, 
\end{align}
\normalsize
Here, we assume general initial concentrations: $x_1(0)= a_1 > 0$, $x_2(0)= a_2 > 0$, and $y_1(0)= b_1 \geq 0$, $y_2(0)= b_2 \geq 0$, $y_3(0)= b_3 \geq 0$, $y_4(0)= b_4 \geq 0$, and $y_5(0)= b_5 \geq 0$ . The coefficients were rounded to 3 decimal places to quote in text, the full precision coefficients are available in the code repository.

\subsection{Single circular limit cycle (linear) }
\label{sec:app_single_cycle_linear}
As described in the remark in Section~\ref{sec:data_defined_dynamics}, Algorithm~\ref{algh:define_attractor}'s padding points $Q_{d,k}^{\pm}$ have the associated vector field $d \mathbf{x}/ dt = \eta\hat{v}_{d, d+1} \mp \eta \hat{n}_{d, d+1}\exp{(\kappa \delta k)}$ with the  exponential dependence on the distance $\delta k$. In this example, we show that replacing the exponential dependence with a linear dependence while training the RNCRN can also produce the desired data-defined circular orbit i.e.  we use the vector field $d \mathbf{x}/ dt = \eta\hat{v}_{d, d+1} \mp \eta \hat{n}_{d, d+1}(\kappa \delta k)$ for padding points.

\begin{figure}[hbt!]
    \centering
    \includegraphics[width=0.67\linewidth]{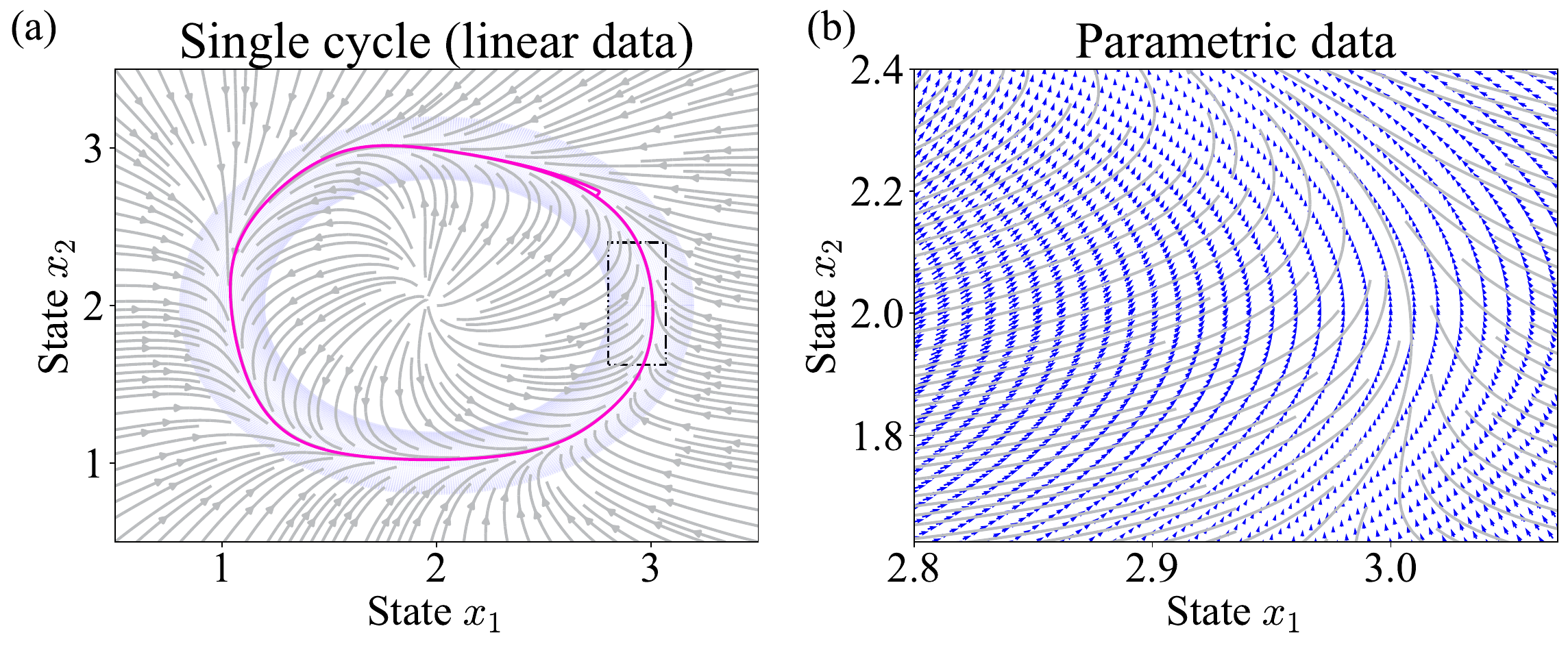}
    \caption{Example RNCRN trained on data to realise a circular attractor using a linear modification of Algorithm~\ref{algh:define_attractor}. In blue the target attractor data was created by sampling points from parametric equation~(\ref{eq:parametric_circle}) then applying a vector field interpolation algorithm, i.e. Algorithm~\ref{algh:define_attractor}, with a linear modification. In grey the reduced RNCRN vector field for a $5$ chemical perceptron RNCRN (see Appendix~\ref{sec:app_single_cycle_linear} for parameters). Panel (a) shows the entire state-space of interest while panel (b) shows a close-up of the state-space to emphasise the training data. The time-trajectory was simulated with $x_1(0)=2.744, x_2(0)=2.693$, $\mu=0.1$, and $y_1(0)=\dots, y_5(0)=0$.  }
    \label{fig:linear_circle}
\end{figure}

After generating data with the aforementioned linear modification to  Algorithm~\ref{algh:define_attractor} on the circular equation~(\ref{eq:parametric_circle}) with $K=20$, $\delta=0.01$,  $\kappa=10$, and $\eta=1$, we trained the RNCRN using Algorithm~\ref{algh:data_def_RNCRN_with_AD}. Figure~\ref{fig:linear_circle} shows the vector field of the reduced RNCRN approximation of the circular repeller orbit. 

This training produced an RNCRN with $M=5$ chemical perceptrons and coefficients $\beta_1= 16.552$, $\beta_2 = 23.165$, and  
\begin{align}
    \boldsymbol{\alpha}_1 &= \begin{pmatrix}
    -8.177 \\
-16.814 \\
-21.754 \\
-7.906 \\
15.501 \\
    \end{pmatrix}, \; \; 
    \boldsymbol{\alpha}_2 = \begin{pmatrix}
    26.539 \\
-22.747 \\
4.566 \\
-6.631 \\
-30.088 \\
    \end{pmatrix}, \;\;
    \boldsymbol{\theta} = \begin{pmatrix}
    59.477 \\
-35.099 \\
-60.028 \\
90.654 \\
54.644 \\
    \end{pmatrix}, \;\; \nonumber \\
    \boldsymbol{\omega_1} &= \begin{pmatrix}
    -33.517 \\
-28.584 \\
39.503 \\
-22.314 \\
0.131 \\
    \end{pmatrix}, \;\;
    \boldsymbol{\omega_2} = \begin{pmatrix}
    -9.963 \\
19.259 \\
-40.894 \\
-17.509 \\
-40.181 \\
    \end{pmatrix}, \nonumber \\
    \boldsymbol{\gamma} &= \begin{pmatrix}
    23.272 \\
27.928 \\
8.608 \\
0.397 \\
20.500 \\
    \end{pmatrix}, \;\;
    \boldsymbol{\tau} = \begin{pmatrix}
    56.930 \\
62.523 \\
61.866 \\
36.392 \\
46.726 \\
    \end{pmatrix},
    \label{eq:coefficents_data_cycle_linear}
\end{align}
where $\boldsymbol{\alpha}_i = (\alpha_{i,1},\alpha_{i,2},\ldots, \alpha_{i,5})^{\top}$ for $i = 1,2$, $\boldsymbol{\omega_i} = 
(\omega_{1,i},\omega_{2,i},\ldots,\omega_{5,i})^{\top}$
for $i = 1, 2$, $\boldsymbol{\theta} = 
(\theta_{1},\theta_{2},\ldots,\theta_{5})^{\top}$,  $\boldsymbol{\gamma} = 
(\gamma_{1},\gamma_{2},\ldots,\gamma_{5})^{\top}$, and  $\boldsymbol{\tau} = 
(\tau_{1},\tau_{2},\ldots,\tau_{5})^{\top}$.
The reduced ODEs are given by 
\begin{align}
\frac{\mathrm{d} \tilde{x}_1}{\mathrm{d} t} & = 
g_1(\tilde{x}_1, \tilde{x}_2) = 16.552  + \tilde{x}_1 \sum_{j=1}^{5}\alpha_{1,j} 
\sigma_{\gamma_j,\tau_j} \left(\omega_{j,1} \tilde{x}_1 + \omega_{j,2} \tilde{x}_2 + \theta_{j} \right), \nonumber \\
\frac{\mathrm{d} \tilde{x}_2}{\mathrm{d} t} & = 
g_2(\tilde{x}_1, \tilde{x}_2) = 23.165 + \tilde{x}_2 \sum_{j=1}^{5}\alpha_{2,j} 
\sigma_{\gamma_j,\tau_j} \left(\omega_{j,1} \tilde{x}_1 + \omega_{j,2} \tilde{x}_2 + \theta_{j} \right), 
\label{eq:single_layer_RRE_reduced_cycle_linear}
\end{align}
 while the full ODEs read
 \small
\begin{align}
    \frac{\mathrm{d} x_1}{\mathrm{d}t} &= 16.552 + x_1\left(\sum_{j=1}^{5} \alpha_{1,j}y_j\right), \;\;
    &&\frac{\mathrm{d} x_2}{\mathrm{d}t} = 23.165  + x_2\left(\sum_{j=1}^{5} \alpha_{2,j}y_j\right), \nonumber \\
    \mu\frac{\mathrm{d} y_1}{\mathrm{d}t} &=  23.272  + \left( \sum_{i=1}^2 \omega_{1,i}x_i + \theta_{1} \right)y_1 - 56.930 y_1^2, 
    \;\;&&\mu\frac{\mathrm{d} y_2}{\mathrm{d}t} =  27.928 + \left( \sum_{i=1}^2\omega_{2,i}x_i +  \theta_{2}\right)y_2 -62.523 y_2^2, \nonumber \\
    \mu\frac{\mathrm{d} y_3}{\mathrm{d}t} &=  8.608 + \left( \sum_{i=1}^2\omega_{3,i}x_i + \theta_{3}\right)y_3 -61.866y_3^2,
    \;\;&&\mu\frac{\mathrm{d} y_4}{\mathrm{d}t} = 0.397  + \left( \sum_{i=1}^2\omega_{4,i}x_i + \theta_{4}\right)y_4 -36.392y_4^2, \nonumber \\
    \mu\frac{\mathrm{d} y_5}{\mathrm{d}t} &= 20.500 + \left( \sum_{i=1}^2\omega_{5,i}x_i + \theta_{5}\right)y_5 -46.726y_5^2, 
\end{align}
\normalsize
Here, we assume general initial concentrations: $x_1(0)= a_1 > 0$, $x_2(0)= a_2 > 0$, and $y_1(0)= b_1 \geq 0$, $y_2(0)= b_2 \geq 0$, $y_3(0)= b_3 \geq 0$, $y_4(0)= b_4 \geq 0$, and $y_5(0)= b_5 \geq 0$. The coefficients were rounded to 3 decimal places to quote in text, the full precision coefficients are available in the code repository.

\subsection{Heart-shaped limit cycle}
\label{sec:app_heart_cycle}
As described in Section~\ref{sec:heart_shaped_cycle} training data generated by Algorithm~\ref{algh:define_attractor} and equation~(\ref{eq:parametric_heart}) was used to train an RNCRN using Algorithm~\ref{algh:data_def_RNCRN_with_AD}. This training produced an RNCRN with $M=10$ chemical perceptrons and coefficients $\beta_1= 32.263 $, $\beta_2 = 0.220$, and 

\begin{align}
    \boldsymbol{\alpha}_1 &= \begin{pmatrix}
    -206.685 \\
68.185 \\
63.034 \\
31.454 \\
-1.046 \\
-69.791 \\
3.335 \\
6.724 \\
-4.768 \\
-0.753 \\
    \end{pmatrix}, \; \; 
    \boldsymbol{\alpha}_2 = \begin{pmatrix} 
    14.421 \\
361.350 \\
42.053 \\
-1596.404 \\
11.659 \\
445.898 \\
-26.598 \\
4.145 \\
9.770 \\
-24.237 \\
    \end{pmatrix}, \;\;
    \boldsymbol{\theta} = \begin{pmatrix}
    399.866 \\
-2885.359 \\
-3238.521 \\
-4674.896 \\
1634.028 \\
-2261.417 \\
1268.967 \\
1348.322 \\
-2966.115 \\
-965.133 \\
    \end{pmatrix}, \;\; \nonumber \\
    \boldsymbol{\omega_1} &= \begin{pmatrix}
    -752.176 \\
-1404.888 \\
-2126.800 \\
-6923.930 \\
-694.304 \\
-1179.076 \\
44.345 \\
-472.745 \\
2317.337 \\
388.654 \\
    \end{pmatrix}, \;\;
    \boldsymbol{\omega_2} = \begin{pmatrix}
    -85.831 \\
-2904.666 \\
-3063.465 \\
-3358.036 \\
-126.843 \\
-2667.009 \\
-551.100 \\
-404.818 \\
-1782.721 \\
29.470 \\
    \end{pmatrix}, \nonumber \\
    \boldsymbol{\gamma} &= \begin{pmatrix}
    102.958 \\
317.511 \\
12.669 \\
422.712 \\
0.056 \\
435.578 \\
0.572 \\
0.087 \\
0.189 \\
10.109 \\
    \end{pmatrix}, \;\;
    \boldsymbol{\tau} = \begin{pmatrix}
    735.035 \\
6155.131 \\
5217.168 \\
2052.469 \\
1814.245 \\
6080.718 \\
2388.553 \\
519.255 \\
5648.693 \\
2210.594 \\
    \end{pmatrix},
    \label{eq:coefficents_data_heart}
\end{align}
where $\boldsymbol{\alpha}_i = (\alpha_{i,1},\alpha_{i,2},\ldots, \alpha_{i,5})^{\top}$ for $i = 1,2$, $\boldsymbol{\omega_i} = 
(\omega_{1,i},\omega_{2,i},\ldots,\omega_{5,i})^{\top}$
for $i = 1, 2$, $\boldsymbol{\theta} = 
(\theta_{1},\theta_{2},\ldots,\theta_{5})^{\top}$,  $\boldsymbol{\gamma} = 
(\gamma_{1},\gamma_{2},\ldots,\gamma_{5})^{\top}$, and  $\boldsymbol{\tau} = 
(\tau_{1},\tau_{2},\ldots,\tau_{5})^{\top}$.
The reduced ODEs are given by 
\begin{align}
\frac{\mathrm{d} \tilde{x}_1}{\mathrm{d} t} & = 
g_1(\tilde{x}_1, \tilde{x}_2) = 32.263 + \tilde{x}_1 \sum_{j=1}^{10}\alpha_{1,j} 
\sigma_{\gamma_j,\tau_j} \left(\omega_{j,1} \tilde{x}_1 + \omega_{j,2} \tilde{x}_2 + \theta_{j} \right), \nonumber \\
\frac{\mathrm{d} \tilde{x}_2}{\mathrm{d} t} & = 
g_2(\tilde{x}_1, \tilde{x}_2) = 0.220  + \tilde{x}_2 \sum_{j=1}^{10}\alpha_{2,j} 
\sigma_{\gamma_j,\tau_j} \left(\omega_{j,1} \tilde{x}_1 + \omega_{j,2} \tilde{x}_2 + \theta_{j} \right), 
\label{eq:single_layer_RRE_reduced_heart}
\end{align}
 while the full ODEs read
\small
\begin{align}
    \frac{\mathrm{d} x_1}{\mathrm{d}t} &= 32.263+ x_1\left(\sum_{j=1}^{10} \alpha_{1,j}y_j\right), \;\;
    &&\frac{\mathrm{d} x_2}{\mathrm{d}t} = 0.220 + x_2\left(\sum_{j=1}^{10} \alpha_{2,j}y_j\right), \nonumber \\
    \mu\frac{\mathrm{d} y_1}{\mathrm{d}t} &= 102.958 + \left( \sum_{i=1}^2 \omega_{1,i}x_i + \theta_{1} \right)y_1 -735.035y_1^2, 
    \;\;&&\mu\frac{\mathrm{d} y_2}{\mathrm{d}t} = 317.511 + \left( \sum_{i=1}^2\omega_{2,i}x_i +  \theta_{2}\right)y_2 -6155.131y_2^2, \nonumber \\
    \mu\frac{\mathrm{d} y_3}{\mathrm{d}t} &= 12.669 + \left( \sum_{i=1}^2\omega_{3,i}x_i + \theta_{3}\right)y_3 -5217.168y_3^2,
    \;\;&&\mu\frac{\mathrm{d} y_4}{\mathrm{d}t} = 422.712 + \left( \sum_{i=1}^2\omega_{4,i}x_i +  \theta_{4}\right)y_4 -2052.469y_4^2, \nonumber \\
    \mu\frac{\mathrm{d} y_5}{\mathrm{d}t} &= 0.056 + \left( \sum_{i=1}^2\omega_{5,i}x_i + \theta_{5}\right)y_5 -1814.245y_5^2, 
    \;\;&&\mu\frac{\mathrm{d} y_6}{\mathrm{d}t} = 435.578 + \left( \sum_{i=1}^2\omega_{6,i}x_i + \theta_{6}\right)y_6 -6080.718y_6^2,  \nonumber \\
     \mu\frac{\mathrm{d} y_7}{\mathrm{d}t} &= 0.572 + \left( \sum_{i=1}^2\omega_{7,i}x_i + \theta_{7}\right)y_7 -2388.553y_7^2, 
    \;\;&&\mu\frac{\mathrm{d} y_8}{\mathrm{d}t} = 0.087 + \left( \sum_{i=1}^2\omega_{8,i}x_i + \theta_{8}\right)y_8 -519.255y_8^2,  \nonumber \\
    \mu\frac{\mathrm{d} y_9}{\mathrm{d}t} &= 0.189  + \left( \sum_{i=1}^2\omega_{9,i}x_i + \theta_{9}\right)y_9 -5648.693y_9^2, 
    \;\;&&\mu\frac{\mathrm{d} y_{10}}{\mathrm{d}t} = 10.109 + \left( \sum_{i=1}^2\omega_{10,i}x_i + \theta_{10}\right)y_{10} -2210.594 y_{10}^2, 
    \label{eq:full_rncrn_heart}
\end{align}
Here, we assume general initial concentrations: $x_1(0)= a_1 > 0$, $x_2(0)= a_2 > 0$, and $y_1(0)= b_1 \geq 0$, $y_2(0)= b_2 \geq 0$, $y_3(0)= b_3 \geq 0$, $y_4(0)= b_4 \geq 0$, $y_5(0)= b_5 \geq 0$, $y_6(0)= b_6 \geq 0$, $y_7(0)= b_7 \geq 0$, $y_8(0)= b_8 \geq 0$, $y_9(0)= b_9 \geq 0$, and $y_{10}(0)= b_{10} \geq 0$ . The coefficients were rounded to 3 decimal places to quote in text, the full precision coefficients are available in the code repository.

\normalsize
\subsection{Multiple limit cycles}
\label{sec:app_multiple_cycle}
As described in Section~\ref{sec:multiple_cycle} training data generated by Algorithm~\ref{algh:define_attractor} and equation~(\ref{eq:parametric_circle}) was used to train an RNCRN using Algorithm~\ref{algh:data_def_RNCRN_with_AD}. This training produced an RNCRN with $M=15$ chemical perceptrons and coefficients $\beta_1= 18.815$, $\beta_2 = 18.135 $, and  
\begin{align}
    \boldsymbol{\alpha}_1 &= \begin{pmatrix}
    0.037 \\
-44.974 \\
-9.160 \\
0.511 \\
-0.021 \\
0.567 \\
-4.428 \\
0.549 \\
-0.238 \\
0.241 \\
-3.833 \\
-14.058 \\
-0.073 \\
-0.177 \\
0.025 \\
    \end{pmatrix}, \; \; 
    \boldsymbol{\alpha}_2 = \begin{pmatrix} 
    0.053 \\
-35.678 \\
-108.873 \\
0.039 \\
0.307 \\
-21.397 \\
13.701 \\
0.520 \\
-0.150 \\
-0.041 \\
-11.530 \\
-2.218 \\
0.108 \\
-0.217 \\
0.298 \\
    \end{pmatrix}, \;\;
    \boldsymbol{\theta} = \begin{pmatrix}
    -2169.556 \\
-63.584 \\
-63.308 \\
-2110.566 \\
-2062.301 \\
330.702 \\
371.345 \\
-2167.958 \\
-2232.450 \\
-2186.657 \\
-2189.514 \\
366.100 \\
-2077.129 \\
-2345.668 \\
-2176.678 \\
    \end{pmatrix}, \;\; \nonumber \\
    \boldsymbol{\omega_1} &= \begin{pmatrix}
    -2613.137 \\
-133.234 \\
-85.431 \\
-2471.896 \\
-2582.940 \\
-3.890 \\
-85.207 \\
-2593.340 \\
-2869.315 \\
-2738.075 \\
-2545.244 \\
-229.477 \\
-2510.477 \\
-2841.934 \\
-2434.409 \\
    \end{pmatrix}, \;\;
    \boldsymbol{\omega_2} = \begin{pmatrix}
    -2439.155 \\
79.236 \\
-283.547 \\
-2319.514 \\
-2296.273 \\
-185.955 \\
-36.065 \\
-2420.028 \\
-2499.550 \\
-2462.773 \\
-2429.244 \\
4.170 \\
-2312.757 \\
-2641.583 \\
-2347.650 \\
    \end{pmatrix}, \nonumber \\
    \boldsymbol{\gamma} &= \begin{pmatrix}
    0.046 \\
48.748 \\
55.087 \\
0.047 \\
0.607 \\
1.108 \\
0.359 \\
0.317 \\
0.203 \\
0.087 \\
0.574 \\
0.549 \\
0.069 \\
0.058 \\
0.043 \\
    \end{pmatrix}, \;\;
    \boldsymbol{\tau} = \begin{pmatrix}
    105.841 \\
632.562 \\
465.905 \\
5.862 \\
114.611 \\
388.573 \\
365.923 \\
28.740 \\
85.871 \\
30.612 \\
2.894 \\
411.963 \\
84.905 \\
75.267 \\
8.467 \\
    \end{pmatrix},
    \label{eq:coefficents_data_multiple_cycle}
\end{align}
where $\boldsymbol{\alpha}_i = (\alpha_{i,1},\alpha_{i,2},\ldots, \alpha_{i,15})^{\top}$ for $i = 1,2$, $\boldsymbol{\omega_i} = 
(\omega_{1,i},\omega_{2,i},\ldots,\omega_{15,i})^{\top}$
for $i = 1, 2$, $\boldsymbol{\theta} = 
(\theta_{1},\theta_{2},\ldots,\theta_{15})^{\top}$,  $\boldsymbol{\gamma} = 
(\gamma_{1},\gamma_{2},\ldots,\gamma_{15})^{\top}$, and  $\boldsymbol{\tau} = 
(\tau_{1},\tau_{2},\ldots,\tau_{15})^{\top}$.
The reduced ODEs are given by 
\begin{align}
\frac{\mathrm{d} \tilde{x}_1}{\mathrm{d} t} & = 
g_1(\tilde{x}_1, \tilde{x}_2) = 18.815  + \tilde{x}_1 \sum_{j=1}^{15}\alpha_{1,j} 
\sigma_{\gamma_j,\tau_j} \left(\omega_{j,1} \tilde{x}_1 + \omega_{j,2} \tilde{x}_2 + \theta_{j} \right), \nonumber \\
\frac{\mathrm{d} \tilde{x}_2}{\mathrm{d} t} & = 
g_2(\tilde{x}_1, \tilde{x}_2) = 18.135   + \tilde{x}_2 \sum_{j=1}^{15}\alpha_{2,j} 
\sigma_{\gamma_j,\tau_j} \left(\omega_{j,1} \tilde{x}_1 + \omega_{j,2} \tilde{x}_2 + \theta_{j} \right), 
\label{eq:single_layer_RRE_reduced_multiple}
\end{align}
 while the full ODEs read
\small
\begin{align}
    \frac{\mathrm{d} x_1}{\mathrm{d}t} &= 18.815+ x_1\left(\sum_{j=1}^{10} \alpha_{1,j}y_j\right), \;\;
    &&\frac{\mathrm{d} x_2}{\mathrm{d}t} = 18.135  + x_2\left(\sum_{j=1}^{10} \alpha_{2,j}y_j\right), \nonumber \\
    \mu\frac{\mathrm{d} y_1}{\mathrm{d}t} &= 0.046  + \left( \sum_{i=1}^2 \omega_{1,i}x_i + \theta_{1} \right)y_1 -105.841y_1^2, 
    \;\;&&\mu\frac{\mathrm{d} y_2}{\mathrm{d}t} = 48.748  + \left( \sum_{i=1}^2\omega_{2,i}x_i +  \theta_{2}\right)y_2 -632.562y_2^2, \nonumber \\
    \mu\frac{\mathrm{d} y_3}{\mathrm{d}t} &= 55.087  + \left( \sum_{i=1}^2\omega_{3,i}x_i + \theta_{3}\right)y_3 -465.905y_3^2,
    \;\;&&\mu\frac{\mathrm{d} y_4}{\mathrm{d}t} = 0.047   + \left( \sum_{i=1}^2\omega_{4,i}x_i +  \theta_{4}\right)y_4 -5.862y_4^2, \nonumber \\
    \mu\frac{\mathrm{d} y_5}{\mathrm{d}t} &= 0.607  + \left( \sum_{i=1}^2\omega_{5,i}x_i + \theta_{5}\right)y_5 -114.611y_5^2, 
    \;\;&&\mu\frac{\mathrm{d} y_6}{\mathrm{d}t} =  1.108 + \left( \sum_{i=1}^2\omega_{6,i}x_i + \theta_{6}\right)y_6 -388.573y_6^2,  \nonumber \\
     \mu\frac{\mathrm{d} y_7}{\mathrm{d}t} &= 0.359  + \left( \sum_{i=1}^2\omega_{7,i}x_i + \theta_{7}\right)y_7 -365.923y_7^2, 
    \;\;&&\mu\frac{\mathrm{d} y_8}{\mathrm{d}t} = 0.317  + \left( \sum_{i=1}^2\omega_{8,i}x_i + \theta_{8}\right)y_8 -28.740y_8^2,  \nonumber \\
    \mu\frac{\mathrm{d} y_9}{\mathrm{d}t} &= 0.203   + \left( \sum_{i=1}^2\omega_{9,i}x_i + \theta_{9}\right)y_9 -85.871y_9^2, 
    \;\;&&\mu\frac{\mathrm{d} y_{10}}{\mathrm{d}t} = 0.087  + \left( \sum_{i=1}^2\omega_{10,i}x_i + \theta_{10}\right)y_{10} - 30.612y_{10}^2, \nonumber \\
    \mu\frac{\mathrm{d} y_{11}}{\mathrm{d}t} &=  0.574  + \left( \sum_{i=1}^2\omega_{11,i}x_i + \theta_{11}\right)y_{11} -2.894y_{11}^2, 
    \;\;&&\mu\frac{\mathrm{d} y_{12}}{\mathrm{d}t} = 0.549  + \left( \sum_{i=1}^2\omega_{12,i}x_i + \theta_{12}\right)y_{12} -411.963y_{12}^2,  \nonumber \\
     \mu\frac{\mathrm{d} y_{13}}{\mathrm{d}t} &= 0.069  + \left( \sum_{i=1}^2\omega_{13,i}x_i + \theta_{13}\right)y_{13} -84.905y_{13}^2, 
    \;\;&&\mu\frac{\mathrm{d} y_{14}}{\mathrm{d}t} = 0.058  + \left( \sum_{i=1}^2\omega_{14,i}x_i + \theta_{14}\right)y_{14} -75.267y_{14}^2,  \nonumber \\
    \mu\frac{\mathrm{d} y_{15}}{\mathrm{d}t} &=  0.043  + \left( \sum_{i=1}^2\omega_{15,i}x_i + \theta_{15}\right)y_{15} -8.467 y_{15}^2, 
    \label{eq:full_rncrn_multiple}
\end{align}
\normalsize
Here, we assume general initial concentrations: $x_1(0)= a_1 > 0$, $x_2(0)= a_2 > 0$, and $y_1(0)= b_1 \geq 0$, $y_2(0)= b_2 \geq 0$, $y_3(0)= b_3 \geq 0$, $y_4(0)= b_4 \geq 0$, $y_5(0)= b_5 \geq 0$, $y_6(0)= b_6 \geq 0$, $y_7(0)= b_7 \geq 0$, $y_8(0)= b_8 \geq 0$, $y_9(0)= b_9 \geq 0$, $y_{10}(0)= b_{10} \geq 0$, $y_{11}(0)= b_{11} \geq 0$, $y_{12}(0)= b_{12} \geq 0$, $y_{13}(0)= b_{13} \geq 0$, $y_{14}(0)= b_{14} \geq 0$, and $y_{15}(0)= b_{15} \geq 0$. The coefficients were rounded to 3 decimal places to quote in text, the full precision coefficients are available in the code repository.

\subsection{Toroidal limit cycle in three dimensions}
\label{app:torus_orbit}

\begin{figure}[hbt!]
    \centering
    \includegraphics[width=\linewidth]{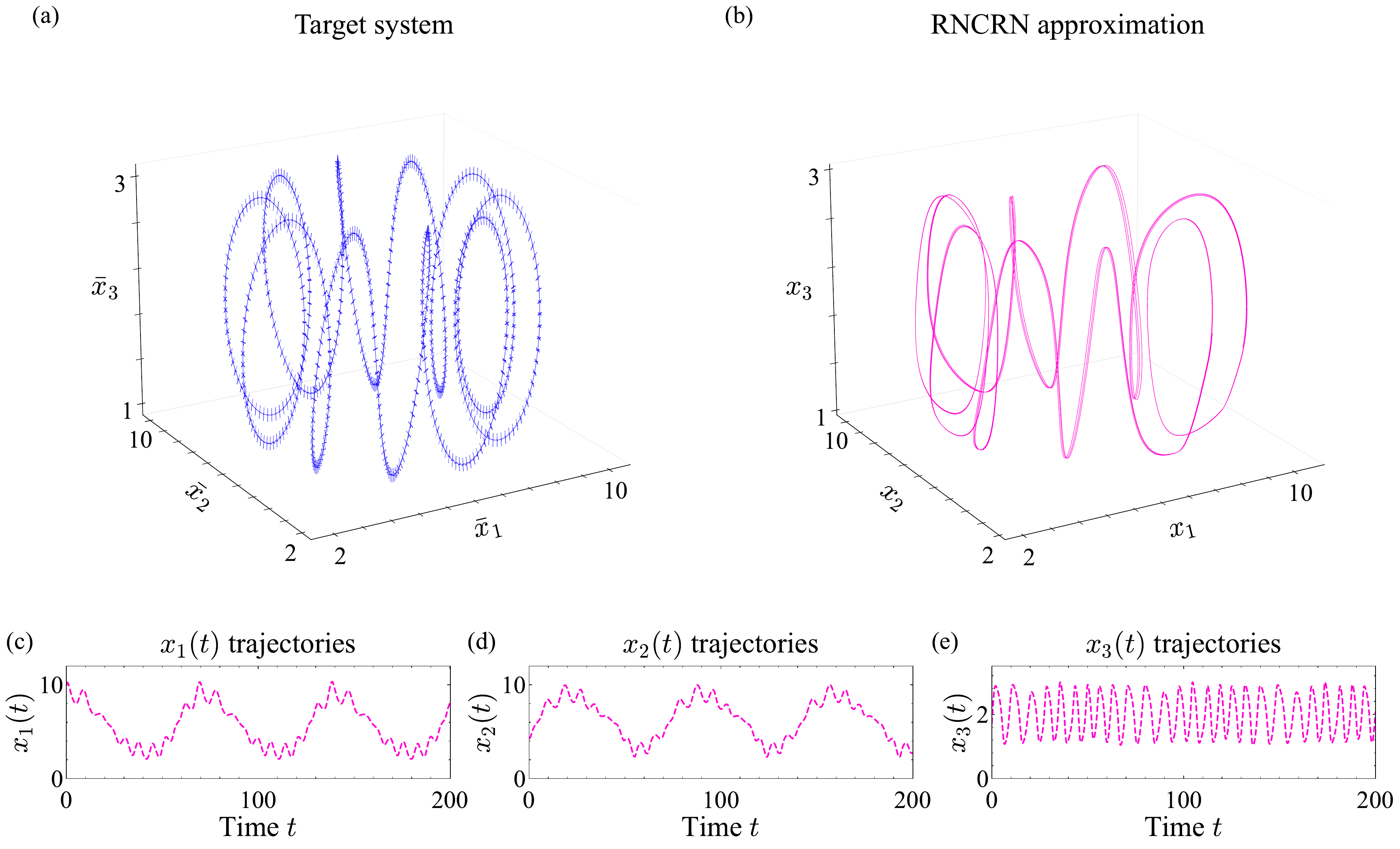}
    \caption{ An RNCRN trained to realise a toroidal attractor without an underlying ODE. Panel (a) shows the target attractor data, in blue, that was created by sampling points from parametric equation~(\ref{eq:parametric_torus}) then applying a  modified vector field interpolation algorithm, i.e. Algorithm~\ref{algh:define_attractor}. Panel (b) shows, in magenta, an example state-space trajectory from the $M=7$ RNCRN approximation with $\mu=0.01$, $x_1(0)=10.126$, $x_2(0)=4.257,$ $x_3(0)=1.703$, and $y_1(0)=\dots, y_7(0)=0$  (the system is fully specified in Appendix~\ref{app:torus_orbit}). Panel (c)--(e) show the same trajectory as in panel (b) but in the time-domain for $x_1(t), x_2(t),$ and $x_3(t)$, respectively.  }
    \label{fig:torus_orbit}
\end{figure}

Let us now show that Algorithm~\ref{algh:define_attractor}, which is formulated in two dimensions, can be generalized into higher dimensions. In particular, consider the parametric 
equation for a curve on a torus 
\begin{align}
    x_1(s) &= 3\cos(s)+\cos(10s)\cos(s)+ 6, \nonumber \\
    x_2(s) &= 3\sin(s) +\cos(10s)\sin(s)+ 6, \nonumber \\
    x_3(s) &= \sin(10s)+2,
    \label{eq:parametric_torus}
\end{align}
where $s\in[0, 2\pi)$. Figure~\ref{fig:torus_orbit}(a) shows 
three-dimensional training data produced from a generalization of Algorithm~\ref{algh:define_attractor} to three dimensions. Using this data and Algorithm~\ref{algh:data_def_RNCRN_with_AD}, we train an RNCRN with $M=7$ chemical perceptrons  which, as demonstrated in Figure~\ref{fig:torus_orbit}(b), numerically displays 
a three-dimensional stable limit cycle
which is approximately of the desired toroidal shape.

The principal modification was of the data generation algorithm was the generalisation of the placement of padding points into more than one perpendicular direction. 

{\bf Higher-dimensional data generation. } Given a list of points in higher-dimensional $P_d \in \mathbb{R}_>^N$ where $N>2$ we modify Algorithm~\ref{algh:define_attractor} such that the unit normal from part (a) is replaced with $N-1$ orthonormal unit vectors that form a basis  with the $\hat{v}_{d,d+1}$ direction. Each of these orthonormal unit vectors produce additional padding points and are used to compute the local vector field approximations. To find the orthonormal basis we recognise that a column vector $\hat{v}_{d,d+1}\in \mathbb{R}>^{N\times 1} $ can be decomposed into an orthonormal matrix $Q\in \mathbb{R}>^{N\times N}$ and a scalar $R\in\mathrm{R}$ via QR decomposition where the first column of $Q$ will be linearly dependent on $\hat{v}_{d,d+1}$. It follows that the orthonormal unit vectors are the $N-1$ remaining column vectors of $Q$.
\\

{\bf RNCRN parameters. } Using the higher-dimensional data-generation modification of Algorithm~\ref{algh:define_attractor} and Algorithm~\ref{algh:data_def_RNCRN_with_AD} for training, we found an $M=7$ RNCRN with coefficients $\beta_1= 1.570$, $\beta_2 = 0.408 $, $\beta_3=6.885,$ and  
\begin{align}
    \boldsymbol{\alpha}_1 &= \begin{pmatrix}
    -5.227 \\
-6.586 \\
0.332 \\
-1.297 \\
1.037 \\
-0.752 \\
7.607 \\
    \end{pmatrix}, \; \; 
    \boldsymbol{\alpha}_2 = \begin{pmatrix} 
    -0.755 \\
-5.934 \\
-3.663 \\
14.209 \\
1.077 \\
1.567 \\
2.768 \\
    \end{pmatrix}, \;\;
    \boldsymbol{\alpha}_3 = \begin{pmatrix} 
    30.463 \\
-19.509 \\
23.421 \\
-20.966 \\
-45.739 \\
4.597 \\
3.169 \\
    \end{pmatrix}, \;\;
    \boldsymbol{\theta} = \begin{pmatrix}
    292.047 \\
-280.508 \\
261.422 \\
-255.093 \\
262.072 \\
-198.768 \\
-139.727 \\
    \end{pmatrix}, \;\; \nonumber \\
    \boldsymbol{\omega_1} &= \begin{pmatrix}
    -39.116 \\
-17.292 \\
-3.625 \\
7.586 \\
-9.729 \\
207.473 \\
-109.965 \\
    \end{pmatrix}, \;\;
    \boldsymbol{\omega_2} = \begin{pmatrix}
    -2.949 \\
-1.592 \\
-36.064 \\
-105.612 \\
-9.325 \\
-327.333 \\
-28.268 \\
    \end{pmatrix}, 
    \boldsymbol{\omega_3} = \begin{pmatrix}
    -53.870 \\
111.653 \\
-32.435 \\
145.520 \\
-67.332 \\
-308.171 \\
344.645 \\
    \end{pmatrix}, \nonumber \\
    \boldsymbol{\gamma} &= \begin{pmatrix}
    5.488 \\
7.535 \\
3.963 \\
9.258 \\
6.007 \\
0.004 \\
5.764 \\
    \end{pmatrix}, \;\;
    \boldsymbol{\tau} = \begin{pmatrix}
    666.189 \\
812.839 \\
496.655 \\
912.664 \\
428.803 \\
713.388 \\
4637.415 \\
    \end{pmatrix},
    \label{eq:coefficents_data_torus}
\end{align}
where $\boldsymbol{\alpha}_i = (\alpha_{i,1},\alpha_{i,2},\ldots, \alpha_{i,7})^{\top}$ for $i = 1,2,3$, $\boldsymbol{\omega_i} = 
(\omega_{1,i},\omega_{2,i},\ldots,\omega_{7,i})^{\top}$
for $i = 1, 2, 3$, $\boldsymbol{\theta} = 
(\theta_{1},\theta_{2},\ldots,\theta_{7})^{\top}$,  $\boldsymbol{\gamma} = 
(\gamma_{1},\gamma_{2},\ldots,\gamma_{7})^{\top}$, and  $\boldsymbol{\tau} = 
(\tau_{1},\tau_{2},\ldots,\tau_{7})^{\top}$.
The reduced ODEs are given by 
\begin{align}
\frac{\mathrm{d} \tilde{x}_1}{\mathrm{d} t} & = 
g_1(\tilde{x}_1, \tilde{x}_2, \tilde{x}_3) =  1.570 + \tilde{x}_1 \sum_{j=1}^{7}\alpha_{1,j} 
\sigma_{\gamma_j,\tau_j} \left(\omega_{j,1} \tilde{x}_1 + \omega_{j,2} \tilde{x}_2  + \omega_{j,3} \tilde{x}_3 + \theta_{j} \right), \nonumber \\
\frac{\mathrm{d} \tilde{x}_2}{\mathrm{d} t} & = 
g_2(\tilde{x}_1, \tilde{x}_2, \tilde{x}_3) =  0.408  + \tilde{x}_2 \sum_{j=1}^{7}\alpha_{2,j} 
\sigma_{\gamma_j,\tau_j} \left(\omega_{j,1} \tilde{x}_1 + \omega_{j,2} \tilde{x}_2 + \omega_{j,3} \tilde{x}_3 + \theta_{j} \right), 
\nonumber \\
\frac{\mathrm{d} \tilde{x}_3}{\mathrm{d} t} & = 
g_3(\tilde{x}_1, \tilde{x}_2, \tilde{x}_3) =   6.885 + \tilde{x}_3 \sum_{j=1}^{7}\alpha_{3,j} 
\sigma_{\gamma_j,\tau_j} \left(\omega_{j,1} \tilde{x}_1 + \omega_{j,2} \tilde{x}_2  + \omega_{j,3} \tilde{x}_3 + \theta_{j} \right), 
\label{eq:single_layer_RRE_reduced_torus}\end{align}
 while the full ODEs read
\small
\begin{align}
    \frac{\mathrm{d} x_1}{\mathrm{d}t} &= 1.570+ x_1\left(\sum_{j=1}^{7} \alpha_{1,j}y_j\right), \;\;
    &&\frac{\mathrm{d} x_2}{\mathrm{d}t} =  0.408 + x_2\left(\sum_{j=1}^{7} \alpha_{2,j}y_j\right), \nonumber \\
    \frac{\mathrm{d} x_3}{\mathrm{d}t} &= 6.885  + x_3\left(\sum_{j=1}^{7} \alpha_{3,j}y_j\right), \;\;&&\mu\frac{\mathrm{d} y_1}{\mathrm{d}t} =  5.488 + \left( \sum_{i=1}^3 \omega_{1,i}x_i + \theta_{1} \right)y_1 -666.189y_1^2, \nonumber \\
    \mu\frac{\mathrm{d} y_2}{\mathrm{d}t} &=   7.535+ \left( \sum_{i=1}^3\omega_{2,i}x_i +  \theta_{2}\right)y_2 -812.839y_2^2, \;\;&&
    \mu\frac{\mathrm{d} y_3}{\mathrm{d}t} =  3.963 + \left( \sum_{i=1}^3\omega_{3,i}x_i + \theta_{3}\right)y_3 -496.655y_3^2, \nonumber \\
    \mu\frac{\mathrm{d} y_4}{\mathrm{d}t} &=  9.258  + \left( \sum_{i=1}^3\omega_{4,i}x_i +  \theta_{4}\right)y_4 -912.664y_4^2, \;\;&&
    \mu\frac{\mathrm{d} y_5}{\mathrm{d}t} = 6.007  + \left( \sum_{i=1}^3\omega_{5,i}x_i + \theta_{5}\right)y_5 -428.803y_5^2, \nonumber \\
    \mu\frac{\mathrm{d} y_6}{\mathrm{d}t} &= 0.004  + \left( \sum_{i=1}^3\omega_{6,i}x_i + \theta_{6}\right)y_6 -713.338y_6^2,  \;\;&&
     \mu\frac{\mathrm{d} y_7}{\mathrm{d}t} = 5.764  + \left( \sum_{i=1}^3\omega_{7,i}x_i + \theta_{7}\right)y_7 -4637.415y_7^2, 
    \label{eq:full_rncrn_torus}
\end{align}
\normalsize
Here, we assume general initial concentrations: $x_1(0)= a_1 > 0$, $x_2(0)= a_2 > 0$, $x_3(0)= a_3 > 0$, and $y_1(0)= b_1 \geq 0$, $y_2(0)= b_2 \geq 0$, $y_3(0)= b_3 \geq 0$, $y_4(0)= b_4 \geq 0$, $y_5(0)= b_5 \geq 0$, $y_6(0)= b_6 \geq 0$, and $y_7(0)= b_7 \geq 0$. The coefficients were rounded to 3 decimal places to quote in text, the full precision coefficients are available in the code repository.

\subsection{Equilibrium-oscillation piecewise system with data-defined oscillations}
\label{app:pulse_drug}
We now detail the construction of the classification-controlled RNCRNs applied to approximate the equilibrium-oscillation piecewise system with data-defined oscillations (as introduced in Section~\ref{sec:targ_pulse_drug}). We use the general method to train a classification-controlled RNCRN, as defined in Algorithm~\ref{algh:class_controlled_RNCRN}, with appropriate modification to use data-defined vector fields.
\\

{\textbf{\textit{Step 1. } Point-wise parametrized RNCRN.}} Two dynamical states were defined: a cyclic attractor and a stable equilibrium. First, using Algorithm~\ref{algh:define_attractor} an attractor was generated for a parametric orbit, equation~(\ref{eq:parametric_circle}), centered on $b_1=b_2=1.5$ with radius $a=1$. Second, using the ODEs $\mathrm{d} x_1/ \mathrm{d}t = 0.5- x_1$ and $\mathrm{d} x_2/ \mathrm{d}t =0.5 - x_2$ the vector field of the stable equilibrium was used over compact subset $\mathbb{K}_1=[0.01, 3], \mathbb{K}_2=[0.01, 3]$. Using the associated control points $r_1 = 1$ and $r_2 = 0.1$ a modified Algorithm 3 (i.e. allowing for data-defined vector fields as in Algorithm~\ref{algh:data_def_RNCRN_with_AD}) was used to find a point-wise parameterized RNCRN that bifurcates between the two target dynamics.

The resulting parametrized RNCRN was found with tolerance $\varepsilon \approx 10^{-1}$  met with an RNCRN with $M=5$ chemical perceptrons and coefficients $\beta_1=0.790, \beta_2=0.858,$ 
\begin{align}
    \boldsymbol{\alpha}_1 &= \begin{pmatrix}
    89.517 \\
-13.080 \\
0.321 \\
-98.540 \\
-69.707 \\
    \end{pmatrix}, \; \; 
    \boldsymbol{\alpha}_2 = \begin{pmatrix} 
    96.384 \\
-50.709 \\
-24.353 \\
-11.676 \\
-124.643 \\
    \end{pmatrix}, \;\;
    \boldsymbol{\theta} = \begin{pmatrix}
    -1654.188 \\
578.636 \\
-2579.316 \\
-3291.187 \\
-1568.960 \\
    \end{pmatrix}, \;\; 
    \boldsymbol{\tau} = \begin{pmatrix}
    38.776 \\
8373.519 \\
3629.027 \\
13631.220 \\
3385.638 \\
    \end{pmatrix}, \;\; \nonumber \\
    \boldsymbol{\omega_1} &= \begin{pmatrix}
    -1357.492 \\
-727.745 \\
-276.335 \\
-2029.546 \\
-239.827 \\
    \end{pmatrix}, \;\;
    \boldsymbol{\omega_2} = \begin{pmatrix}
    -1128.933 \\
-802.213 \\
694.332 \\
295.107 \\
-775.896 \\
    \end{pmatrix},
    \boldsymbol{\psi} = \begin{pmatrix}
    2380.343 \\
1605.154 \\
1198.530 \\
2737.624 \\
-8382.229 \\
    \end{pmatrix},
    \boldsymbol{\gamma} = \begin{pmatrix}
    75.071 \\
8.557 \\
5.288 \\
73.001 \\
70.181 \\
    \end{pmatrix},
    \label{eq:coefficents_point_controlled_pulse}
\end{align}
where $\boldsymbol{\alpha}_i = (\alpha_{i,1},\alpha_{i,2},\ldots, \alpha_{i,5})^{\top}$ for $i = 1,2$, $\boldsymbol{\omega_i} = 
(\omega_{1,i},\omega_{2,i},\ldots,\omega_{5,i})^{\top}$
for $i = 1, 2$, $\boldsymbol{\theta} = 
(\theta_{1},\theta_{2},\ldots,\theta_{5})^{\top}$, $\boldsymbol{\tau} = 
(\tau_{1},\tau_{2},\ldots,\tau_{5})^{\top}$,  $\boldsymbol{\psi} = 
(\psi_{1},\psi_{2},\ldots,\psi_{5})^{\top}$, and $\boldsymbol{\gamma} = 
(\gamma_{1},\gamma_{2},\ldots,\gamma_{5})^{\top}$. The reduced ODEs are given by 
\begin{align}
\frac{\mathrm{d} \tilde{x}_1}{\mathrm{d} t} & = 
g_1(\tilde{x}_1, \tilde{x}_2; \tilde{r}) =\beta_1  + \tilde{x}_1 \sum_{j=1}^{5}\alpha_{1,j} 
\sigma_{\gamma_j, \tau_j} \left(\omega_{j,1} \tilde{x}_1 + \omega_{j,2} \tilde{x}_2 + \psi_j \tilde{r} + \theta_{j} \right), \nonumber \\
\frac{\mathrm{d} \tilde{x}_2}{\mathrm{d} t} & = 
g_2(\tilde{x}_1, \tilde{x}_2; \tilde{r}) = \beta_2 + \tilde{x}_2 \sum_{j=1}^{5}\alpha_{2,j} 
\sigma_{\gamma_j, \tau_j} \left(\omega_{j,1} \tilde{x}_1 + \omega_{j,2} \tilde{x}_2 + \psi_j \tilde{r}+ \theta_{j} \right), 
\label{eq:single_layer_RRE_reduced_point_pulse}
\end{align}
 while the full ODEs read
 \small
\begin{align}
    \frac{\mathrm{d} x_1}{\mathrm{d}t} &= \beta_1 + x_1\left(\sum_{j=1}^{5} \alpha_{1,j}y_j\right), \;\;
    &&\frac{\mathrm{d} x_2}{\mathrm{d}t} = \beta_2 + x_2\left(\sum_{j=1}^{5} \alpha_{2,j}y_j\right), \nonumber \\
    \mu\frac{\mathrm{d} y_1}{\mathrm{d}t} &= \gamma_1 + \left( \sum_{i=1}^2 \omega_{1,i}x_i + \psi_1 r +\theta_{1} \right)y_1 -\tau_1 y_1^2, 
    \;\;&&\mu\frac{\mathrm{d} y_2}{\mathrm{d}t} = \gamma_2 + \left( \sum_{i=1}^2\omega_{2,i}x_i + \psi_2 r +  \theta_{2}\right)y_2 -\tau_2 y_2^2, \nonumber \\
    \mu\frac{\mathrm{d} y_3}{\mathrm{d}t} &= \gamma_3 + \left( \sum_{i=1}^2\omega_{3,i}x_i +  \psi_3 r + \theta_{3}\right)y_3 -\tau_3y_3^2,
    \;\;&&\mu\frac{\mathrm{d} y_4}{\mathrm{d}t} = \gamma_4 + \left( \sum_{i=1}^2\omega_{4,i}x_i + \psi_4r + \theta_{4}\right)y_4 -\tau_4y_4^2, \nonumber \\
    \mu\frac{\mathrm{d} y_5}{\mathrm{d}t} &= \gamma_5 + \left( \sum_{i=1}^2\omega_{5,i}x_i +\psi_5 r + \theta_{5}\right)y_5 -\tau_5y_5^2, 
    &&\frac{\mathrm{d} r}{\mathrm{d}t} = 0 
    \label{eq:full_rncrn_point_pulse}
\end{align}
\normalsize
Here, we assume general initial concentrations: $x_1(0)= a_1 \in [0.01, 3]$, $x_2(0)= a_2 \in [0.01, 3]$, and $y_1(0)= b_1 \geq 0$, $y_2(0)= b_2 \geq 0$, $y_3(0)= b_3 \geq 0$, $y_4(0)= b_4 \geq 0$, $y_5(0)= b_5 \geq 0$, and $r(0) \geq 0$. The coefficients were rounded to 3 decimal places to quote in text, the full precision coefficients are available in the code repository.
\\

{\textbf{\textit{Step 2.}  Quasi-static classification CRN.}} We now demonstrate the training of the classification component of the classification-controlled RNCRN described in Figure~\ref{fig:sac_pulse_target}(b). We define a target map from the input species $\Lambda_1$ to the bifurcation species as 
\begin{align}
    \bar{r}(\bar{\lambda_1}) = \begin{cases}
        1 &\text{ for all } \bar{\lambda}_1 \in [0, 0.25], \\
        0.1 &\text{ for all }\bar{\lambda}_1 > 0.25,
    \end{cases}
    \label{eq:target_Glu_R}
\end{align}
we approximate the target map, equation~(\ref{eq:target_Glu_R}),  by training a small feed-forward neural network with the chemical activation function, $\sigma_{\gamma}$, i.e.
\begin{align}
    \tilde{z}_1 & = \sigma_{\gamma}\left( \omega_{1}^{(z)}\tilde{\lambda_1} + \theta_1^{(z)}\right)\nonumber \\
    \tilde{z}_2 & = \sigma_{\gamma}\left( \omega_{2}^{(z)}\tilde{\lambda_1} + \theta_2^{(z)}\right)\nonumber \nonumber \\
    \tilde{z}_3 & = \sigma_{\gamma}\left( \omega_{3}^{(z)}\tilde{\lambda_1} + \theta_3^{(z)}\right)\nonumber \nonumber \\
    \tilde{r}& = \sigma_{\gamma}\left(\omega_{1}^{(r)}\tilde{z}_1 + \omega_{2}^{(r)}\tilde{z}_2 + \omega_{3}^{(r)}\tilde{z}_3 + \theta_{0}^{(r)}\right)
    \label{eq:feed_forward_pulse_neural_CRN}
\end{align}
an RRE with a steady state equivalent to equation~(\ref{eq:feed_forward_pulse_neural_CRN}) can be constructed at equilibrium from a series of chemical-perceptrons\cite{anderson_reaction_2021}. We introduce the same timescale separation parameter, $\mu$, as in the RNCRN so that we have the RRE 
\begin{align}
    \frac{\mathrm{d} \lambda_1}{\mathrm{d}t} &= 0, \nonumber  && \lambda_1(0) \geq 0\\
    \mu\frac{\mathrm{d} z_1}{\mathrm{d}t} &= \gamma + \left( \omega_{1}^{(z)}\lambda_1 + \theta_1^{(z)}\right)z_1-z_1^2, &&z_1(0) = c_1 \geq 0, \nonumber \\
    \mu\frac{\mathrm{d} z_2}{\mathrm{d}t} &= \gamma +  \left( \omega_{2}^{(z)}\lambda_1 + \theta_2^{(z)}\right) z_2 -z_2^2, &&z_2(0) = c_2 \geq 0, \nonumber \\
    \mu\frac{\mathrm{d} z_3}{\mathrm{d}t} &= \gamma +  \left( \omega_{3}^{(z)}\lambda_1 + \theta_3^{(z)}\right) z_3 -z_3^2, &&z_3(0) = c_3 \geq 0, \nonumber \\
    \mu\frac{\mathrm{d} r}{\mathrm{d}t} &= \gamma +\left(\omega_{1}^{(r)}z_1 + \omega_{2}^{(r)}z_2 + \omega_{3}^{(r)}z_3 + \theta_{0}^{(r)}\right)r -r^2, &&r(0) = d \geq 0.
    \label{eq:feed_forward_pulse_neural_CRN_RRE}
\end{align}
Using standard backpropogation techniques\cite{rumelhart_learning_1986} to train equation~(\ref{eq:feed_forward_pulse_neural_CRN}) using data sampled from equation~(\ref{eq:target_Glu_R}) we were able to find parameter values $\gamma = 1$, $\omega_{1}^{(z)}=-3.899$, $\omega_{2}^{(z)}=-4.832$, $\omega_{3}^{(z)}=-5.089$,  $\theta_1^{(z)}=10.786$, $\theta_2^{(z)}=11.864$, $\theta_3^{(z)}=12.527$, $\omega_{1}^{(r)}=5.960$, $\omega_{2}^{(r)}=-5.255$, $\omega_{3}^{(r)}=-7.505$, and $\theta_{0}^{(r)}=0.436$. 
\\

{\textbf{\textit{Step 3}. Dynamical approximation.}} We now combined the RREs to form the complete classification-controlled RNCRN. Using the rates constants defined in equation~(\ref{eq:coefficents_point_controlled_pulse}) and equation~(\ref{eq:feed_forward_pulse_neural_CRN_RRE}) such that the RREs are,
\begin{align}
    \frac{\mathrm{d} \lambda_1}{\mathrm{d}t} &= 0,
    &&\frac{\mathrm{d} x_1}{\mathrm{d}t} =\beta_1+ x_1\left(\sum_{j=1}^{5} \alpha_{1,j}y_j\right), \;\;
    \nonumber \\ \frac{\mathrm{d} x_2}{\mathrm{d}t}& = \beta_2 + x_2\left(\sum_{j=1}^{5} \alpha_{2,j}y_j\right),  && \mu\frac{\mathrm{d} z_1}{\mathrm{d}t} = \gamma + \left( \omega_{1}^{(z)}\lambda_1 + \theta_1^{(z)}\right)z_1-z_1^2, \nonumber \\
    \mu\frac{\mathrm{d} z_2}{\mathrm{d}t} &= \gamma +  \left( \omega_{2}^{(z)}\lambda_1 + \theta_2^{(z)}\right) z_2 -z_2^2, 
    &&\mu\frac{\mathrm{d} z_3}{\mathrm{d}t} = \gamma +  \left( \omega_{3}^{(z)}\lambda_1 + \theta_3^{(z)}\right) z_3 -z_3^2, \nonumber \\
    \mu\frac{\mathrm{d} r}{\mathrm{d}t} &= \gamma +\left(\omega_{1}^{(r)}z_1 + \omega_{2}^{(r)}z_2 + \omega_{3}^{(r)}z_3 + \theta_{0}^{(r)}\right)r -r^2, &&
    \mu\frac{\mathrm{d} y_1}{\mathrm{d}t} = \gamma_1 + \left( \sum_{i=1}^2 \omega_{1,i}x_i + \psi_1 r +\theta_{1} \right)y_1 -\tau_1 y_1^2, 
    \nonumber \\ \mu\frac{\mathrm{d} y_2}{\mathrm{d}t} &= \gamma_2 + \left( \sum_{i=1}^2\omega_{2,i}x_i + \psi_2 r +  \theta_{2}\right)y_2 -\tau_2 y_2^2, && \mu\frac{\mathrm{d} y_3}{\mathrm{d}t} = \gamma_3 + \left( \sum_{i=1}^2\omega_{3,i}x_i +  \psi_3 r + \theta_{3}\right)y_3 -\tau_3y_3^2,
    \nonumber \\ \mu\frac{\mathrm{d} y_4}{\mathrm{d}t} &= \gamma_4 + \left( \sum_{i=1}^2\omega_{4,i}x_i + \psi_4 r + \theta_{4}\right)y_4 -\tau_4y_4^2,    &&\mu\frac{\mathrm{d} y_5}{\mathrm{d}t} = \gamma_5 + \left( \sum_{i=1}^2\omega_{5,i}x_i +\psi_5 r + \theta_{5}\right)y_5 -\tau_5y_5^2. 
    \label{eq:full_rncrn_class_pulse}
\end{align}
The reduced vector field i.e. equation~(\ref{eq:full_rncrn_class_pulse}) as $\mu \rightarrow0$ is 
\begin{align}
\tilde{\lambda_1}(t) &= \tilde{\lambda_1}(0), \nonumber \\
    \tilde{z}_1 & = \sigma_{\gamma}\left( \omega_{1}^{(z)}\tilde{\lambda_1} + \theta_1^{(z)}\right)\nonumber \\
    \tilde{z}_2 & = \sigma_{\gamma}\left( \omega_{2}^{(z)}\tilde{\lambda_1} + \theta_2^{(z)}\right)\nonumber \nonumber \\
    \tilde{z}_3 & = \sigma_{\gamma}\left( \omega_{3}^{(z)}\tilde{\lambda_1} + \theta_3^{(z)}\right)\nonumber \nonumber \\
    \tilde{r}& = \sigma_{\gamma}\left(\omega_{1}^{(r)}\tilde{z}_1 + \omega_{2}^{(r)}\tilde{z}_2 + \omega_{3}^{(r)}\tilde{z}_3 + \theta_{0}^{(r)}\right), \nonumber \\
\frac{\mathrm{d} \tilde{x}_1}{\mathrm{d} t} & = 
g_1(\tilde{x}_1, \tilde{x}_2; r) = 4.251  + \tilde{x}_1 \sum_{j=1}^{5}\alpha_{1,j} 
\sigma_{\gamma_j, \tau_j} \left(\omega_{j,1} \tilde{x}_1 + \omega_{j,2} \tilde{x}_2 + \psi_j r + \theta_{j} \right), \nonumber \\
\frac{\mathrm{d} \tilde{x}_2}{\mathrm{d} t} & = 
g_2(\tilde{x}_1, \tilde{x}_2; r) = 5.494 + \tilde{x}_2 \sum_{j=1}^{5}\alpha_{2,j} 
\sigma_{\gamma_j, \tau_j} \left(\omega_{j,1} \tilde{x}_1 + \omega_{j,2} \tilde{x}_2 + \psi_j r+ \theta_{j} \right). 
\label{eq:reduced_class_control_pulse}
\end{align}

\section{Appendix: Training algorithms}
\label{app:training}
We now present various versions of the training algorithms used throughout this manuscript to train single-layer RNCRNs, parameterized RNCRNs, and classification-controlled RNCRNs.

\subsection{Algorithm 2. Two-step algorithm for training \emph{parametrized RNCRNs} to approximate bifurcations.}
\label{app:training_with_static_species}
In Section~\ref{sec:inherit_bifs_main} and Appendix~\ref{sec:app_bif}, we present examples of parametrized RNCRN trained by Algorithm~\ref{algh:para_RNCRN}.

\begin{table}[h]
\renewcommand\tablename{Algorithm}
\hrule
\vskip 2.5 mm
Fix a target system~(\ref{eq:target_ODE_w_general_static_species}),  target executive species compact sets 
$\mathbb{K}_1, \mathbb{K}_2, \ldots, \mathbb{K}_N \subset (0,+\infty)$ and target parameter species compact sets 
$\mathbb{L}\subset (0,+\infty)^L$.
Fix also the rate coefficients $\beta_1,\beta_2,\ldots,\beta_N \ge 0$ 
and $\gamma_1, \ldots, \gamma_M > 0$ in the parametrized RNCRN system~(\ref{eq:single_layer_RNCRN_static_param}). 
\begin{enumerate}
\item[\textbf{(a)}] \textbf{Quasi-static approximation}.
Fix a tolerance $\varepsilon > 0$.
Fix also the number of perceptrons $M \ge 1$. Using the backpropagation algorithm~\cite{rumelhart_learning_1986}, 
find the coefficients $\alpha_{i,j}^*, \theta_{j}^*, \omega_{j,i}^*, \psi_{j,l}^*$ 
for $i = 1, 2, \ldots, N$, $j = 1,2,\ldots,M$,  $l = 1,2,\ldots,L$,
such that (mean-square) distance 
between 
$(f_i(x_1,x_2,\ldots,x_N, \lambda_1, \ldots, \lambda_L) - \beta_i)/x_i$ and 
$\sum_{j=1}^M \alpha_{i,j}^* 
\sigma_{\gamma_j} \left(\sum_{k=1}^N 
\omega_{j,k}^* x_k + \sum_{l=1}^L 
\psi_{j,l}^* \lambda_l+ \theta_{j}^* \right)$
is within the tolerance for $(x_1,x_2,\ldots,x_N, \lambda_1, \ldots, \lambda_L) \in 
 \mathbb{K}_1 \times \mathbb{K}_2 \times \ldots \times\mathbb{K}_N\times \mathbb{L}$. If the tolerance $\varepsilon$ is not met,
then repeat step \textbf{(a)} with $M + 1$.

\item[\textbf{(b)}] \textbf{Dynamical approximation}.
Substitute $\alpha_{i,j} = \alpha_{i,j}^*$,
$\theta_{j} = \theta_{j}^*$, $\omega_{j,i} = \omega_{j,i}^*$, $\psi_{j,l} = \psi_{j,l}^*$ into the parametrized RNCRN~(\ref{eq:single_layer_RNCRN_static_param}). 
Fix the initial conditions $x_1(0),x_2(0),\ldots,x_M(0) \ge 0$, 
$y_1(0),y_2(0),\ldots,y_M(0) \ge 0$, ($\lambda_1(0), \lambda_2(0),\ldots, \lambda_L(0)) \in \mathbb{L}$, and time $T > 0$.
Fix also the speed $0 < \mu \ll 1$ of the perceptrons.
Numerically solve the target system~(\ref{eq:target_ODE_w_general_static_species})
and the parameterized RNCRN system~(\ref{eq:single_layer_RNCRN_static_param}) 
over the desired interval $[0, T]$.
Time $T > 0$ must be such that 
$\bar{x}_i(t), x_i(t)  \in \mathbb{K}_i$
for all $t \in [0,T]$ for $i = 1, 2,\ldots, N$.
If $\bar{x}_i(t)$ and $x_i(t)$ are sufficiently close
according to a desired criterion for all $i$, 
then terminate the algorithm.
Otherwise, repeat step \textbf{(b)} with a smaller $\mu$.
If no desirable $\mu$ is found, then go back to
step \textbf{(a)} and choose a smaller $\varepsilon$.
\end{enumerate}
\hrule
\caption{Algorithm 1 from~\cite{dack_recurrent_2025} is modified to include static executive species (i.e. parameter species).  In this modified training procedure, $\lambda_1, \dots, \lambda_L$ and $\psi_{j,l}$ are treated as additional inputs to the artificial neural network component of the algorithm in the same way as $x_1, \dots, x_M$ and $\omega_{j,i}$, and then one minimizes over the $x_1, \dots, x_M$ the target functions as usual. This is equivalent to removing any feedback to $\lambda_l$ and so setting $\mathrm{d} \lambda_l/\mathrm{d}t = 0$. See the accompanying code examples~\cite{Dack_github_2026} for more details.}
\label{algh:para_RNCRN}
\end{table}
\newpage

\subsection{Algorithm~\ref{algh:para_RNCRN_with_AD}. Two-step algorithm for training \emph{parametrized RNCRNs} to approximate piecewise systems.}
\label{app:alg_para_piece}
 In Section~\ref{sec:dynamical_piecewises} and Appendix~\ref{app:dynamical_piecewises}, we trained example systems using a modified version of Algorithm~\ref{algh:para_RNCRN} to approximate piecewise systems. Note, in this section we also trained the \emph{parameterized RNCRN} with automatic differentiation (AD) ~\cite{bradbury_jax_2026,deepmind_deepmind_2020} presented here, for clarity, as Algorithm~\ref{algh:para_RNCRN_with_AD}. This expanded algorithm uses the most general form of the parameterized RNCRN
\begin{align}
    \frac{\mathrm{d}x_i}{\mathrm{d}t} &= \beta_i + x_i\sum_{j=1}^M\alpha_{i,j} y_j,   &&\text{ for } i = 1, \dots, N, \nonumber\\
    \frac{\mathrm{d}\lambda_l}{\mathrm{d}t} &= 0,  && \text{ for }l= 1, \dots, L, \nonumber \\
     \frac{\mathrm{d}y_j}{\mathrm{d}t} &= \frac{\gamma_j}{\mu} + \left(\sum_{i=1}^N\frac{\omega_{j,i}}{\mu}x_i + \sum_{l=1}^L\frac{\psi_{j,l}}{\mu}\lambda_l+ \frac{\theta_{j}}{\mu} \right)y_j - \frac{\tau_j}{\mu}y_j^2, && \text{ for } j = 1, \dots, M,
    \label{eq:single_layer_RNCRN_static_param_w_AD}
\end{align}
where $x_i = x_i(t) \geq 0$, $y_j = y_j(t) \geq 0$,  $\lambda_l = \lambda_l(t) = \lambda_l(0) \geq 0$, $\beta_i\geq0$, $\alpha_{i,j}\in \mathbb{R}$, $\gamma_j > 0$, $\tau_j > 0$, $\omega_{ji} \in \mathbb{R}$, $\theta_{j} \in \mathbb{R}$, and $\psi_{j,l}\in \mathbb{R}$.

 These AD techniques are just an alternative method for optimizing the reduced vector field so they do not change the conceptual purpose of the quasi-static approximation step. However, they allow for optimization over more reaction coefficients in the RNCRN. The use of AD techniques allowed the formulation of the optimization function over the reduced vector field directly, \textit{i.e. } $g_i(\tilde{x}_1, \tilde{x}_2, \dots, \tilde{x}_N) \approx f_i (\tilde{x}_1, \tilde{x}_2, \dots, \tilde{x}_N)$ rather than having to optimize over a transformed target vector field $(f_i(\tilde{x}_1,\tilde{x}_2,\ldots,\tilde{x}_N) - \beta_i)/\tilde{x}_i \approx \sum_{j=1}^M \alpha_{i,j}^* 
\sigma_{\gamma} \left(\sum_{k=1}^N 
\omega_{j,k}^* \tilde{x}_k + \theta_{j}^* \right) $. Therefore, several rate constants that we used to fix arbitrarily prior to the quasi-static approximation are now free to be optimized (i.e. $\beta_i$ and $\gamma$). Additionally, we allow the constant production reaction, $\gamma_j>0$, and the self-sequestration reaction, $\tau_j>0$, to take independent values for each reaction $j=1,2,\dots, M$.

Including these new parameters induces a slightly different chemical activation function in the reduced vector field that is defined as 
\begin{align}
    \sigma_{\gamma_j, \tau_j}(A) = \frac{1}{2\tau_j}\left( A^2 + \sqrt{A^2 + 4\tau_j\gamma_j }\right).
    \label{eq:full_chemical_activation_fn}
\end{align}

\begin{table}[h]
\renewcommand\tablename{Algorithm}
\hrule
\vskip 2.5 mm
Fix a target system~(\ref{eq:target_general_piecewise_dynamics}),  target executive species compact sets 
$\mathbb{K}_1, \mathbb{K}_2, \ldots, \mathbb{K}_N \subset (0,+\infty)$, and the disjoint sets $\mathbb{L}_1,
\mathbb{L}_2, \ldots,\mathbb{L}_P \subset (0, +\infty)^L$ in the parameter species space corresponding to conditional target functions $f_i^{(1)}, \ldots, f_i^{(P)}$.
\begin{enumerate}
\item[\textbf{(a)}] \textbf{Quasi-static approximation}.
Fix a tolerance $\varepsilon > 0$.
Fix also the number of perceptrons $M \ge 1$. Using the backpropagation algorithm~\cite{rumelhart_learning_1986} implemented via automatic differentiation~\cite{bradbury_jax_2026}, 
find the coefficients $\alpha_{i,j}^*, \theta_{j}^*, \omega_{j,i}^*, \psi_{j,l}^*, \beta_i^*, \gamma_j^*, \tau_j^*$ 
for $i = 1, 2, \ldots, N$, $j = 1,2,\ldots,M$,  $l = 1,2,\ldots,L$,
such that the (mean-square) distance for all $p=1,\ldots, P$ 
between 
$f_i^{(p)}(x_1,x_2,\ldots,x_N)$ and 
$\beta_i + x_i\sum_{j=1}^M \alpha_{i,j}^* 
\sigma_{\gamma_j, \tau_j} \left(\sum_{k=1}^N 
\omega_{j,k}^* x_k + \sum_{l=1}^L 
\psi_{j,l}^* \lambda_l + \theta_{j}^* \right)$
is within the tolerance for $(x_1,x_2,\ldots,x_N, \lambda_1, \ldots, \lambda_L) \in 
 \mathbb{K}_1 \times \mathbb{K}_2 \times \ldots \times\mathbb{K}_N\times \mathbb{L}_p$. If the tolerance $\varepsilon$ is not met,
then repeat step \textbf{(a)} with $M + 1$.

\item[\textbf{(b)}] \textbf{Dynamical approximation}.
Substitute $\alpha_{i,j} = \alpha_{i,j}^*$,
$\theta_{j} = \theta_{j}^*$, $\omega_{j,i} = \omega_{j,i}^*$, $\psi_{j,l} = \psi_{j,l}^*, \beta_i = \beta_i^*, \gamma_j =  \gamma_j^*, \tau_j = \tau_j^* $ into the parametrized RNCRN~(\ref{eq:single_layer_RNCRN_static_param_w_AD}). 
Fix the initial conditions $x_1(0),x_2(0),\ldots,x_M(0) \ge 0$, 
$y_1(0),y_2(0),\ldots,y_M(0) \ge 0$, time $T > 0$, and  $(\lambda_1(0), \lambda_2(0),\ldots, \lambda_L(0) )$ in any $\mathbb{L}_p$.
Fix also the speed $0 < \mu \ll 1$ of the perceptrons.
Numerically solve the target system~(\ref{eq:target_general_piecewise_dynamics})
and the parameterized RNCRN system~(\ref{eq:single_layer_RNCRN_static_param_w_AD}) 
over the desired interval $[0, T]$.
Time $T > 0$ must be such that 
$\bar{x}_i(t), x_i(t)  \in \mathbb{K}_i$
for all $t \in [0,T]$ for $i = 1, 2,\ldots, N$.
If $\bar{x}_i(t)$ and $x_i(t)$ are sufficiently close
according to a desired criterion for all $i$, 
then terminate the algorithm.
Otherwise, repeat step \textbf{(b)} with a smaller $\mu$.
If no desirable $\mu$ is found, then go back to
step \textbf{(a)} and choose a smaller $\varepsilon$.
\end{enumerate}
\hrule
\caption{Two-step algorithm for training the \emph{parametrized RNCRN} to approximate piecewise systems. See the accompanying code examples~\cite{Dack_github_2026} for more details.}
\label{algh:para_RNCRN_with_AD}
\end{table}
\newpage

\subsection{Algorithm~\ref{algh:class_controlled_RNCRN}. Algorithm for training the \emph{classification-controlled RNCRN}.}
\label{app:class_controlled_algo}

\begin{table}[h]
\renewcommand\tablename{Algorithm}
\hrule
\vskip 2.5 mm
Fix a target system~(\ref{eq:target_general_piecewise_dynamics}),  target executive species compact sets 
$\mathbb{K}_1, \mathbb{K}_2, \ldots, \mathbb{K}_N \subset (0,+\infty)$, and the disjoint sets $\mathbb{L}_1,
\mathbb{L}_2, \ldots,\mathbb{L}_P \subset (0, +\infty)^L$ in the parameter species space corresponding to conditional target functions $f_i^{(1)}, \ldots, f_i^{(P)}$.
\begin{enumerate}
\item[\textbf{(a)}] \textbf{Point-wise parametrized RNCRN}. Apply Algorithm~\ref{algh:para_RNCRN_with_AD} to find a $L=1$ parametrized RNCRN that bifurcates between the $P$ conditional target functions using $P$ points, $r_1, \ldots, r_p$, selected arbitrarily from $(0, +\infty)$. Substitute the resulting parameters into~(\ref{eq:modular_sac_RRE}) as $\alpha_{i,j} = \alpha_{i,j}$,
$\theta_{j}^{(y)}=\theta_{j}$, $\omega_{j,i}^{(y)} = \omega_{j,i} , \psi_{j} = \psi_{j,1}, \beta_i = \beta_i, \gamma_j = \gamma_j$, and $ \tau_j=\tau_j$.

\item[\textbf{(b)}] \textbf{Quasi-static classification CRN}.
Fix a tolerance $\varepsilon > 0$.
Fix also the number of perceptrons $K \ge 1$ and $\gamma > 0 $. Using the backpropagation algorithm~\cite{rumelhart_learning_1986} 
find the coefficients $\omega_{k,l}^{*(z)}, \theta_{k}^{*(z)}, \omega_{k}^{*(r)},  \theta_{0}^{*(r)}$
for $l = 1, 2, \ldots, L$, $k = 1,2,\ldots,K$, 
such that (mean-square) distance 
between 
$ r_p$ and 
$\sigma_\gamma \left(\sum_{k=1}^K \omega_{k}^{*(r)} \sigma_\gamma  \left(\sum_{l=1}^L \omega^{*(z)}_{k,l} \lambda_l + \theta_{k}^{*(z)} \right) + \theta_{0}^{*(r)} \right)$ 
is within the tolerance for all for all $(\lambda_1,\ldots, \lambda_L)\in\mathbb{L}_p$ for all $p=1,\dots,P$. If the tolerance $\varepsilon$ is not met,
then repeat step \textbf{(b)} with $K + 1$.

\item[\textbf{(c)}] \textbf{Dynamical approximation}.
Substitute $\omega_{k,l}^{(z)} = \omega_{k,l}^{*(z)}, \theta_{k}^{(z)} = \theta_{k}^{*(z)}, \omega_{k}^{(r)} = \omega_{k}^{*(r)},  \theta_{0}^{(r)} = \theta_{0}^{*(r)}$ into~(\ref{eq:modular_sac_RRE_0}) and couple with~(\ref{eq:modular_sac_RRE}) with parameters found in \textbf{(a)}. 
Fix the initial conditions $x_1(0),x_2(0),\ldots,x_M(0) \ge 0$,  
$y_1(0),y_2(0),\ldots,y_M(0) \ge 0$, $z_1(0),z_2(0),\ldots,z_K(0) \ge 0$, $r(0) \ge 0$, time $T > 0$, and  $(\lambda_1(0), \lambda_2(0),\ldots, \lambda_L(0) )$ in any $\mathbb{L}_p$.
Fix also the speed $0 < \mu \ll 1$ of the perceptrons.
Numerically solve the target system~(\ref{eq:target_general_piecewise_dynamics})
and the classification-controlled RNCRN system~(\ref{eq:modular_sac_RRE})--(\ref{eq:modular_sac_RRE_0}) 
over the desired interval $[0, T]$.
Time $T > 0$ must be such that 
$\bar{x}_i(t), x_i(t)  \in \mathbb{K}_i$
for all $t \in [0,T]$ for $i = 1, 2,\ldots, N$.
If $\bar{x}_i(t)$ and $x_i(t)$ are sufficiently close
according to a desired criterion for all $i$, 
then terminate the algorithm. Otherwise, repeat step \textbf{(b)} with a smaller $\mu$.
If no desirable $\mu$ is found, then go back to
step \textbf{(a)} and choose a smaller $\varepsilon$.
\end{enumerate}
\hrule
\caption{An algorithm to train \emph{classification-controlled RNCRN} to approximate piecewise systems of the form~(\ref{eq:target_general_piecewise_dynamics}). See the accompanying code examples~\cite{Dack_github_2026} for more details.}
\label{algh:class_controlled_RNCRN}
\end{table}

\subsection{Algorithm~\ref{algh:data_def_RNCRN_with_AD}. Two-step algorithm for training the \emph{RNCRN} to approximate data-defined limit cycles.}
\label{app:data_def_RNCRN_with_AD}
 In Section~\ref{sec:data_defined_dynamics} and Appendix~\ref{app:data_defined_attractors}, we trained example systems using a modified versions of Algorithm 1 from \cite{dack_recurrent_2025} on data from Algorithm~\ref{algh:define_attractor} in this manuscript.  For clarity, we present this as Algorithm~\ref{algh:data_def_RNCRN_with_AD} and note it involved both 
 automatic differentiation techniques and, due to not having an ODE system to validate trajectories, an abridged dynamical approximation step with an informal inspection-based termination.
 
We allow the constant production reaction, $\gamma_j>0$, and the self-sequestration reaction, $\tau_j>0$, to take independent values for each reaction $j=1,2,\dots, M$. The full RNCRN in this situation is 
\begin{align}
    \frac{\mathrm{d}x_i}{\mathrm{d}t} &= \beta_i + x_1\sum_{j=1}^M\alpha_{i,j}y_j,  &&  x_i(0) = a_i, &&&\text{ for } i = 1, \dots, N, \nonumber\\
     \frac{\mathrm{d}y_j}{\mathrm{d}t} &= \frac{\gamma_j}{\mu}+ \left(\sum_{i=1}^N\frac{\omega_{j,i}}{\mu}x_i +\frac{\theta_{j}}{\mu} \right)y_j - \frac{\tau_j}{\mu}y_j^2, &&  y_j(0) = b_j, &&&\text{ for } j = 1, \dots, M.
     \label{eq:full_rncrn_w_AD}
\end{align}

Including these new parameters induces a slightly different chemical activation function in the reduced vector field that is defined as in~(\ref{eq:full_chemical_activation_fn}).

\begin{table}[h]
\renewcommand\tablename{Algorithm}
\hrule
\vskip 2.5 mm
Fix target data generated by Algorithm~\ref{algh:define_attractor} (or otherwise) as $\mathbb{X} =\{ X^{(1)}, \ldots, X^{(D+1)}\}$, the set of points in executive species space, and $\mathbb{\dot{X}} = \{ \dot{X}^{(1)}, \ldots, \dot{X}^{(D+1)}\}$, the associated vector field for each point. Where $X^{(d)}\in \mathbb{R}_{\geq0}^{N}$ and $\dot{X}^{(d)}\in \mathbb{R}^{N}$ for $d=1, 2, \ldots, D+1$ and $i = 1, 2, \ldots, N$. 
\begin{enumerate}
\item[\textbf{(a)}] \textbf{Quasi-static approximation}.
Fix a tolerance $\varepsilon > 0$.
Fix also the number of perceptrons $M \ge 1$. Using the backpropagation algorithm~\cite{rumelhart_learning_1986} implemented via automatic differentiation~\cite{bradbury_jax_2026}, 
find the coefficients $\alpha_{i,j}^*, \theta_{j}^*, \omega_{j,i}^*, \beta_i^*, \gamma_j^*, \tau_j^*$ 
for $i = 1, 2, \ldots, N$, $j = 1,2,\ldots,M$, 
such that (mean-square) distance 
between 
$ \dot{X}^{(d)}_i$ and 
$\beta_i + x_i^{(d)}\sum_{j=1}^M \alpha_{i,j}^* 
\sigma_{\gamma_j, \tau_j} \left(\sum_{k=1}^N 
\omega_{j,k}^* x_k^{(d)} + \theta_{j}^* \right)$
is within the tolerance for all for $d=1, 2, \ldots, D+1$ and $i = 1, 2, \ldots, N$. If the tolerance $\varepsilon$ is not met,
then repeat step \textbf{(a)} with $M + 1$.

\item[\textbf{(b)}] \textbf{Dynamical approximation}.
Substitute $\alpha_{i,j} = \alpha_{i,j}^*$,
$\theta_{j} = \theta_{j}^*$, $\omega_{j,i} = \omega_{j,i}^*, \beta_i = \beta_i^*, \gamma_j =  \gamma_j^*, \tau_j = \tau_j^* $ into the RNCRN~(\ref{eq:full_rncrn_w_AD}). 
Fix the initial conditions $x_1(0),x_2(0),\ldots,x_M(0) \ge 0$,  
$y_1(0),y_2(0),\ldots,y_M(0) \ge 0$, and time $T > 0$.
Fix also the speed $0 < \mu \ll 1$ of the perceptrons.
Numerically solve the RNCRN system~(\ref{eq:full_rncrn_w_AD}) 
over the desired interval $[0, T]$.
If, by inspection, the time-trajectory implements the intended dynamics then terminate the algorithm. Otherwise, repeat step \textbf{(b)} with a smaller $\mu$.
If no desirable $\mu$ is found, then go back to
step \textbf{(a)} and choose a smaller $\varepsilon$.
\end{enumerate}
\hrule
\caption{An algorithm for training an RNCRN to approximate dynamics from vector field data. Termination of this algorithm is based on an informal inspection of the time-trajectories as there is no ground-truth time-trajectories for these systems.  See the accompanying code examples~\cite{Dack_github_2026} for more details.}
\label{algh:data_def_RNCRN_with_AD}
\end{table}

\end{document}